\documentclass[]{aa}  

\usepackage{graphicx}
\usepackage{txfonts}
\usepackage{multirow}
\usepackage{dirtytalk}
\usepackage{amsmath}
\usepackage{xcolor}
\usepackage{soul}
\usepackage{enumitem}
\usepackage{booktabs}
\usepackage[rightcaption]{sidecap}
\usepackage{comment}
\usepackage{microtype}
\microtypesetup{protrusion=true,expansion=true}

\def \arcmin      {\text{$^\prime$}}
\def \arcsec      {\text{$^{\prime\prime}$}}

\usepackage{hyperref}
\hypersetup{
    colorlinks=false,
    linkcolor=blue,
    filecolor=magenta,      
    urlcolor=cyan,
    pdftitle={Overleaf Example},
    pdfpagemode=FullScreen,
    }

\newcommand{\appropto}{\mathrel{\vcenter{
  \offinterlineskip\halign{\hfil$##$\cr
    \propto\cr\noalign{\kern2pt}\sim\cr\noalign{\kern-2pt}}}}}
\begin{document} 

   \title{The MIni-Halos in MIni-Clusters (MIMIC) project}

   \subtitle{New insights into the origin of mini-halos from a LOFAR, uGMRT, and MeerKAT study of diffuse radio emission in Abell~496 and ZwCl~235}

   \author{F. Ubertosi
          \inst{1,2}\fnmsep\thanks{Contact: \texttt{francesco.ubertosi2@unibo.it}}
          \and
          S. Giacintucci\inst{3}
          \and
          A. Botteon\inst{2}
          \and
          A. Bonafede\inst{1,2}
          \and 
          T. Pasini\inst{2}
          \and
          M. Balboni\inst{1,4}
          \and
          N. Biava\inst{2}
          \and
          G.
          Brunetti\inst{2}
          \and
          T. Clarke\inst{3}
          \and
          V. Cuciti\inst{1,2}
          \and
          F. de Gasperin\inst{2}
          \and
          P. Domínguez-Fernández\inst{13,5}
          \and
          H. Edler\inst{6,7}
          \and
          F. Gastaldello\inst{4}
          \and
          M. Gitti\inst{1,2}
          \and
          K. Kolokythas\inst{8,9}
          \and
          G. V. Pignataro\inst{1}
          \and
          C. J. Riseley\inst{10,11}
          \and
          R. J. van Weeren\inst{12}
          \and
          T. Venturi\inst{2}
          \and
          J. Zuhone\inst{5}
          }

  \institute{Dipartimento di Fisica e Astronomia, Università di Bologna, via Gobetti 93/2, I-40129 Bologna, Italy,
         \and
             Istituto Nazionale di Astrofisica - Istituto di Radioastronomia (IRA), via Gobetti 101, I-40129 Bologna, Italy,
        \and
        U.S. Naval Research Laboratory, 4555 Overlook Avenue SW, Code 7213, Washington, DC 20375, USA,
        \and
        INAF - IASF Milano, via A. Corti 12, 20133 Milano, Italy,
        \and
        Center for Astrophysics | Harvard \& Smithsonian, 60 Garden Street, Cambridge, MA 02138, USA,
        \and
        ASTRON, Netherlands Institute for Radio Astronomy, Oude Hoogeveensedijk 4, 7991 PD Dwingeloo, The Netherlands,
        \and
        Hamburger Sternwarte, University of Hamburg, Gojenbergsweg 112, D-21029, Hamburg, Germany,
        \and
        Centre for Radio Astronomy Techniques and Technologies, Department of Physics and Electronics, Rhodes University,
        P.O. Box 94, Makhanda 6140, South Africa
        \and
        South African Radio Astronomy Observatory, 2 Fir Street, Observatory 7925, South Africa
        \and
        Astronomisches Institut der Ruhr-Universit\"{a}t Bochum (AIRUB), Universit\"{a}tsstra{\ss}e 150, 44801 Bochum, Germany,
        \and 
        Ruhr Astroparticle and Plasma Physics Center (RAPP Center), 44780 Bochum, Germany
        \and
        Leiden Observatory, Leiden University, PO Box 9513, 2300 RA Leiden, The Netherlands
        \and
        Instituto de Ciencias Nucleares, Universidad Nacional Aut\'onoma de M\'exico, A. P. 70-543 04510 D. F. Mexico 
             }

   \date{Received 28 April 2026 / Accepted 19 October 2026}

 
  \abstract
   {}
   {We aim at investigating the presence of diffuse radio emission in low-mass (M$_{500}\leq3\times10^{14}$~M$_{\odot}$), relaxed galaxy clusters.}
   {We present a pilot study for the MIMIC (MIni-halos in MIni-Clusters) project, focusing on Abell~496 and ZwCl~235, using LOFAR (54 MHz, 144 MHz), uGMRT (400 MHz, 650 MHz), and MeerKAT (1.28 GHz) data, as well as \textit{Chandra} X-ray data.}
   {
    We detect diffuse radio emission in the central $\sim$100~kpc of both Abell~496 and ZwCl~235, spatially confined within the sloshing-induced cold fronts seen in X-ray images. The diffuse sources satisfy the classification criteria for radio mini-halos and represent the least powerful known to date, with 1.4~GHz radio powers of $(2.2\pm0.2)\times10^{22}$~W/Hz in Abell~496 and $(1.3\pm0.3)\times10^{22}$~W/Hz in ZwCl~235. The mini-halos have steep spectra ($-1.9\leq\alpha\leq-1.4$), and in Abell~496 we find outward steepening of the spectral index. 
    In both clusters, the radio and X-ray surface brightness are correlated over the diffuse source extent, with sub-linear or nearly linear slopes depending on frequency.
    Separately, we present the properties of two extended ($>200$~kpc) tailed radio galaxies in both clusters.
}
   {The steep spectra, sub-linear radio–X-ray correlations, and confinement within cold fronts suggest that contribution from fossil plasma advected from the central radio galaxy and plausible turbulent re-acceleration are at play in Abell~496 and ZwCl~235. These two systems allow us to extend mini-halo scaling relations toward the low-power regime. We confirm a strong correlation between the mini-halo radio power and the cool core X-ray luminosity, with $P_{1.4} \propto L_{X,70}^{1.5\pm0.2}$. We find, for the first time, tentative indications that mini-halos with steeper radio spectra may preferentially be associated with lower core X-ray luminosities and lower radio power. We speculate that mini-halos may originate from a combination of hadronic and leptonic reacceleration mechanisms, whose relative importance varies with the properties of the host cluster: higher core X-ray luminosities at a given mass - reflecting higher gas densities -- might make the contribution of hadronic interactions more substantial.
   }


   \maketitle
%

\section{Introduction}\label{sec:intro}

In recent years, increasing attention has been devoted to diffuse radio emission in galaxy clusters, thanks to sensitive radio telescopes that revealed a growing number of cluster-scale diffuse sources (e.g., \citealt{vanweeren2019,knowles2022,botteon2022}). The origin of diffuse radio emission extending for hundreds of kpc remains an open question. 
The synchrotron cooling timescale of the electrons ($\sim10^{7-8}$~yr) is shorter than the typical particle transport timescales ($\sim$10$^{8-9}$~yr) and, as such, diffusion processes alone would fail to explain the 100s~kpc -- Mpc size of diffuse sources (e.g., \citealt{jaffe1977,bagchi2002}). Potential solutions to this \say{slow diffusion problem} involve additional mechanisms -- such as turbulent reacceleration of particles or in-situ (hadronic) production of electrons (e.g., \citealt{brunetti2014} for a review). These mechanisms have been extensively investigated in massive galaxy clusters (with M$_{500}\gtrapprox5\times10^{14}$~M$_{\odot}$), which satisfy the conditions to produce and sustain diffuse synchrotron radio sources (e.g., \citealt{giacintucci2017, vanweeren2019,knowles2019}). In the case of unrelaxed galaxy clusters, energetic mergers are known to be the origin of diffuse radio sources -- most notably radio halos and radio relics \citep{vanweeren2019}; in these cases shock and turbulent reacceleration are thought to be the dominant mechanisms. Similarly, relaxed galaxy clusters can show diffuse radio emission. The dense \say{cool-core} at the center of these systems is likely to host radio mini-halos (80\% of relaxed clusters with M$_{500}\geq6\times10^{14}$~M$_{\odot}$, see \citealt{giacintucci2017}), that are diffuse radio sources surrounding the central brightest cluster galaxy (BCG), with sizes of $\sim100-500$~kpc (e.g., \citealt{gitti2002,giacintucci2014,giacintucci2017,giacintucci2019,trehaeven2023}). Mini-halos are typically confined within sloshing-induced cold fronts (e.g., \citealt{giacintucci2017}), which supports their connection with weak turbulence in the central regions, and therefore a turbulent re-acceleration scenario for their origin (see \citealt{bravi2016}). At the same time, the radio surface brightness profiles are typically more peaked than the X-ray surface brightness ones, supporting an hadronic origin of the relativistic particles \citep{ignesti2020}. Therefore, their origin is still debated. The active galactic nucleus (AGN) in the BCG may also contribute to the origin of mini-halos by injecting electrons into the intracluster medium (ICM). These particles could be advected by gas motions and provide a reservoir of seed electrons for re-acceleration \citep{2021ApJ...914...73Z,vazza2024,dominguezfernandez2024}, while AGN-driven jets and bubbles may also generate turbulence in the cluster core \citep{sotira2025,bellomi2025}. In addition, the AGN may supply cosmic-ray protons relevant for hadronic scenarios \citep{pfrommer2004}.
\\ \indent In connection to this point, an open question on mini-halos is related to whether their radio powers scale with cluster mass (e.g., \citealt{kolokythas2025}). Literature studies have found contrasting evidence of such a scaling (see e.g., \citealt{giacintucci2019,laferriere2020}), and rather pointed to the connection of mini-halo power with the global or core cluster X-ray luminosities (\citealt{gitti2002,gitti2004,bravi2016,gitti2018,giacintucci2019,laferriere2020}). For radio halos in unrelaxed clusters, the observed scaling of radio power with mass and dynamics is consistent with turbulent re-acceleration models (e.g., \citealt{cassano2013,2025A&A...695A.180B}). Whether giant radio halos and mini-halos share similar scalings (and thus, origin of the relativistic particles) remains debated (e.g., \citealt{cassano2008,biava2024}). A potential source of uncertainty in establishing the existence of a radio power -- mass relation for mini-halos is that most known mini-halos are hosted in clusters with $M_{500}\geq5\times10^{14}$~M$_{\odot}$; only a few are known for $M_{500}\sim3-5\times10^{14}$~M$_{\odot}$ \citep{giacintucci2019,laferriere2020,kolokythas2025}, and evidence for mini-halos at lower cluster masses is still unclear\footnote{The mini-halo cluster 2A~0335+096 lies slightly above or below $3\times10^{14}$~M$_{\odot}$ depending on the reference ($3.4\times10^{14}$~M$_{\odot}$, \citealt{piffaretti2011}, or $2.3\times10^{14}$~M$_{\odot}$, \citealt{planck_2016_IIDatarelease}).} (see also \citealt{kolokythas2025}). 
\\\indent Therefore, populating the low mass regime of relaxed galaxy clusters with mini-halos could help in advancing our understanding of the origin and powering mechanism of these diffuse radio sources. The issue with low-mass clusters is that the radio emission is, on average, fainter. Scaling relations between the radio power of halos and mini-halos, and the X-ray luminosity or mass of the host cluster predict that at masses lower than $10^{14.5}$~M$_{\odot}$, the expected radio power at 1.4~GHz is $10^{22} - 10^{23}$~W/Hz \citep{giacintucci2019,laferriere2020,cuciti2021}. Over a scale of $\leq$100~kpc and for $z\leq0.15$, this implies a radio surface brightness of the order of $\geq 0.05~\mu$Jy/arcsec$^{2}$ around 1 GHz or $0.5~\mu$Jy/arcsec$^{2}$ around 100~MHz (for $\alpha=-1$), detectable only nowadays with new sensitive radio telescopes (e.g., MeerKAT, \citealt{jonas2016}; LOFAR, \citealt{vanHaarlem2013}; and uGMRT, \citealt{gupta2017}). 


\begin{figure*}
\sidecaption
    \centering
    \includegraphics[width=0.495\linewidth]{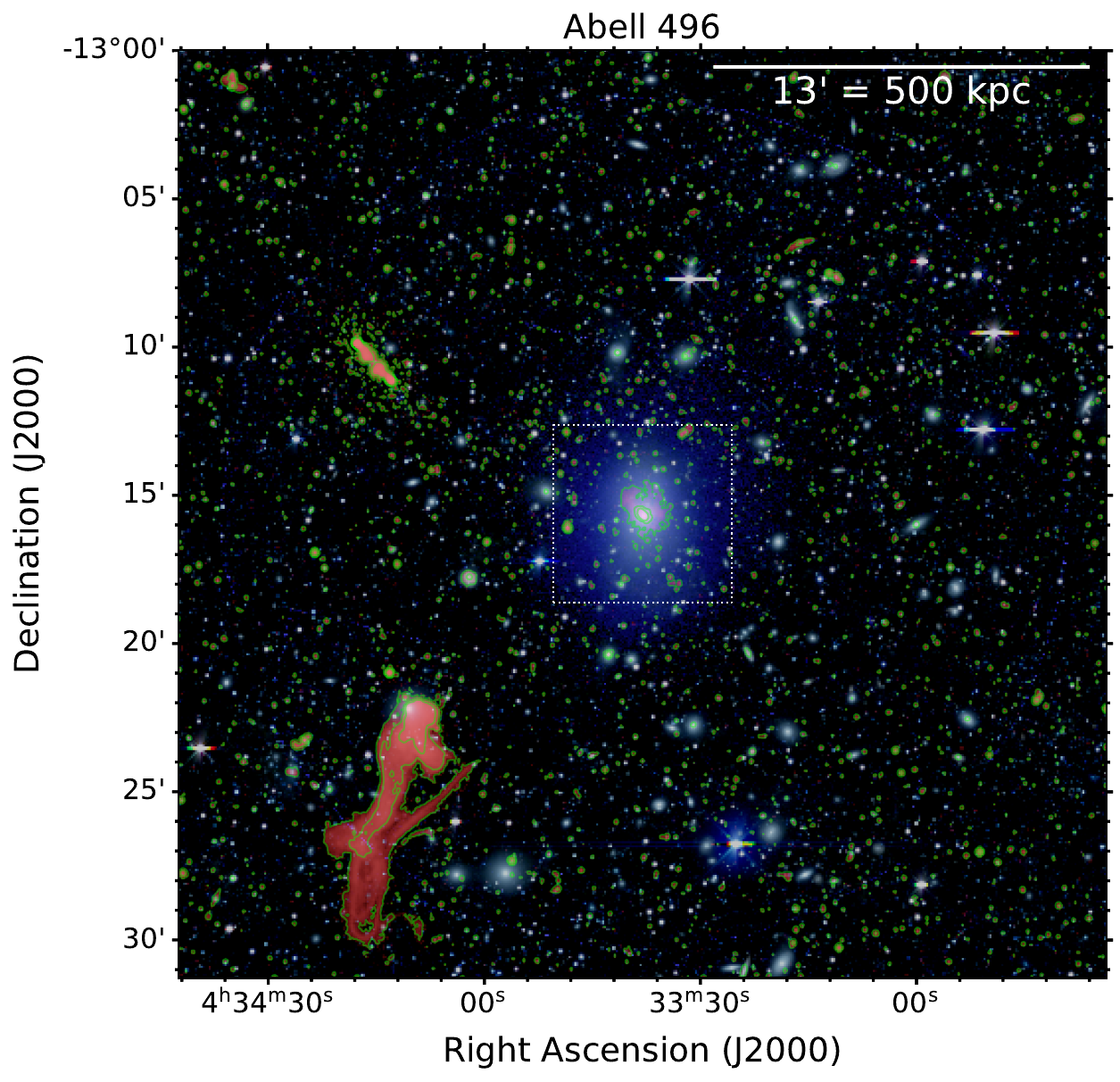}
    \includegraphics[width=0.495\linewidth]{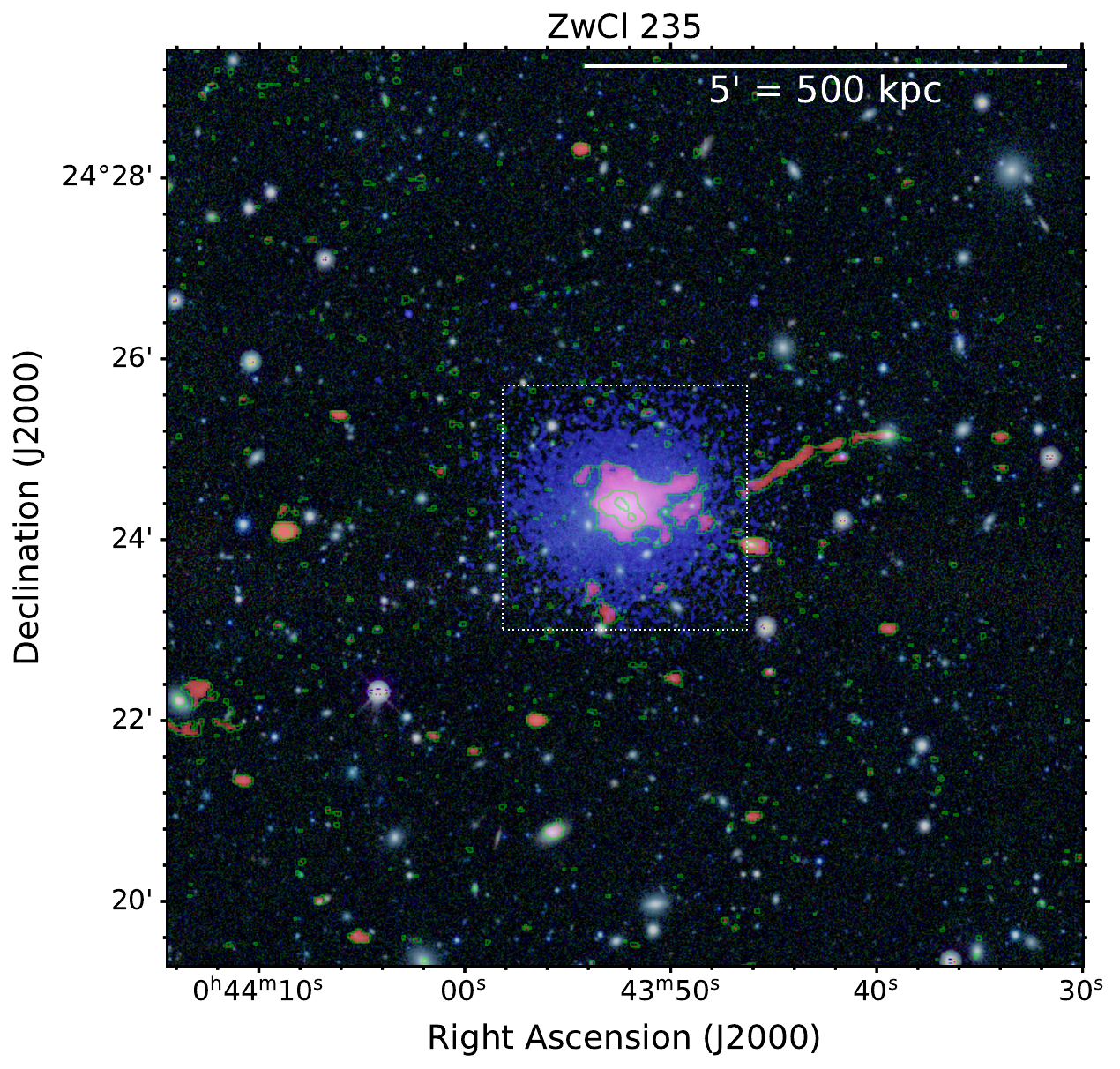}
    \caption{\textit{Left:} Color-composite image of Abell 496. Red traces radio emission detected with MeerKAT at L band, blue represents the X-ray emission from {\it Chandra} (convolved to a resolution of 20$\arcsec{\rm \sim13~kpc}$), and white is optical emission from the DESI survey \citep{desi2019}. We note the centrally-concentrated X-ray emission from the cool core and the presence of a spectacular, large radio galaxy to the south-east (see Appendix \ref{app:tailed}). The green contours are drawn at $3,\,30,\,300\sigma_{\rm rms}$ of the MeerKAT L band image shown in red (see Tab.~\ref{tab:radioimages}). The white dotted box encloses the central $3\arcmin$ of Abell~496, where our analysis is focused. \textit{Right}: Color-composite image of ZwCl 235. Red traces radio emission detected with LOFAR HBA, blue represents the X-ray emission from {\it Chandra} (convolved to a resolution of 3$\arcsec{\rm \sim5~kpc}$ -- this is smaller than the one adopted for Abell~496 to account for the higher redshift of ZwCl~235), and white is optical emission from the DESI survey \citep{desi2019}. A tailed radio galaxy is present west of the cool core (see Appendix \ref{app:tailed}). The green contours are drawn at $3,\,30,\,300\sigma_{\rm rms}$ of the LOFAR HBA image shown in red (see Tab.~\ref{tab:radioimages}). The white dotted box encloses the central $1\arcmin$ of ZwCl~235, where our analysis is focused.}
    \label{fig:largeA496Z235}
\end{figure*}

In order to investigate the presence of mini-halos in low-mass galaxy clusters,  we selected systems from the BCS \citep{ebeling1998} and Reflex \citep{bohringer2004} surveys with: 
\begin{enumerate}[label=(\alph*), noitemsep, leftmargin=*,topsep=0pt, nosep]
\item X-ray flux $\geq9\times 10^{-12}$ erg/cm$^{2}$/s (tracing a hot halo; e.g., \citealt{ettori2013});
\item H$\alpha$ luminosity of the BCG $\geq10^{40}$ erg/s (to select only cool cores, e.g., \citealt{2008ApJ...687..899R}; see \citealt{fornasiero2025} for a similar selection);
\item Mass M$_{500}$ between $1.5 - 3\times10^{14}$~M$_{\odot}$ (from \citealt{piffaretti2011,mcxc_ii}).
\end{enumerate}
The selected 13 systems (see Appendix \ref{app:mimics}) form the \say{MIMIC} (MIni-halos in MIni-Clusters) project sample.  

In this work, we present a pilot investigation of ZwCl 235 and Abell 496, two systems at the higher end of the mass range, to demonstrate the potential of this project based on targeted new radio observations. In the future, we will expand this study to the full sample using newly awarded MeerKAT L band observations as well as archival data. Abell~496 and ZwCl 235 are nearby, low-mass galaxy clusters ($M_{500} \sim 2.8\times10^{14}$~M$_{\odot}$, \cite{piffaretti2011}, $z<0.1$; see Appendix \ref{app:mimics}). They share some very similar properties: their central BCG is a radio galaxy with evidence of ongoing AGN activity in the core, and of aging radio lobes (tracing past outbursts) on tens of kpc scales (see \citealt{Hogan15b,ubertosi23AA,Ubertosi2024c,ubertosi2024} for both of them); based on X-ray observations, they both have a typical cool core, with central temperature $kT\sim$3~keV for Abell~496 \citep{2009ApJS..182...12C}, and $kT\sim2.5$~keV for ZwCl~235 \citep{ubertosi23AA}; and ultimately, they both host multiple nested cold fronts up to $\sim100$~kpc from the center, tracing ongoing ICM sloshing \citep{2012MNRAS.420.3632R,ubertosi23AA}.
\par Throughout this work we assume a $\Lambda$CDM cosmology with H$_{0}$=70~km~s$^{-1}$~Mpc$^{-1}$, $\Omega_{\text{m}}=0.3$, and $\Omega_{\Lambda}=0.7$. At the redshift of Abell~496 ($z=0.033$), this gives a scale of 0.656 kpc~arcsec$^{-1}$ and a luminosity distance $D_L = 144.5$~Mpc; at the redshift of ZwCl~235 ($z=0.083$), this cosmology gives a scale of 1.56 kpc~arcsec$^{-1}$ and a luminosity distance $D_L = 377.7$~Mpc. We define the radio spectral index $\alpha$ with the following convention: $S_{\nu} \propto \nu^{\alpha}$, where $S_{\nu}$ is the flux density at a frequency $\nu$. Unless otherwise stated, uncertainties are reported at 1$\sigma$.

\section{Data and analysis techniques}\label{sec:datareduction}

We consider multi-frequency radio observations from different facilities; for ZwCl 235 we analyze LOFAR Low Band Antenna (LBA, 54 MHz), LOFAR High Band Antenna (HBA, 144 MHz), and new uGMRT Band 3 (400 MHz) observations. For Abell 496 we analyze new uGMRT Band 4 (650 MHz) and new MeerKAT L band (1280 MHz) observations. We provide extensive information on the data calibration techniques we adopted to obtain calibrated visibilities in Appendix \ref{app:datareduction}.

Aiming for a radio and X-ray comparison, we also employ the {\it Chandra} X-ray images of the thermal gas in Abell~496 and ZwCl~235 presented in \citet{Ubertosi2024c} and \citet{ubertosi23AA}, respectively. We describe the procedures used to reduce the data and to produce X-ray images in Appendix~\ref{app:datareduction}. The images in the 0.5 - 7 keV energy band are used to explore the radio and X-ray spatial connection and surface brightness correlations (see Sect.~\ref{subsec:xray}).

\subsection{Radio data analysis}\label{subsec:analysisdata}
Imaging of the calibrated datasets for Abell 496 and ZwCl 235 was performed in \texttt{WSClean} v3.2, testing different weighting and deconvolution schemes. A summary of the parameters of the images presented in this work is reported in Tab. \ref{tab:radioimages}. When measuring flux densities $S_{\nu}$ from the images at frequency $\nu$, we computed the uncertainties $\delta S_{\nu}$ as:
\begin{equation}\label{eq:uncertflux}
    \delta S_{\nu} = \sqrt{N_{beam}\sigma_{\text{rms}}^{2} + (f_{\%}S_{\nu})^{2}},
\end{equation}
where $N_{beam}$ is the ratio between the area within which the flux density is computed and the beam area, $\sigma_{\text{rms}}$ is the rms noise, and $f_{\%}$ is the relative uncertainty on the flux density scale. We assumed $f_{\%} = 6\%$ (LOFAR LBA, \citealt{deGasperin2023}), $f_{\%} = 10\%$ (LOFAR HBA, \citealt{Shimwell22}), $f_{\%} = 10\%$ (uGMRT Band 3 and Band 4, \citealt{chandra2004}), and $f_{\%} = 5\%$ (MeerKAT L band, \citealt{knowles2022}). 

\begin{figure*}[ht!]
    \centering
    \includegraphics[width=0.8\linewidth]{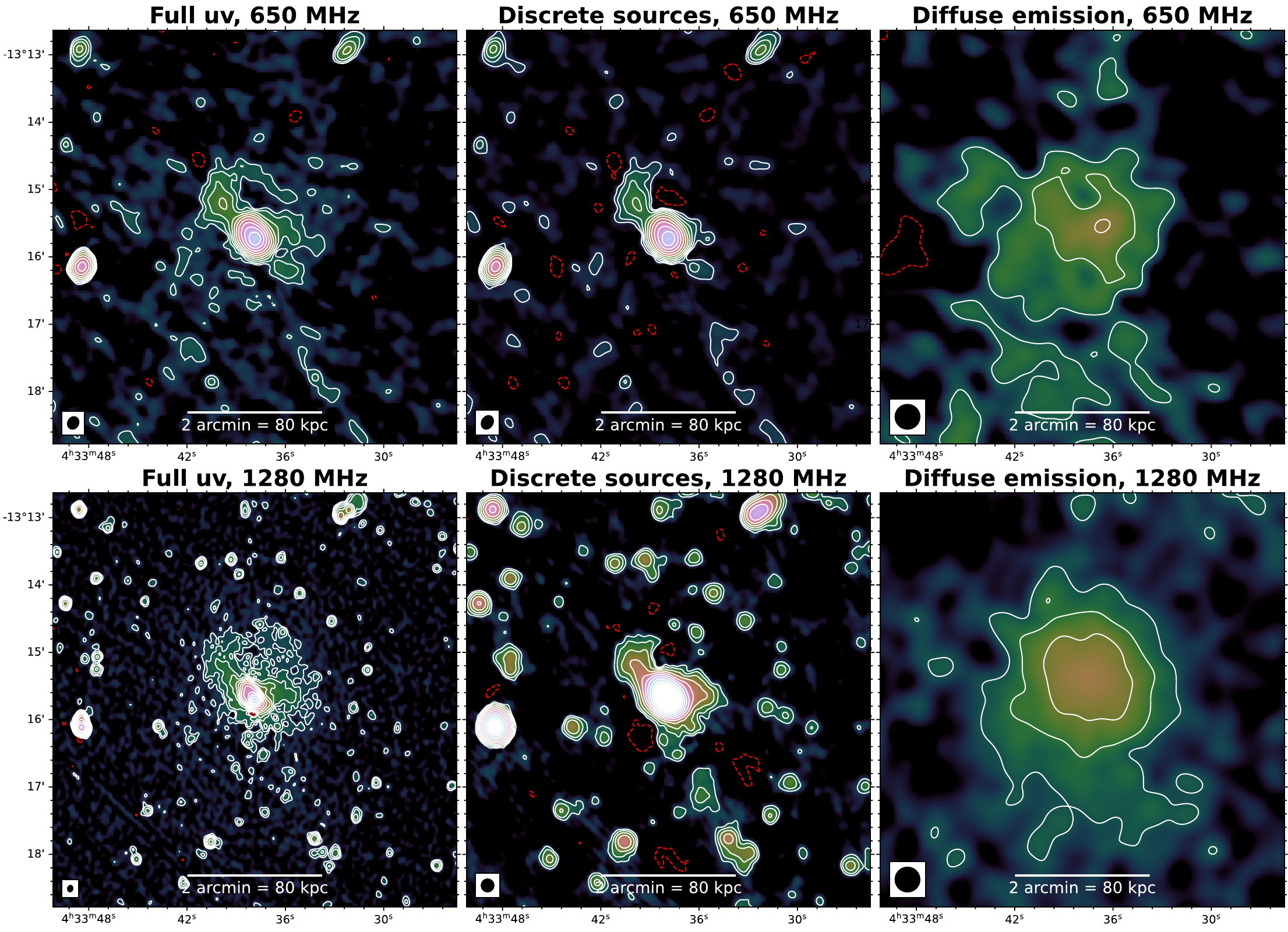}
    \caption{Radio observations of Abell~496. {\it Top panels:} uGMRT Band 4 (550 MHz - 750 MHz) images of the central $\sim5^{\prime}$ of Abell~496. {\it Bottom panels}: MeerKAT L band (856 MHz - 1712 MHz) images of the central $\sim5^{\prime}$ of Abell~496. From left to right, both rows: image obtained by imaging the full uv-range of visibilities ({\it left panels}), by imaging with an inner uv-cut of 2~k$\lambda$, to select discrete radio sources less than $L'=100\arcsec\sim70~{\rm kpc}$({\it central panels}), and after subtracting the discrete sources model from the full-uv dataset (\textit{right panels}). Details of the images, including the rms noise and angular resolution, are given in Tab. \ref{tab:radioimages}. In each panel, the color scale is logarithmic and starts from half the rms noise up to the peak in the image. Contours are drawn starting at 3$\sigma_{rms}$ and increase by a factor of 2 (the $-3\sigma_{rms}$ contour is shown in dashed red). We note that diffuse emission is already present in the full-uv images, but is detected at low significance due to the higher angular resolution and very large dynamic range ($\geq10^{5}$).}
    \label{fig:total_A496}
\end{figure*}

\begin{figure*}[ht!]
    \centering
    \includegraphics[width=0.8\linewidth]{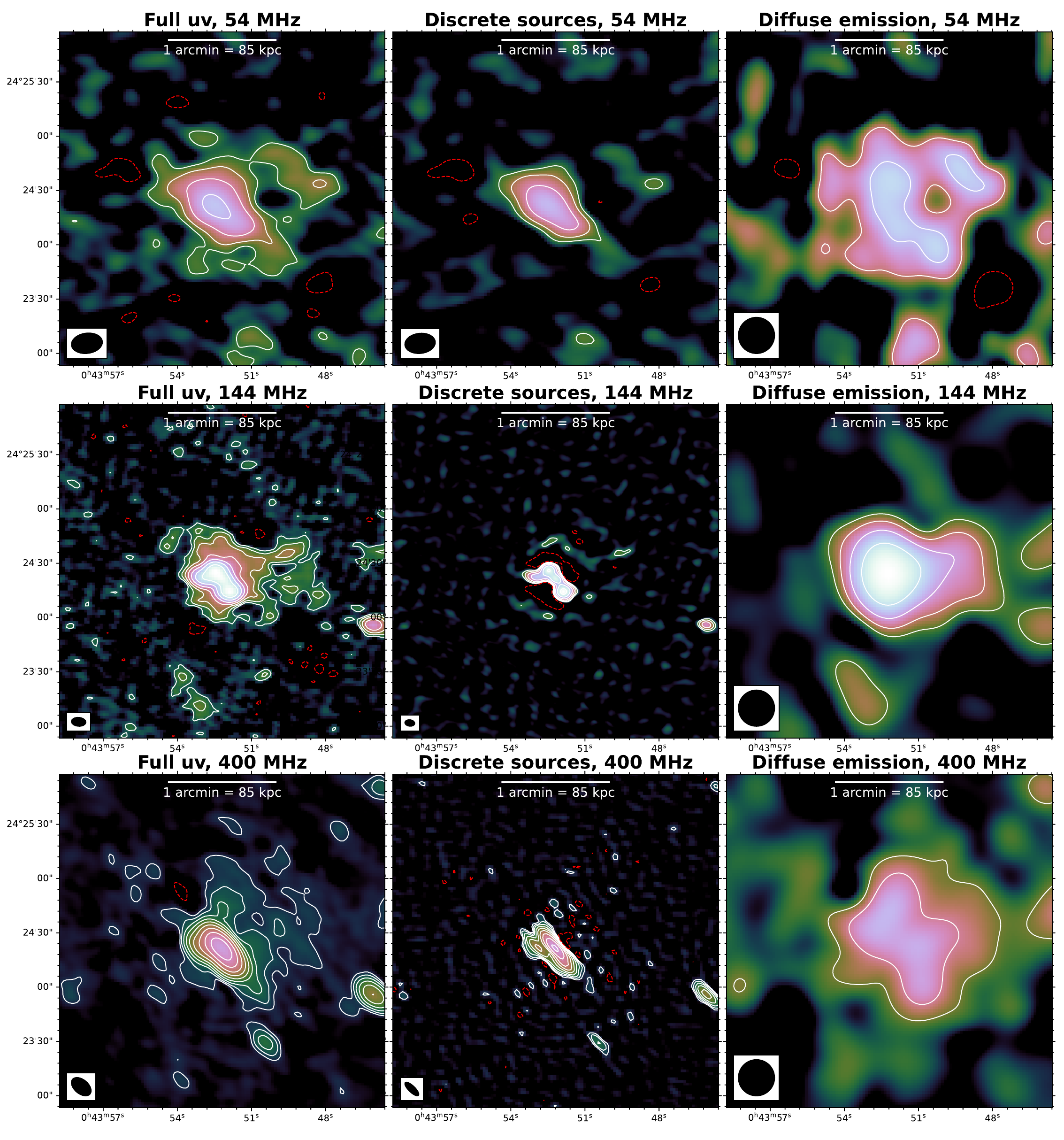}
    \caption{Radio observations of ZwCl 235. {\it Top:} LOFAR LBA (42 MHz - 66 MHz) images of the central $\sim3^{\prime}$ of ZwCl~235. {\it Middle:} LOFAR HBA (120 MHz - 168 MHz) images of the central $\sim3^{\prime}$ of ZwCl~235. {\it Bottom}: uGMRT Band 3 (300 MHz - 500 MHz) images of the central $\sim3^{\prime}$ of ZwCl~235. From left to right, all rows: image obtained by imaging the full uv-range of visibilities ({\it left panels}), by imaging with an inner uv-cut of 5~k$\lambda$, to select discrete radio sources less than $L'=40\arcsec\sim60~{\rm kpc}$ ({\it central panels}), and after subtracting the discrete sources model from the full-uv dataset (\textit{right panels}). Details of the images, including the rms noise and angular resolution, are given in Tab. \ref{tab:radioimages}. In each panel, the color scale is logarithmic and starts from half the rms noise up to the peak in the image. Contours start at 3$\sigma_{rms}$ and increase by a factor of 2 (the $-3\sigma_{rms}$ contour is shown in dashed red).}
    \label{fig:total_z235}
\end{figure*}

For both objects, images at intermediate angular resolution and sensitivity (Briggs weighting with \texttt{robust = -0.5}) of the full field of view were obtained using the full range of baselines. A wide-field version of these images from MeerKAT L band data (for Abell 496) and from LOFAR HBA data (for ZwCl 235) is combined with X-ray and optical images in Fig. \ref{fig:largeA496Z235}. Zoomed-in versions of the radio images at all frequencies are shown in the left panels of Fig. \ref{fig:total_A496} for Abell 496 and of Fig. \ref{fig:total_z235} for ZwCl 235. These images show radio emission on scales of $\sim$100~kpc at the position of the BCG, some extended sources around the field, and point sources throughout the field of view. 

\subsection{Subtraction of discrete radio sources}\label{subsubsec:subtraction}
Since our goal is to study diffuse radio emission, we first subtracted the discrete radio sources. This was achieved by producing images from data with a lower uv-cut, chosen to preserve emission only up to a given physical scale $L$ (which needs to be chosen \say{ad-hoc} in each cluster). The model derived from these "discrete-source" images was then subtracted from the visibilities in the uv domain, yielding residual visibilities that span the full uv-range but are sensitive only to radio emission on scales $L'\geq L$. This approach has been widely adopted in previous studies of cluster-scale diffuse radio emission (e.g., \citealt{botteon2022}); however, it is particularly critical in the case of cool-core clusters (e.g., \citealt{bonafede2023}), since these typically host a radio-loud BCG (e.g., \citealt{Hogan15b}), often displaying extended radio jets. These can reach spatial scales of 50$-$100 kpc, close to those of radio mini-halos (100$-$500 kpc), making the separation between the two components especially challenging. 
Our multi-frequency datasets provide both the angular resolution needed to identify AGN-related structures and the sensitivity required to detect faint, diffuse emission. In addition, previous studies of our targets \citep{ubertosi23AA,Ubertosi2024c} already provide constraints on the spatial scales of the discrete-source emission.

The BCG of Abell 496 hosts a radio-bright AGN ($\sim$100 mJy at 1-2 GHz, \citealt{Hogan15b}). This galaxy underwent at least three AGN outbursts in the last 0.1 Gyr with the oldest visible outburst appearing as ultra-steep spectrum ($\alpha\sim-3$) radio lobes on the northeast - southwest sides of the BCG \citep{Ubertosi2024c}. Based on previous 150 MHz and 330 MHz GMRT images, the full extent of the central radio galaxy reaches $L\sim100\arcsec{\rm \sim70~kpc}$ \citep{Ubertosi2024c}. This is also confirmed from the full-uv images of the target from the new MeerKAT and uGMRT data shown in the left panels of Fig. \ref{fig:total_A496}. Therefore, we produced a model of the discrete radio sources by separately imaging the Band 4 and L band data with a lower uv-cut of 2~k$\lambda$ (equivalent to 100$\arcsec{\rm \sim70~kpc}$) and setting \texttt{robust = +0.5}. We adopted this relatively natural weighting in the case of Abell~496 to better recover the flux density of the most extended and diffuse radio lobes associated with the BCG, at the expense of angular resolution. The resulting images (central panels of Fig. \ref{fig:total_A496}) correctly show radio emission from the central radio galaxy and discrete point sources only. Then, we subtracted this model from the original visibilities, effectively producing a new dataset where only the emission on scales $L'\geq100\arcsec\sim70$~kpc is sampled. The resulting visibilities were imaged with \texttt{robust = +0.5} and by applying a Gaussian taper of 20$\arcsec{\sim13~{\rm kpc}}$, to maximize the sensitivity to the diffuse emission. The resulting images, at a common angular resolution of 22${\arcsec\sim14~{\rm kpc}}$, are shown in the right panels of Fig.~\ref{fig:total_A496}.

A relatively bright radio galaxy ($150$~mJy at 144~MHz, \citealt{birzan2020}) is present at the center of ZwCl~235, with jets extending for 20$\arcsec$ ($\sim$30~kpc) on each side of the BCG \citep{ubertosi23AA}. This gives a scale of $L = 40"$ ($\sim$60~kpc) for the selection of discrete radio emission (based on the left panels of Fig. \ref{fig:total_z235}, the extent of the radio lobes does not vary with frequency). Therefore, we produced a model of the discrete radio sources by imaging the LBA, HBA, and Band 3 data with a lower uv-cut of 5~k$\lambda$ (equivalent to $40\arcsec\sim60$~kpc) and with \texttt{robust = -0.5}, obtaining the images shown in the central panels of Fig. \ref{fig:total_z235}. In this case, a more uniform weighting was adopted to achieve higher angular resolution, since the AGN-related emission in ZwCl~235 is relatively less resolved than in Abell~496 and does not require enhanced sensitivity to extended structures. The images show radio emission from the central radio galaxy and discrete point sources only. Then, we subtracted this model from the original visibilities, producing a new dataset where only the emission on scales $L'\geq40\arcsec\sim60$~kpc is sampled. The images obtained from the residual dataset with \texttt{robust = +0.5} are shown at a common resolution of 20$\arcsec\sim31~{\rm kpc}$ in the right panels of Fig. \ref{fig:total_z235}.

\section{Results}\label{sec:results}

\subsection{Diffuse radio emission in two low-mass cool cores}\label{subsec:diffuseemission}
A plethora of radio structures are visible in the two low-mass clusters (Fig. \ref{fig:largeA496Z235}). Within the core of Abell~496, the radio lobes of the central AGN dominate the radio emission, although hints of more extended emission are already visible (see also the left panels in Fig. \ref{fig:total_A496} and \ref{fig:total_z235}). On larger scales, a spectacular and large radio source is visible to the bottom left of Fig. \ref{fig:largeA496Z235} (left panel). This corresponds to the tailed radio galaxy PKS 0431-134 \citep{feretti1992}. In ZwCl 235, an elongated radio structure is visible on the west side of the cluster (see Fig. \ref{fig:largeA496Z235}, right panel); its rightmost tip sits on top of an elliptical member galaxy of ZwCl~235 \citep{rykoff2016}, suggesting this is also a tailed radio galaxy. We discuss the main properties of these sources in Appendix \ref{app:tailed}. In the following, we focus on the two cool cores. 
\\\indent In the discrete sources-subtracted images of Abell~496 and ZwCl~235 (right panels of Fig. \ref{fig:total_A496} and \ref{fig:total_z235}, respectively) we clearly identify diffuse radio emission at all frequencies. The existence of a diffuse source in these clusters has not been previously reported. In Abell~496, \citet{Ubertosi2024c} previously reported a non-detection of diffuse emission, but argued that a diffuse source below the sensitivity of the data available at the time ($4-12\,\mu{\rm Jy\,arcsec^{-2}}$ depending on frequency) could still be present. The new uGMRT and MeerKAT \say{Diffuse emission} images reach deeper sensitivities of $0.05 - 0.2\,\mu{\rm Jy\,arcsec^{-2}}$ (Tab.~\ref{tab:radioimages}) and indeed reveal evidence for a diffuse radio source. The diffuse emission (best visible in the MeerKAT L band image) is relatively amorphous, and seems to be elongated towards the south. It shows a radius of $r\sim50$~kpc {($\sim75\arcsec$)} in east-west, and of $r\sim100$~kpc {($\sim150\arcsec$)} in north-south. The shape of the central contour in the uGMRT Band 4 image resembles the morphology of the central radio galaxy, possibly suggesting that some residual discrete source emission might be present. In ZwCl~235, diffuse emission beyond the extent of the radio galaxy is visible from 54 MHz to 400 MHz; we note again some potential residual discrete source emission from the central BCG in the 54 MHz image. The overall shape is amorphous, but the 144 MHz image suggests a larger extent west of the cluster center ($r\sim55~{\rm kpc}\sim35\arcsec$ in north-south, and $r\sim80~{\rm kpc}\sim50\arcsec$ in east-west). We also note that there might be some residual RFI noise at 144 MHz (the subtle NW-SE linear pattern in Fig.~\ref{fig:total_z235}, middle left panel, possibly due to imperfect calibration). The local non-Gaussian rms noise excess associated with this feature is $\sim30\%$ in the \say{Full uv} image, and $\sim20\%$ in the \say{Diffuse emission} image. Given that the diffuse emission at 144~MHz is detected at high significance ($\geq3,6,12...\times\sigma_{\rm rms}$), and its integrated flux density uncertainty is dominated by the relative uncertainty on the flux density scale (see Table~\ref{tab:fluxtable} and below), this residual noise pattern, therefore, is not affecting the robustness of the detection of the diffuse radio emission.

Since the spatial scales of any central diffuse emission and the AGN structures in both Abell 496 and ZwCl 235 may be comparable, we performed a series of tests to assess how the flux densities and spectral indices of the recovered AGN emission and diffuse emission vary depending on the choice of the inner uv-cut adopted for the discrete-source model. These tests, shown in Appendix~\ref{app:testuv}, support our choices of 2~k$\lambda$ and 5~k$\lambda$ as the proper filtering scales for discrete source emission in Abell~496 and ZwCl~235, respectively.

\begin{figure*}[ht!]
    \centering
    \includegraphics[width=\linewidth]{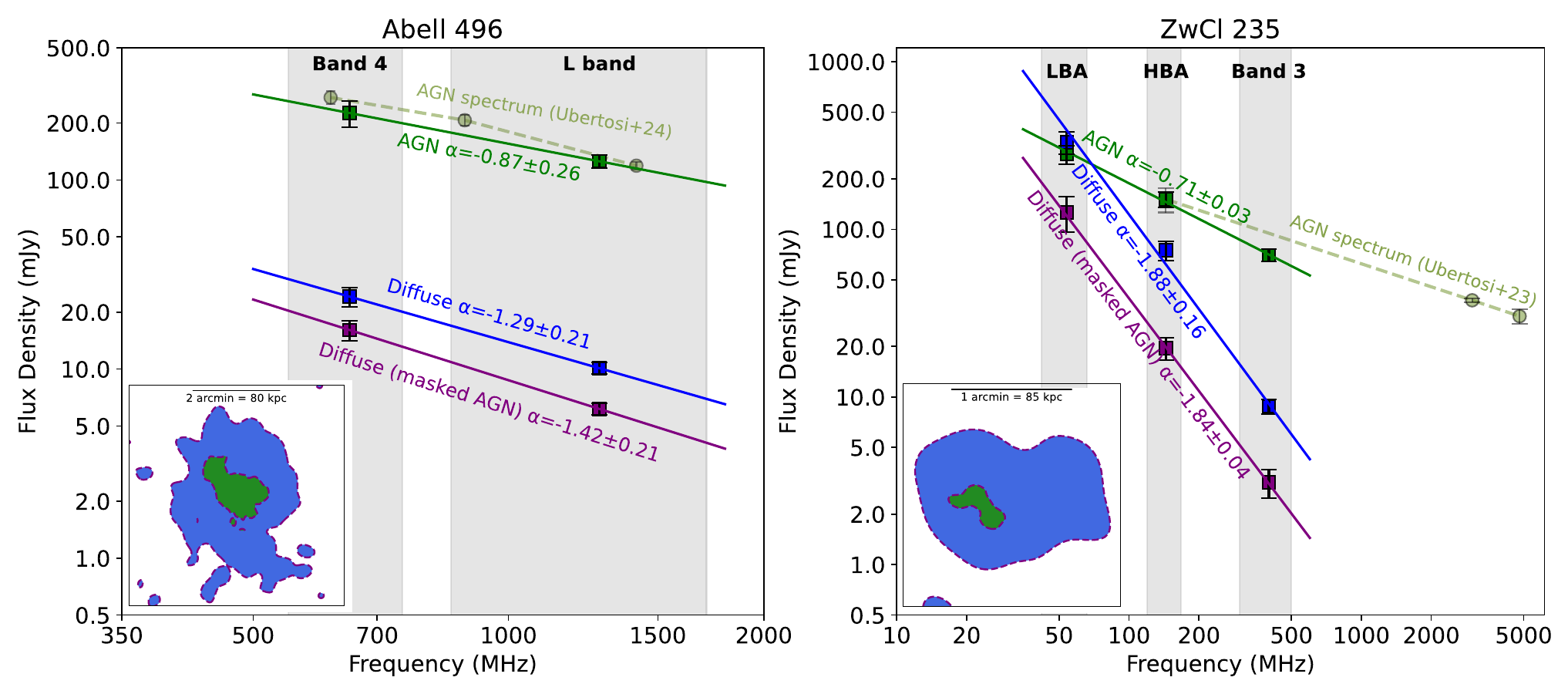}

    \caption{Synchrotron spectra of radio sources in Abell~496 (left) and ZwCl~235 (right). In both plots, we show the integrated spectrum of the central AGN (green), of the diffuse radio emission from the source-subtracted image (blue), and of the diffuse radio emission after masking the region where the central AGN is located (purple). We also highlight the frequency bands where the flux density measurements were taken. For Abell~496, the spectral slopes were computed analytically, while for ZwCl~235, they represent the best-fit to the three points. The bottom-left insets show the regions used to extract the different spectra and were obtained by plotting the contours shown in the images of Fig. \ref{fig:total_A496} for Abell~496 and Fig. \ref{fig:total_z235} for ZwCl~235. 
    For both systems, we also overlay in olive green the integrated radio spectrum of the BCG from the literature. For Abell~496, we plot the available flux density measurements from \citet{Ubertosi2024c} over the same frequency range as our measurements, while for ZwCl~235 the available data from \citet{ubertosi23AA} cover a higher frequency range. As noted in the main text, these flux density measurements represent integrated emission from multiple AGN outbursts, or the superposition of lobes and core components.
    }
    \label{fig:spix}
\end{figure*}

We measured the total flux density (above 3$\sigma$ confidence) of the central radio galaxy (from the \say{Discrete sources} image) and diffuse emission (from the \say{Diffuse emission} image) in both clusters as a function of frequency. We note that images of a given target or structure at different frequencies were produced using the same Briggs robust parameter (Tab.~\ref{tab:radioimages}). By definition, the \say{Discrete sources} images were generated using a common inner \textit{uv} cut. The \say{Diffuse emission} images were produced using the same \textit{uv} taper, convolved to a common restoring beam, and with a common inner \textit{uv} cut of $200\lambda$. We also measured the flux density of the diffuse emission in a region that excludes the central radio galaxy, to verify the effect of any potential contamination from the central radio galaxy. Specifically, we masked out the region enclosed by the 3$\sigma$ contour of the central radio galaxy from the area used to measure the diffuse emission. The diffuse-emission flux density was then integrated only over the remaining region, without extrapolating or interpolating the emission across the masked area (therefore, the resulting flux densities likely underestimate total ones). The spectra (and masks) are shown in Fig. \ref{fig:spix}. The flux density measurements are reported in Tab.~\ref{tab:fluxtable}. For Abell~496, we analytically compute the spectral index between 650 MHz and 1280 MHz. For ZwCl~235, we fit the three flux densities to measure the spectral index. 

As evident in Fig. \ref{fig:spix}, in Abell~496 we measure an AGN spectral index of $\alpha_{650}^{1280} = -0.87\pm0.16$. For reference, using the integrated literature flux density measurements in this frequency range for the radio BCG of Abell~496 in Table 4 of \citet{Ubertosi2024c} ($S_{617\text{MHz}} = 273\pm22$~mJy, $S_{888\text{MHz}} = 207\pm14$~mJy, and $S_{1415\text{MHz}}=119\pm6$~mJy), we would obtain spectral indices of $\alpha_{617}^{888} = -0.76\pm0.29$ and $\alpha_{888}^{1415} = -1.18\pm0.18$, clearly in agreement with our new estimate (see the olive green points in Fig.~\ref{fig:spix}, left).
For clarity, we note that the integrated flux density arises from the superposition of radio emission from three different outbursts (see \citealt{Ubertosi2024c}). The spectral index of the diffuse emission is, instead, steeper, $\alpha_{650}^{1280} = -1.42\pm0.18$. We do not observe an appreciable change in the spectral index if the region corresponding to the AGN is included ($\alpha_{650}^{1280} = -1.29\pm0.17$). In ZwCl~235, the integrated spectral index of the radio galaxy determined from fitting to the 54, 144 and 400 MHz measurements, $\alpha_{fit}^{54-400} = -0.68\pm0.03$, is very close to the typical injection index of relativistic electrons in radio AGN \citep{Harwood13}. For reference, using the integrated flux density measurements for the radio BCG of ZwCl~235 from \citet{ubertosi23AA} at higher frequencies ($S_{144\text{MHz}} = 151\pm24$~mJy, $S_{3000\text{MHz}} = 37.7\pm1.0$~mJy, and $S_{4800\text{MHz}}=30.3\pm3.0$~mJy, see the olive green points in Fig.~\ref{fig:spix}, left), we would obtain a spectral index of $\alpha_{144}^{3000} = -0.46\pm0.05$ and $\alpha_{3000}^{4800} = -0.47\pm0.22$. These values are flatter than the low-frequency spectral index estimated in this work, but in agreement with the emergence of a flat-spectrum core component at high frequencies \citet{ubertosi23AA}. The diffuse emission shows a steeper $\alpha_{fit}^{54-400} = -1.85\pm0.03$ (consistent, again, with the spectral index obtained by including the AGN region, $\alpha_{fit}^{54-400} = -1.77\pm0.15$), which can be described as an ultra-steep spectrum. We have also inspected spectral index maps, and found them to be very patchy (also due to the relatively coarse angular resolution of $20\arcsec$, especially for the case of ZwCl~235) and less informative than the integrated spectra discussed above or the point-to-point analysis presented below. For clarity, we stress that in both systems the AGN-masked and AGN-included spectral indices differ by only $\Delta\alpha\lesssim0.1$, corresponding to $\lesssim0.5\sigma$ given the associated uncertainties, and are therefore statistically consistent with one another. While the masked-AGN spectral indices shift in the sense expected if the region corresponding to the AGN retains some residual, flatter-spectrum emission associated with the radio galaxy itself, the low significance of this shift prevents any reliable interpretation.

\subsection{Thermal and non-thermal connection: the X-ray comparison}\label{subsec:xray}

\begin{figure*}[ht!]
    \centering
    \includegraphics[width=0.32\linewidth]{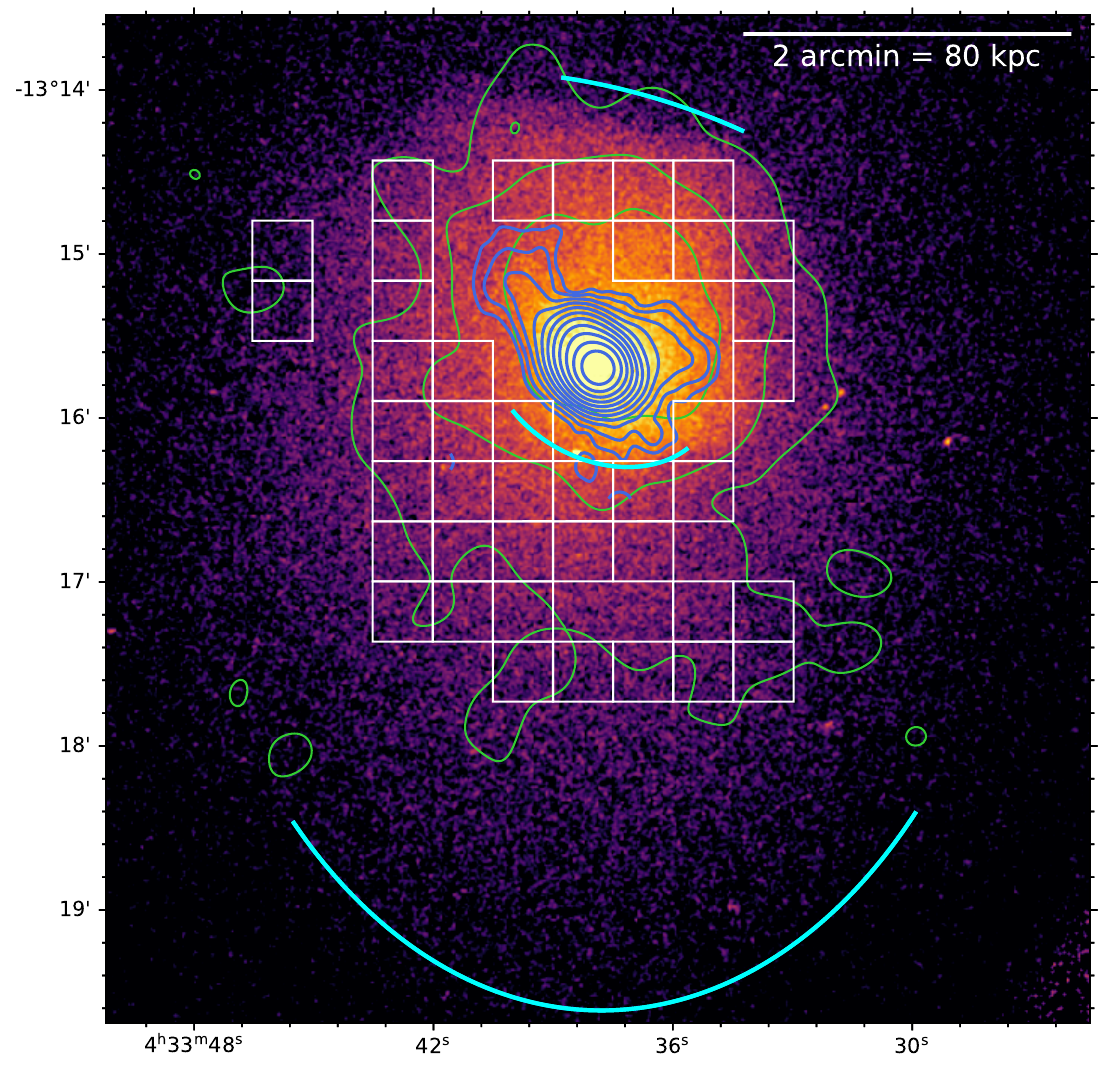}
    \includegraphics[width=0.32\linewidth]{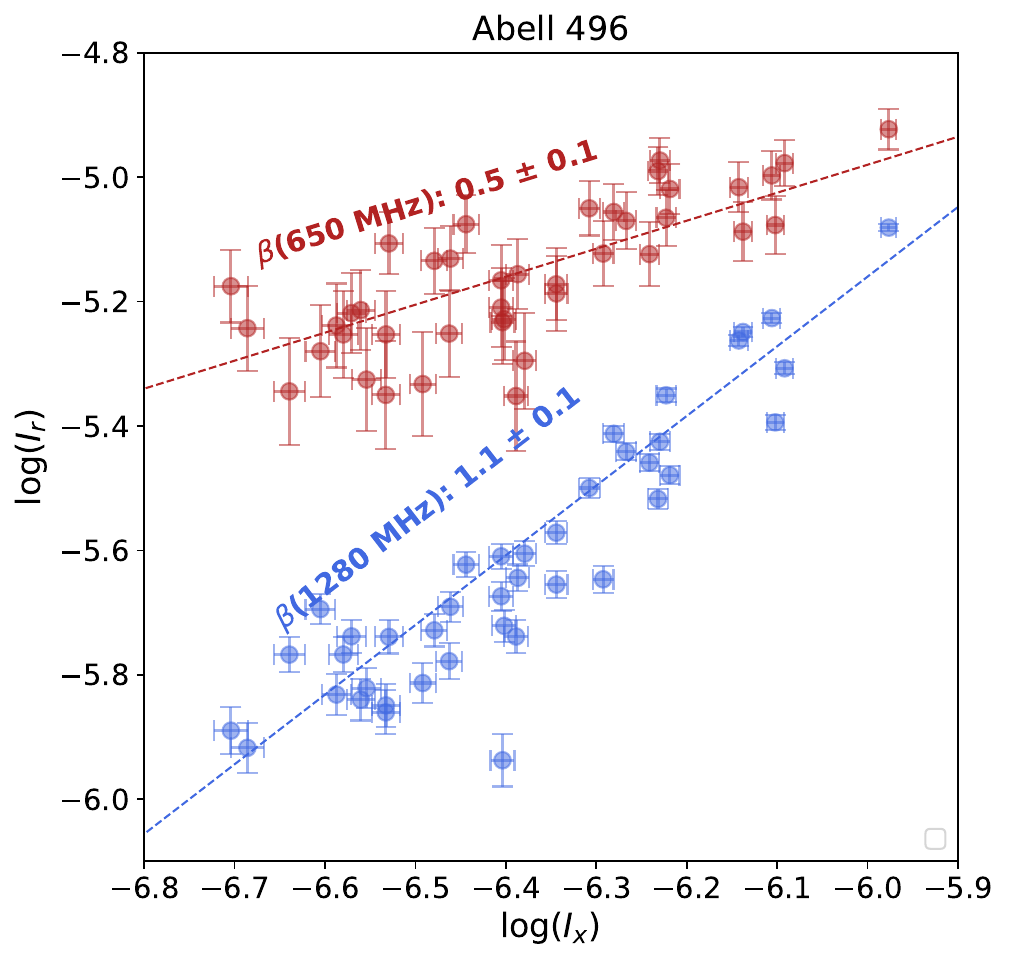}
    \includegraphics[width=0.32\linewidth]{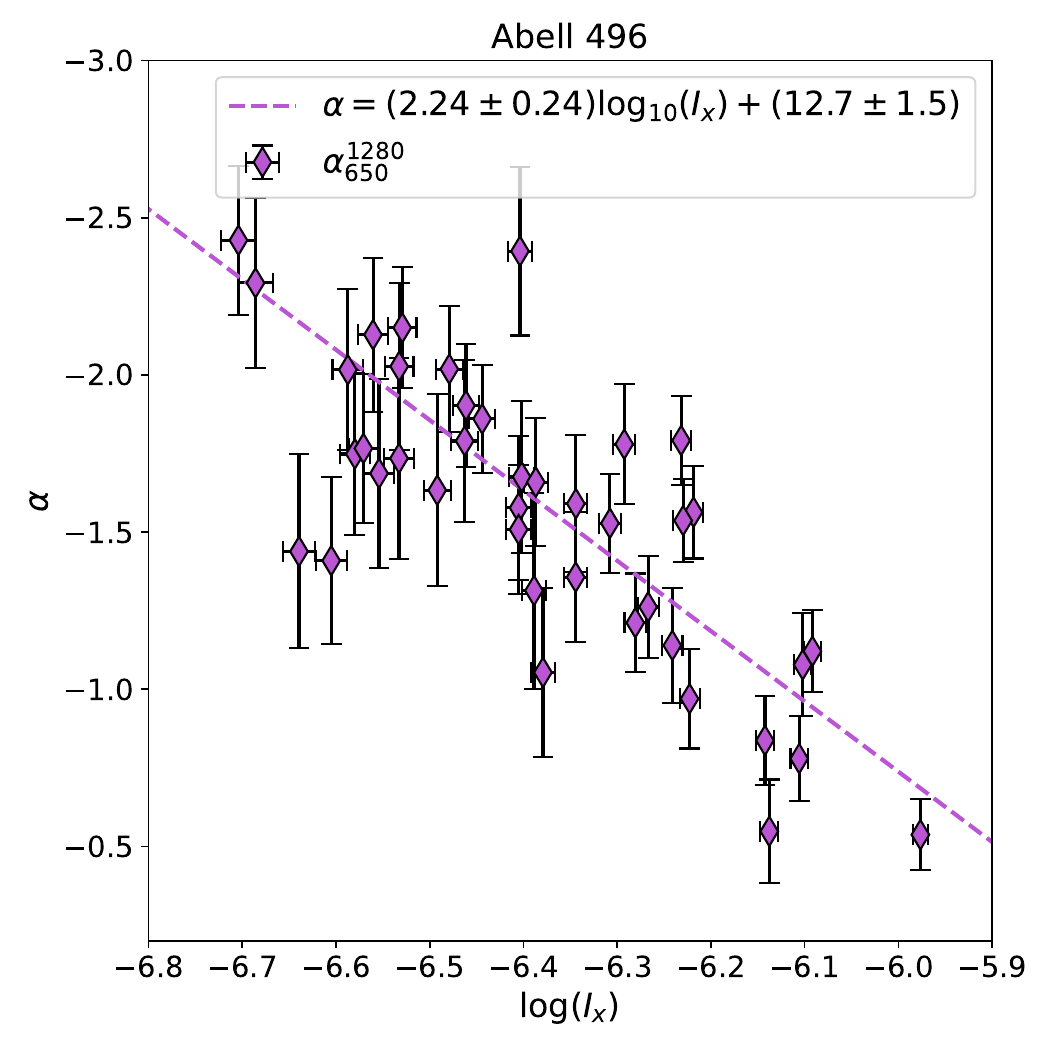}
    \caption{Thermal and non-thermal interaction in Abell~496. {\it Left}: 0.5 - 7 keV {\it Chandra} image of Abell~496, with green contours of the diffuse emission at 1280 MHz from MeerKAT data (as in Fig. \ref{fig:total_A496}, right). Cyan arcs show cold fronts in the ICM. 
    Blue contours trace the emission from the central radio galaxy, while the white boxes show the regions used to study the $I_r/I_x$ and $\alpha/I_x$ relations of the other panels. {\it Center}: point-to-point analysis of the X-ray and radio surface brightness of the diffuse radio emission at 650 MHz (red) and 1280 MHz (blue). The dashed lines represent the best fit to the two scatter plots, with the best-fit slopes annotated with matching color on the plot. {\it Right}: point-to-point analysis of the 650 MHz - 1280 MHz spectral index and X-ray surface brightness. The dashed line represents the best fit of the $\alpha/I_x$ relation (see \ref{subsec:xray} for details).}
    \label{fig:radioxA496}
\end{figure*}
\begin{figure*}[ht!]
    \centering
    \includegraphics[width=0.32\linewidth]{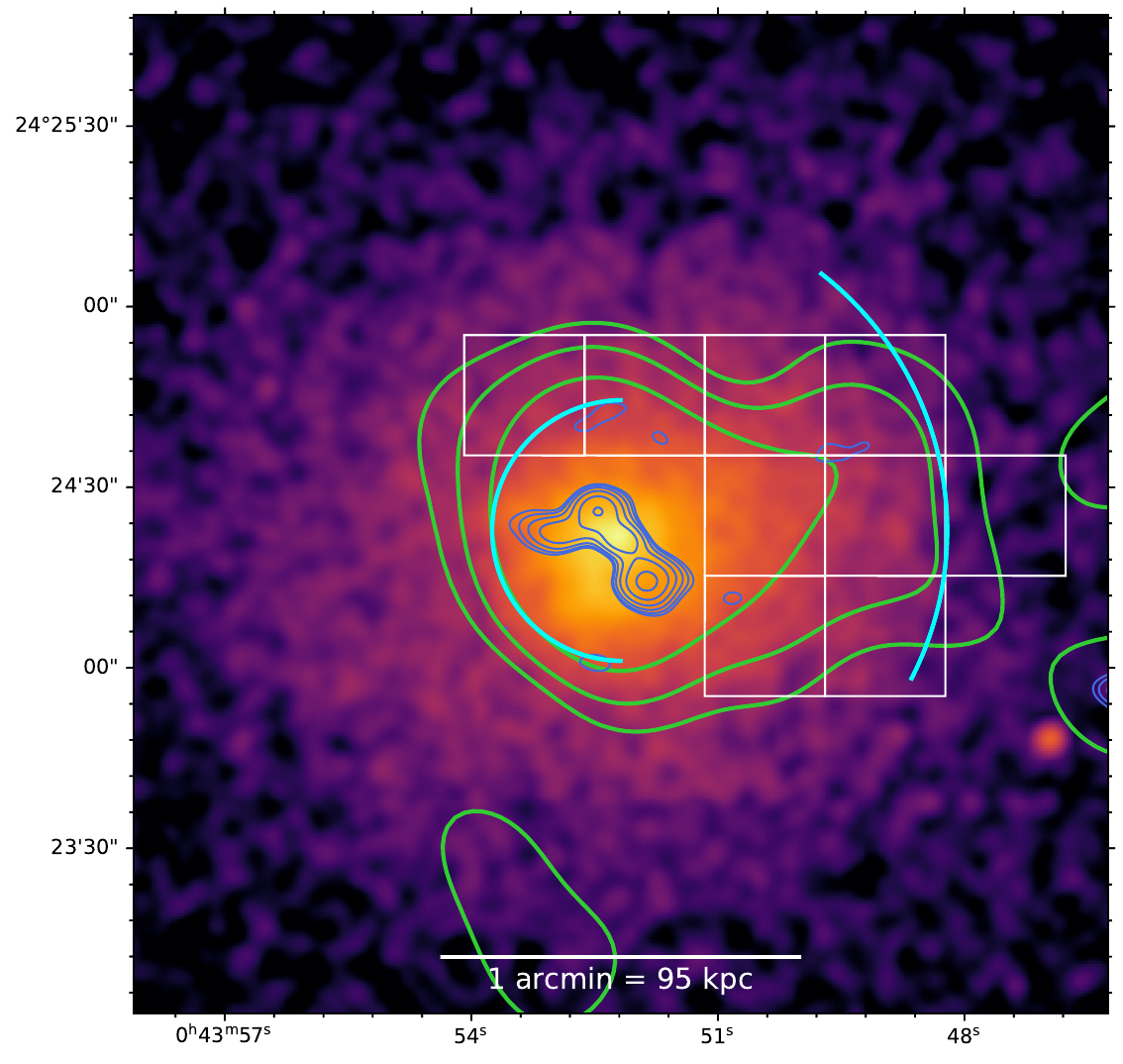}
    \includegraphics[width=0.32\linewidth]{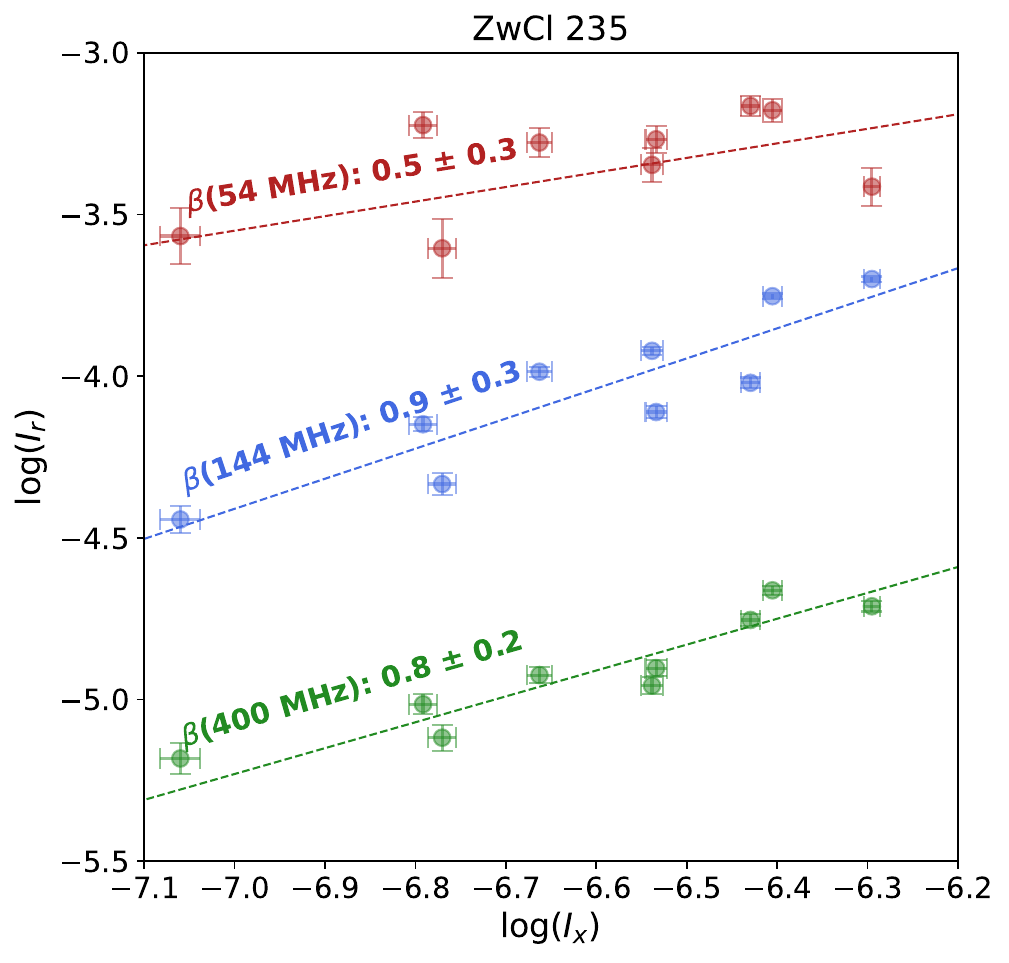}
    \includegraphics[width=0.32\linewidth]{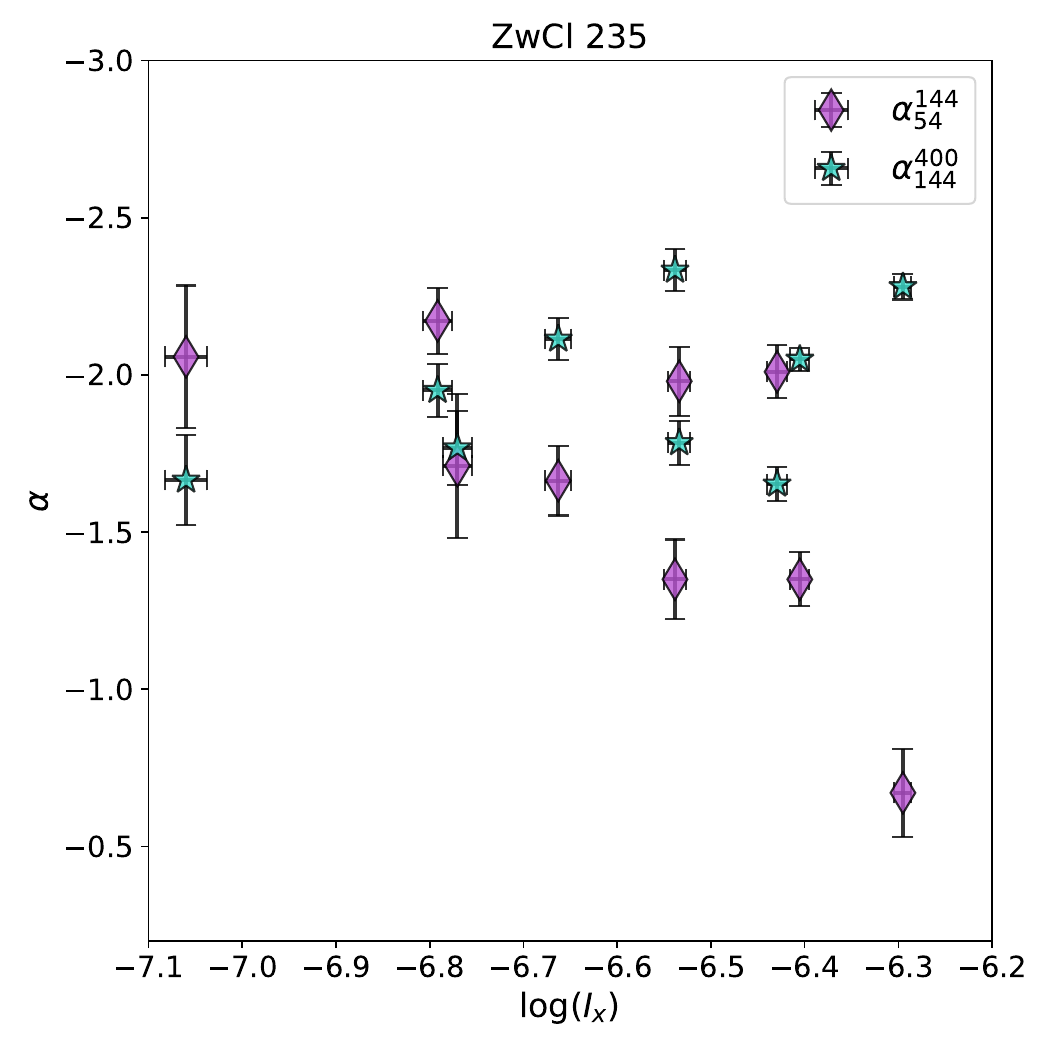}
    \caption{Thermal and non-thermal interaction in ZwCl~235. {\it Left}: 0.5 - 7 keV {\it Chandra} image of ZwCl~235, with green contours of the diffuse emission at 144 MHz from LOFAR HBA data (as in Fig. \ref{fig:total_z235}, right). Cyan arcs show two cold fronts in the ICM \citep{ubertosi23AA}. Blue contours trace the emission from the central radio galaxy, while the white boxes show the regions used to study the $I_r/I_x$ and $\alpha/I_x$ relations of the other panels. {\it Center}: point-to-point analysis of the X-ray and radio surface brightness of the diffuse radio emission at 54 MHz (red), 144 MHz (blue), and 400 MHz (green). The dashed lines represent the best fit to the scatter plots, with the best-fit slopes annotated with matching color on the plot. {\it Right}: point-to-point analysis of the 54 MHz - 144 MHz spectral index, the 144 MHz - 400 MHz spectral index, and X-ray surface brightness. No best-fit lines are overlaid due to the lack of a statistically meaningful correlation (see \ref{subsec:xray} for details).}
    \label{fig:radioxZ235}
\end{figure*}

We show in Fig. \ref{fig:radioxA496} (left) the contours of the diffuse radio emission at 1280 MHz in Abell~496 on the 0.5 - 7 keV {\it Chandra} exposure-corrected, background subtracted image of the ICM. We also overlay the position of the known cold fronts (see \citealt{2012MNRAS.420.3632R}).
A similar figure for ZwCl~235 is presented in Fig.~\ref{fig:radioxZ235}, where we plot the contours of the diffuse radio emission at 144~MHz and the position of known cold fronts (from \citealt{ubertosi23AA}) on the X-ray {\it Chandra} image. In both cases, the diffuse radio source is bounded by cold fronts. 
\\\indent The link between the diffuse radio emission and the underlying X-ray emitting gas can be further investigated by studying the relation between the radio and X-ray surface brightness, $I_{\rm R}\propto I_{\rm X}^{\beta}$ (e.g., \citealt{govoni2001,ignesti2020,riseley2022}). The slope of the correlation indicates how the distribution of cosmic-ray electrons and magnetic fields (related to $I_{\rm R}$) compares with the distribution of the density and temperature of the ICM (related to $I_{\rm X}$).
We investigated this topic with a similar approach for both clusters. 
\subsubsection{Radio-X-ray connection in Abell~496}\label{subsub:radioXA496} 
We extracted the radio surface brightnesses from the 650~MHz and 1280~MHz images shown in Fig. \ref{fig:total_A496} (right) using boxes of 22$\arcsec$ size (${\sim14~{\rm kpc}}$; equal to the beam FWHM of the images), and selecting only the regions where the diffuse radio source is detected at $\geq2\sigma_{rms}$ in both maps (for a similar approach, see e.g., \citealt{pignataro2024c}). We also excluded the boxes corresponding to the region where the central radio galaxy is present, to both avoid contamination from residual unsubtracted radio emission, and from AGN feedback footprints (X-ray cavities) in the thermal gas. The regions are overlaid in white on the left panel of Fig. \ref{fig:radioxA496}. We also extracted the X-ray surface brightness from the same boxes after convolving the X-ray image with a Gaussian of 22$\arcsec$ FWHM (so that the radio and X-ray maps are at the same resolution). The resulting points are plotted at each radio frequency  in the central panel of Fig. \ref{fig:radioxA496}. Following the approach of e.g., \citet{riseley2022}, we performed a fit of the points in the log-log space using LinMix\footnote{See the \href{https://linmix.readthedocs.io/en/latest/src/linmix.html}{Linmix webpage}.}, which returned the best fit lines overlaid on the scatter plot. The values of $I_{\rm R}$ and $I_{\rm X}$ are clearly positively correlated at both frequencies. Specifically, we find Spearman ranks and p-values of $\rho_s^{1280\text{MHz}} = 0.89$ ($\leq0.01\%$) and $\rho_s^{650\text{MHz}} = 0.77$ ($\leq0.01\%$); we find Pearson ranks and probabilities of $\rho_p^{1280\text{MHz}} = 0.92$ ($\leq0.01\%$) and $\rho_p^{650\text{MHz}} = 0.77$ ($\leq0.01\%$). The scatter of the data around the best-fit relations is 0.14~dex and 0.09~dex at 650 and 1280 MHz, respectively. The best fit slopes vary with frequency, showing a sub-linear $\beta_{650} = 0.5\pm0.1$ and a nearly linear $\beta_{1280} =1.1\pm0.1$. These slopes indicate that the X-ray brightness declines faster than the radio one, on average. The availability of two frequencies also allows us to explore the relation between the spectral index $\alpha$ and $I_x$ (see e.g., \citealt{riseley2022}). For each point, we used the $I_r$ at 650 MHz and 1280 MHz to measure a spectral index, that we plot as a function of $I_x$ in the right panel of Fig. \ref{fig:radioxA496}. We see a strong trend of flattening spectral index for higher X-ray surface brightness, that is well fit with a power-law of slope $\alpha\propto (2.2\pm0.2)\log_{10}I_{x}$. This trend points to an outward steepening of the synchrotron spectral index. 

\subsubsection{Radio-X-ray connection in ZwCl~235}\label{subsub:radioXZ235} 
We extracted the radio surface brightnesses from the 54~MHz, 144~MHz, and 400 MHz images shown in Fig. \ref{fig:total_z235} (right) using boxes of 20$\arcsec$ size (${\sim31~{\rm kpc}}$; the beam FWHM), and selecting only the regions above $\geq2\sigma_{rms}$ significance in the three maps. We also excluded the boxes corresponding to the region where the central radio galaxy is present. Due to the relatively small angular size of the diffuse emission in ZwCl~235, there are only 9 regions where we can measure the radio surface brightness at three frequencies (left panel of Fig. \ref{fig:radioxZ235}). The X-ray surface brightness was extracted from the same boxes after convolving the X-ray image with a Gaussian of 20$\arcsec$ FWHM. The resulting points are plotted in the central panel of Fig. \ref{fig:radioxZ235}. Even though the statistics is limited, the radio and X-ray surface brightness seem correlated at all frequencies. We find Spearman ranks and probabilities of $\rho_s^{54\text{MHz}} = 0.40$ ($28\%$), $\rho_s^{144\text{MHz}} = 0.85$ ($0.3\%$), and $\rho_s^{400\text{MHz}} = 0.95$ ($0.1\%$); Pearson ranks and probabilities of $\rho_p^{54\text{MHz}} = 0.55$ ($12\%$), $\rho_p^{144\text{MHz}} = 0.90$ ($0.1\%$), and $\rho_p^{400\text{MHz}} = 0.93$ ($\leq0.1\%$). The best-fit slopes are sublinear, specifically, we measure $\beta_{54} = 0.5\pm0.3$, $\beta_{144} =0.9\pm0.3$, and $\beta_{400} =0.8\pm0.2$. The scatter of the data around the best-fit relations is 0.16~dex, 0.10~dex, and 0.07~dex at 54, 144, and 400 MHz, respectively. Testing different grids and sizes of the boxes (e.g., for the 144~MHz image we could use boxes of $\sim$10$\arcsec\sim15~{\rm kpc}$) returns consistent sub-linear values of the slopes. 
We note that the slope of the radio–X-ray correlation is known to be frequency-dependent, with lower-frequency observations tracing less energetic electron populations (see also \citealt{vazza2024}) and typically yielding more sub-linear relations for large halos (e.g., \citealt{rajpurohit2021,riseley2022}); despite probing different frequency ranges, both Abell~496 and ZwCl~235 consistently show nearly-linear or sub-linear slopes. We also tried to explore the relation between the spectral index $\alpha_{54}^{144}$ and $\alpha_{144}^{400}$, and $I_x$ (right panel of Fig. \ref{fig:radioxZ235}; see also \citealt{riseley2022}). No statistically meaningful trends are visible (Spearman ranks $\leq0.5$), although there might be some flattening of the $\alpha_{54}^{144}$ spectral index at increasing $I_x$. The lack of a strong $\alpha$-$I_x$ correlation may also be influenced by the relatively large physical size of the boxes, 20$\arcsec\sim$30~kpc, which might be too coarse for a diffuse radio source of total size $\leq100$~kpc.

\section{Discussion}
\subsection{Classification of the diffuse radio emission}\label{subsec:minihalonature}
In this subsection, we show that the diffuse radio sources in Abell~496 and ZwCl~235 can be interpreted as radio mini-halos. To this end, we summarize the classification criteria originally defined in \citet{giacintucci2017}: mini-halos were defined as centrally located radio sources that: (1) are not morphologically connected to the central AGN or shaped as radio jets or lobes; (2) have a minimum radius of 50~kpc; and (3) have a maximum radius smaller than 0.2\,R$_{500}$\footnote{These numbers refer to radio observations at frequencies typically between 0.6 - 1.4 GHz. We note that lower-frequency or deeper observations have since revealed diffuse radio emission extending to larger radii in some systems; however, a change in the radio surface-brightness profile at $\geq$0.2\,R$_{500}$ often allows the inner emission to be identified as a distinct mini-halo component (e.g., \citealt{savini2018,biava2024,vanweeren2024,vanWeeren2026}).}. The diffuse sources in Abell~496 and ZwCl~235 satisfy the formal mini-halo classification criteria. They are located at the centers of their host cluster and do not display prominent jet or lobe morphologies (although some low-level residual AGN-related structures remain visible after uv-filtering, see Sect.~\ref{subsec:diffuseemission}). Their spatial extents are also consistent with those expected for mini-halos: using the minimum and maximum radii of the diffuse emission reported in Sect.~\ref{subsec:diffuseemission}, and adopting the definition of the average mini-halo radius $R_{\mathrm{MH}} = \sqrt{R_{\mathrm{max}} \times R_{\mathrm{min}}}$ from \citet{giacintucci2017}, we derive $R_{\mathrm{MH}} = 70$~kpc (${\sim106\arcsec}$) for Abell~496 and $R_{\mathrm{MH}} = 65$~kpc (${\sim42\arcsec}$) for ZwCl~235. These correspond to $\sim$0.07\,R$_{500}$ for both clusters, where R$_{500}$ is taken from \citet{mcxc_ii}. 
In addition, some qualitative features that characterize most mini-halos have been identified in the literature (e.g., \citealt{gitti2004,giacintucci2014,giacintucci2017,giacintucci2019,laferriere2020,riseley2022,giacintucci2024,biava2024}), including: the association with a dynamically relaxed galaxy cluster\footnote{We note, however, that exceptions to this property exist, see e.g., Abell~1413 \citep{riseley2023} or Abell~2142 \citep{bruno2023}.}; and a morphological connection with sloshing cold fronts\footnote{See \citet{biava2024} and \citet{bonafede2023} for exceptions.}. Since the diffuse radio emission in both Abell~496 and ZwCl~235 is spatially confined within the cold fronts detected in the X-ray data (Sect. \ref{subsec:xray}), and the clusters are typical relaxed cool cores (e.g., \citealt{Sanderson2006} for Abell~496; \citealt{ubertosi23AA} for ZwCl~235), there is additional support for their interpretation as radio mini-halos.

\indent Overall, our results indicate that diffuse radio emission is not a unique trait of massive systems. This point had already been made for unrelaxed clusters (e.g., \citealt{knowles2016,paul2021}), but it was still unclear for relaxed clusters (see the discussion in \citealt{giacintucci2017}). The only confirmed low-mass cluster to host a mini-halo was 2A~0335.0+096, hosted in a cluster of mass 2.3$\times$10$^{14}$~M$_{\odot}$ (from the Planck Catalog, \citealt{planck_2016_IIDatarelease}) or 3.4$\times$10$^{14}$~M$_{\odot}$ (from the MCXC catalog, \citealt{piffaretti2011,mcxc_ii}). Some candidates were recently reported in \citet{kolokythas2025}: Abell~4038 (1.5$\times$10$^{14}$~M$_{\odot}$), J0314.3–4525 (1.7$\times$10$^{14}$~M$_{\odot}$), and J0351.1–8212 (2.5$\times$10$^{14}$~M$_{\odot}$). Our confirmed detections in Abell~496 and ZwCl~235 (M$_{500}\sim$2.8$\times$10$^{14}$~M$_{\odot}$) provide robust new evidence that mini-halos can form in low-mass, relaxed clusters. This picture will be further tested with forthcoming MeerKAT observations of other MIMIC systems (Table~\ref{tab:mimics}).

\subsection{Origin of the mini-halos Abell~496 and ZwCl~235}
A phenomenological classification as mini-halos based on global properties does not necessarily imply a common origin for the diffuse emission, which may vary among sources that meet the above criteria. The spectral properties of both systems offer new interesting perspectives on this point. The diffuse source in both systems has a steep spectral index ($-1.4$ in Abell~496, and $-1.8$ in ZwCl~235), placing these among the steepest-spectrum mini-halos (see Tab.~\ref{tab:minihalo_scalings}). 
Moreover, in Abell~496 we observe an outward steepening of the spectral index, as demonstrated by the strong correlation between $\alpha$ and $I_{x}$ (Fig. \ref{fig:radioxA496}). In ZwCl~235 there is visual indication of a possible flattening of $\alpha_{54}^{144}$ at higher X-ray surface brightness (Fig. \ref{fig:radioxZ235}), although this trend is not statistically significant given the limited number of independent measurements. The flattening of the spectrum toward regions of higher $I_{x}$ in Abell~496 may be related to a decreasing magnetic field strength with radial distance (e.g., \citealt{brunetti2001,bonafede2022}; see also \citealt{hoang2025}). Alternatively if the particles are being re-accelerated, the mechanism may be more efficient in central regions (as proposed for radio halos, e.g., \citealt{rajpurohit2021}, or for the extended halos surrounding some mini-halos in relaxed clusters, e.g., \citealt{riseley2022,hoang2025}\footnote{See, however, the nearly constant spectral index with radius of the extended halo around the mini-halo in A1795 from \citealt{vanWeeren2026}.}). 

The key question is therefore which mechanism is responsible for the origin of these diffuse radio sources.  The advent of new radio facilities (e.g., MeerKAT, uGMRT, ASKAP, and LOFAR) has led to the discovery of a wide variety of diffuse radio sources; in cool cores, some of these may meet some properties of mini-halos but have an intrinsically different origin (e.g., \citealt{rotella2025}). In this context, it is worth commenting on the minimum radius of 50~kpc adopted to define a radio mini-halo. This threshold was discussed in  \citet{giacintucci2014,giacintucci2017} to identify the spatial scale beyond which particle transport from the central AGN is inefficient and in-situ particle production or re-acceleration is required. We find this point worth revisiting because, although Abell~496 and ZwCl~235 are not the smallest mini-halos, they have the steepest spectra (see Tab.~\ref{tab:minihalo_scalings}). We thus reconsider these calculations in the specific cases of Abell~496 and ZwCl~235, to understand whether particle re-acceleration is at all required for their diffuse radio sources. 
Following \citet{giacintucci2014}, the timescale taken by electrons to be advected at the Alfvénic speed to a radius of 70~kpc (average $R_{\rm MH}$ for the two clusters) in a magnetic field of 5-10 $\mu$G is $t_{\rm A} = 7 \times 10^{8}$~yr (with a speed of $\sim$100~km/s). 
For similar magnetic field strengths, the synchrotron cooling timescales at 400 MHz and 1280 MHz (the highest frequencies at which we observe the mini-halos in ZwCl~235 and Abell~496, respectively) are $t_{\rm syn} = 1-2\times10^{8}$~yr and $t_{\rm syn} = 0.5-1\times10^{8}$~yr. The ratio $t_{\rm A}/t_{\rm syn} \gg 1$ therefore indicates that pure advection at Alfvénic speed is insufficient to explain the size of the diffuse radio sources. 
Sloshing-induced subsonic motions of the thermal gas could provide a more efficient transport mechanism (e.g., \citealt{dominguezfernandez2024}). 
However, even in this case, the relevant velocities remain limited to a fraction of the sound speed ($v_{\rm slosh} \lesssim 0.3\,c_{\rm s}$ e.g., \citealt{zuhone2022}), corresponding to $\sim 200-300~{\rm km~s^{-1}}$ for the 3 keV plasma in our two systems. This implies $t_{\rm slosh} = 2-3\times10^{8}$~yr for a distance of 70~kpc and, in turn, $t_{\rm slosh}/t_{\rm syn} \gtrsim 1$. In other terms, we find that advecting electrons over $\sim70$~kpc within their radiative lifetime would require speeds in the range $\sim400-1200\,{\rm km~s^{-1}}$ (depending on the observing frequency and magnetic field). These values are generally higher than typical sloshing velocities. Support for this argument also comes from very recent XRISM observations of the cool core of Abell~496, which revealed a slow line-of-sight bulk velocity of $69^{+25}_{-20}~{\rm km~s^{-1}}$ \citep{Veronica2026} within a single Resolve pointing ($3\arcmin$ field of view, comparable to the size of the mini-halo).  
Overall, particle transport alone is unlikely to explain the full spatial extent of the mini-halos, and some level of re-acceleration is very likely required, even accounting for the most favorable sloshing conditions. While a contribution from aged plasma advected from the central AGN is still expected, the spectral properties argue against a purely passive ageing scenario: in Abell~496, the spectral index of the oldest radio bubbles ($\alpha = -3$; \citealt{Ubertosi2024c}) is significantly steeper than that of the diffuse emission ($\alpha = -1.4$), implying that, in the absence of re-acceleration, any plasma advected from the central AGN would exhibit much steeper spectra and be undetectable at GHz frequencies. The very steep spectrum observed in ZwCl~235 ($\alpha = -1.8$) may instead indicate a larger relative contribution from aged, weakly re-accelerated plasma.

These considerations point to a scenario in which additional mechanisms play some role in sustaining the diffuse radio emission. A dominant hadronic origin of the relativistic particles in these two systems is strongly disfavoured, primarily due to the steep spectral indices, which would require unrealistically large non-thermal energy densities, comparable to the thermal energy of the ICM (see \citealt{pfrommerensslin2003,brunetti2008,macario2010,brunetti2014} for radio halos -- an even more significant issue for mini-halos, given their $\sim50\times$ higher emissivity, e.g., \citealt{cassano2008}). The nearly linear or sub-linear radio–X-ray surface brightness correlations observed in both clusters from low to high frequencies (see Sect.~\ref{subsec:xray}, Fig. \ref{fig:radioxA496} \& Fig. \ref{fig:radioxZ235}), steep spectral indices, and steepening of the mini-halo spectrum with radius (the latter is robustly seen in Abell~496 only) could instead be suggestive of turbulent re-acceleration scenarios (e.g., \citealt{govoni2001,ignesti2020,riseley2022}). This could be the case in relaxed clusters where not only sloshing, but also AGN feedback, may contribute to injecting turbulence in the thermal gas at the center (e.g., \citealt{bellomi2025}). 
In the next subsection, we extend this discussion to the overall population of mini-halos, combining our results on low-mass clusters with literature results on higher-mass systems.

\begin{figure*}[ht!]
    \centering
    \includegraphics[width=0.32\linewidth]{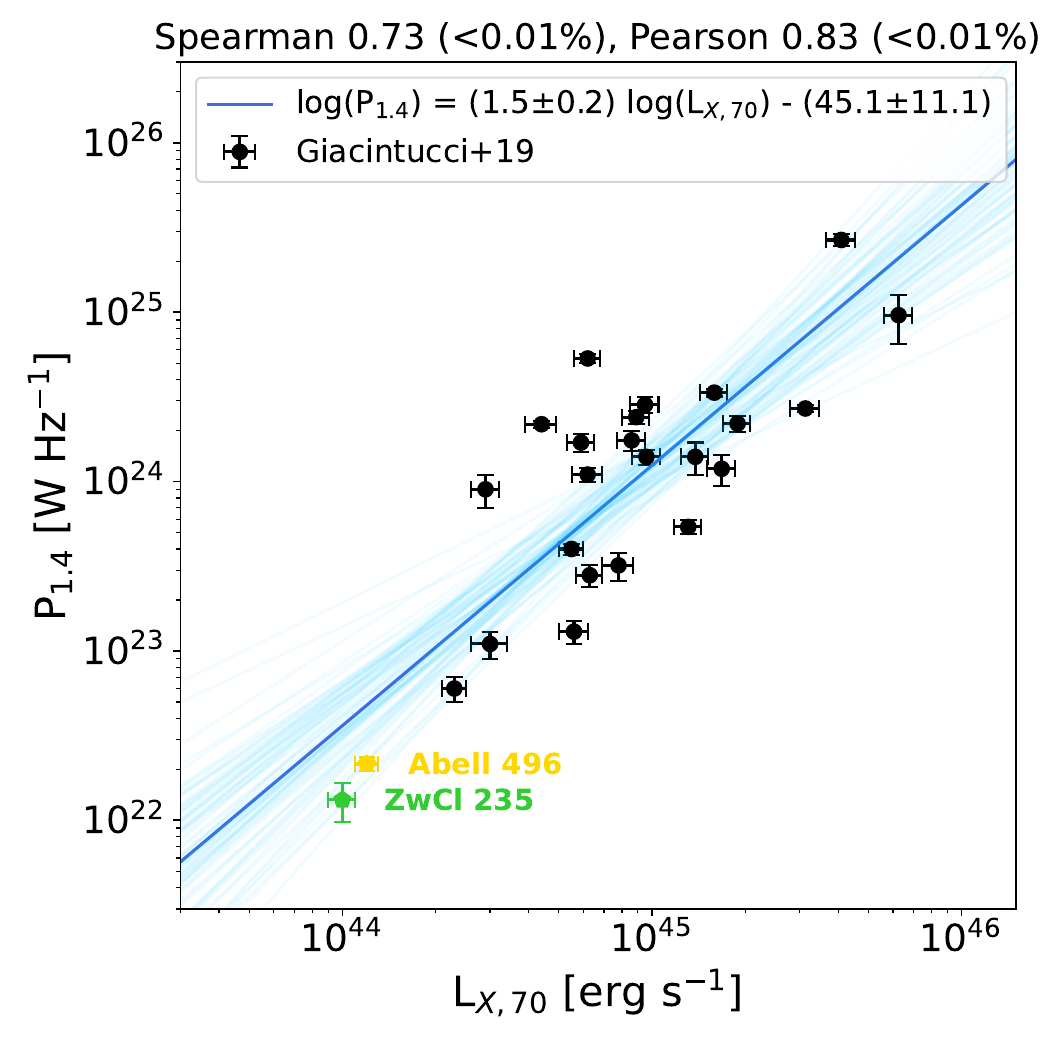}
    \includegraphics[width=0.32\linewidth]{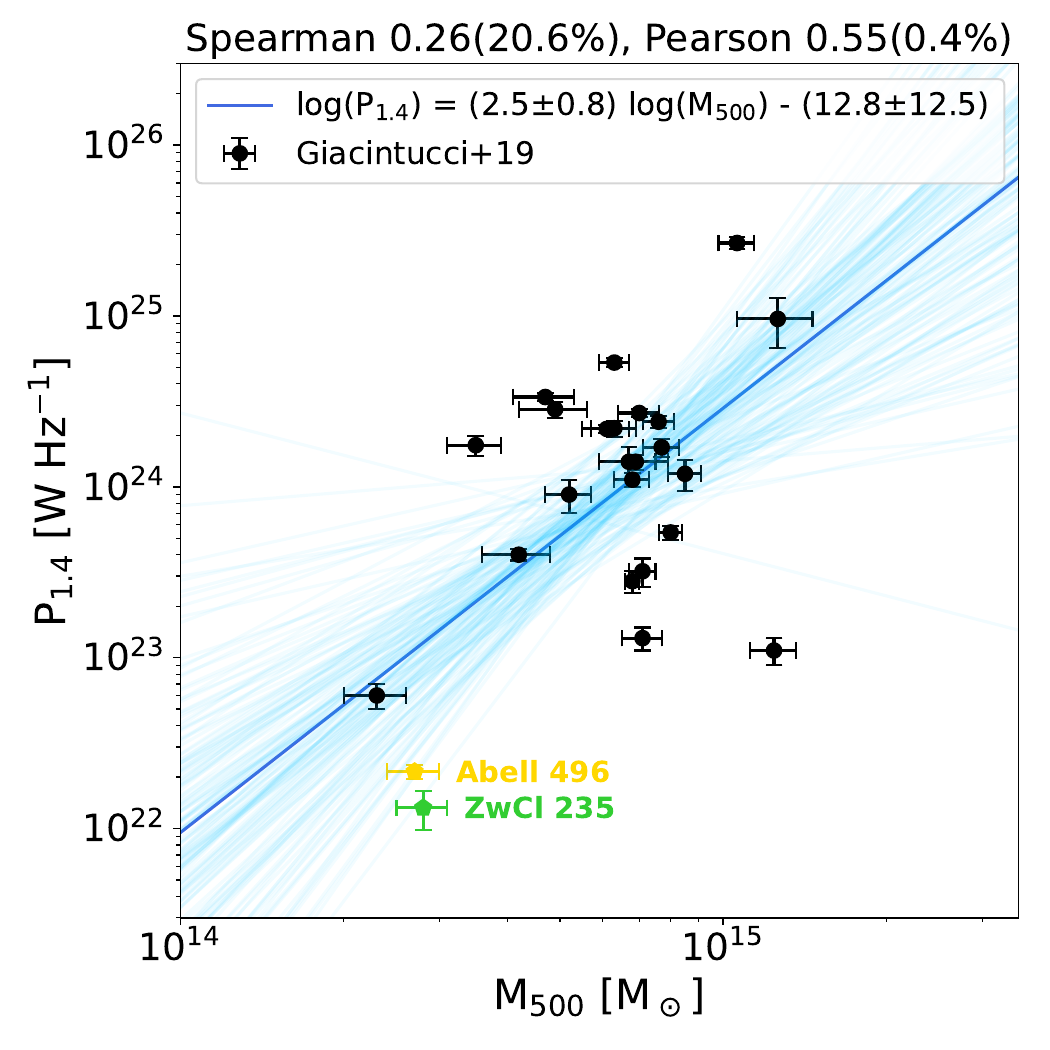}
    \includegraphics[width=0.32\linewidth]{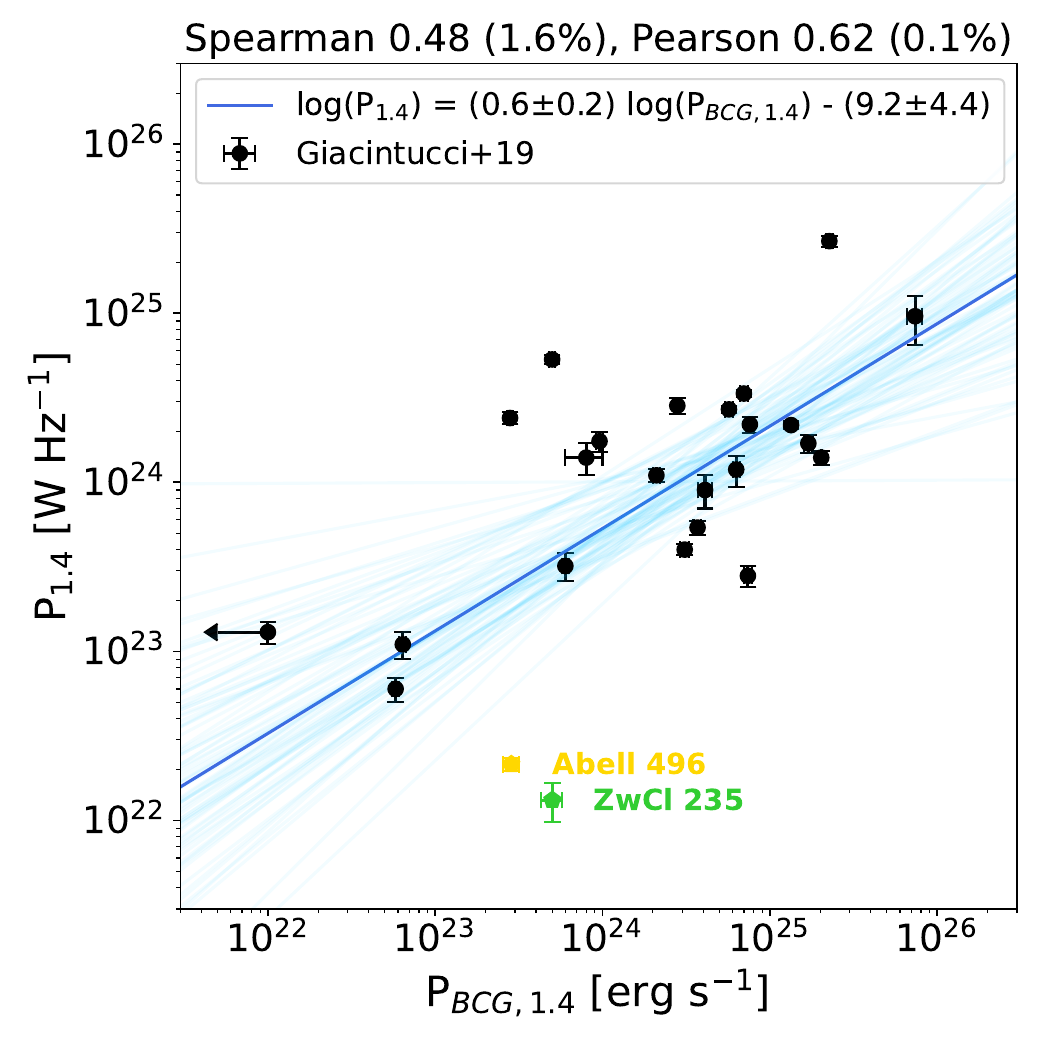}\\
    \includegraphics[width=0.32\linewidth]{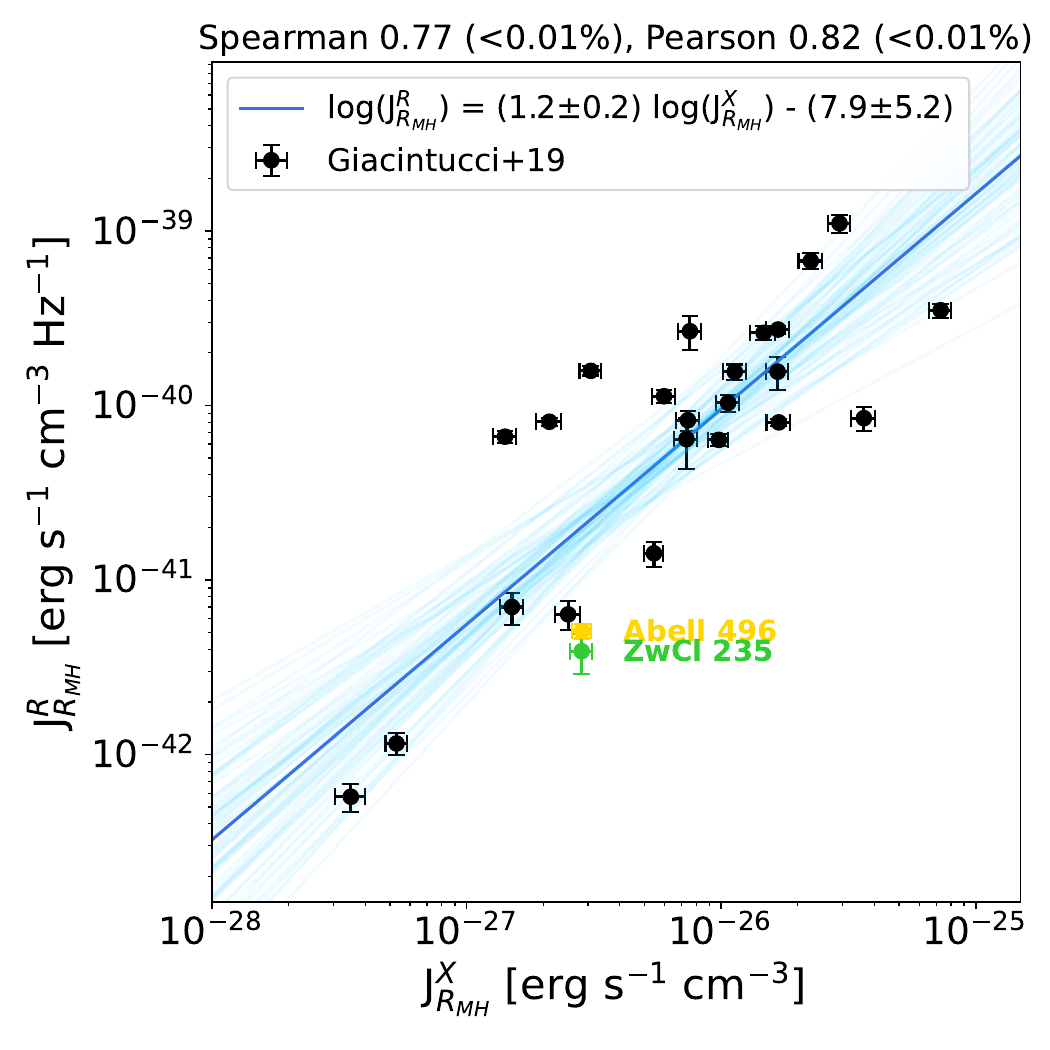}
    \includegraphics[width=0.32\linewidth]{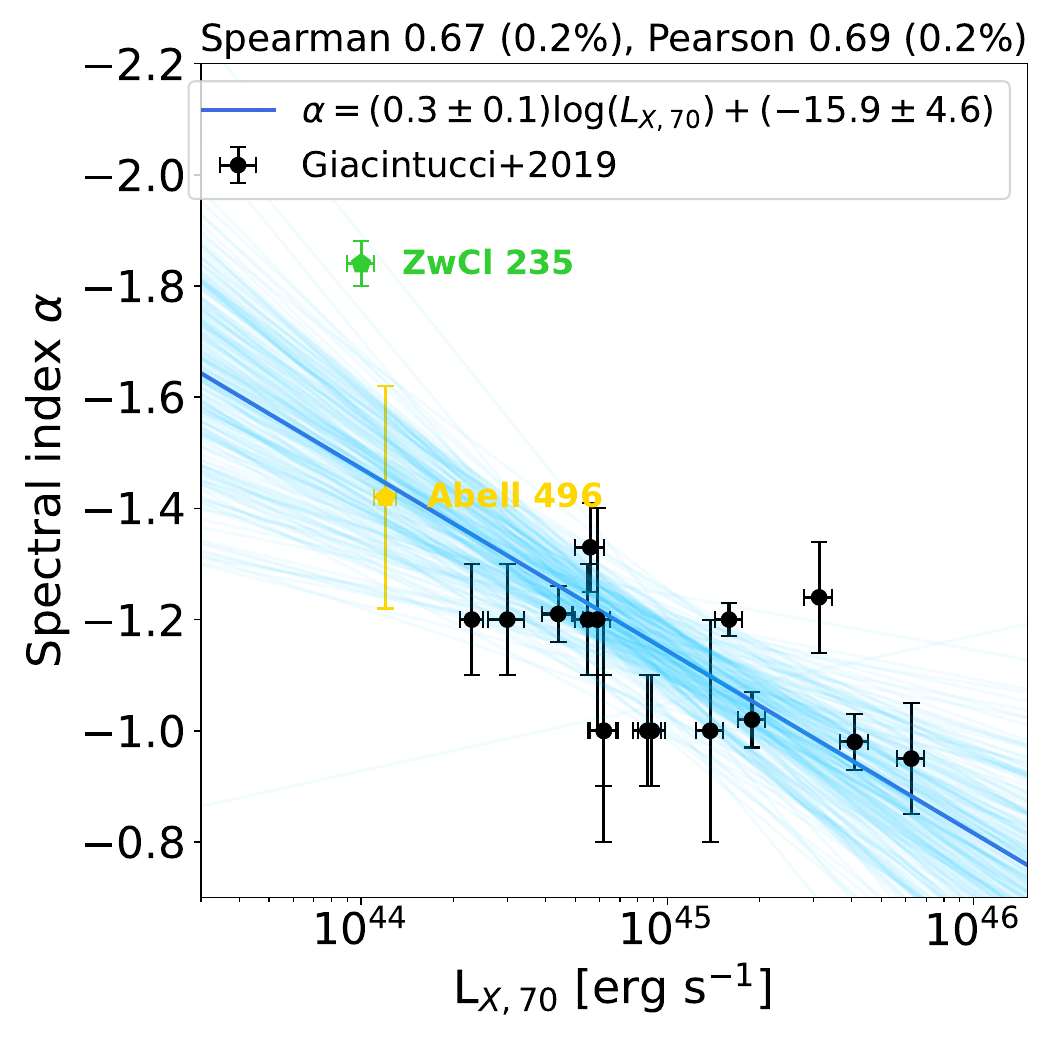}
    \includegraphics[width=0.32\linewidth]{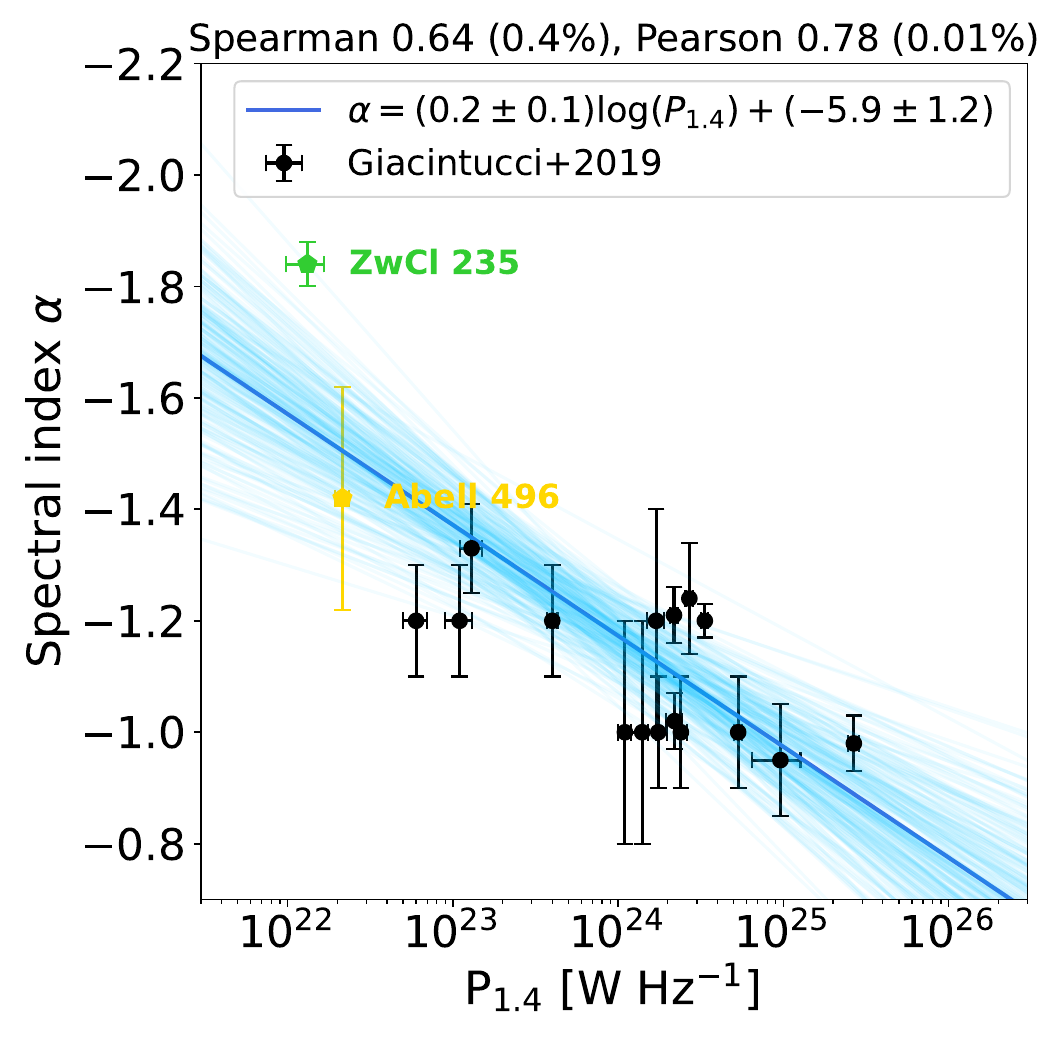}
    
    \caption{{\it Top panels:} Radio power of mini-halos versus X-ray bolometric luminosity of the ICM within 70~kpc (\textit{left}), 
    versus M$_{500}$ (see Tab.~\ref{tab:minihalo_scalings}; \textit{middle}), and versus the radio power of the BCG at 1.4 GHz (\textit{right}). \textit{Bottom left:} Radio average emissivity at 1.4 GHz as a function of the X-ray average emissivity, both computed within the mini-halo radius $R_{\mathrm{MH}}$ (see Sect~\ref{subsec:discscalingrelations}). {\it Bottom middle and right panels:} mini-halo spectral index (when available; from Table~\ref{tab:minihalo_scalings}) of systems versus X-ray bolometric luminosity of the ICM within 70~kpc 
    and versus the radio power of mini-halos. 
    The data from these plots is reported in Tab.~\ref{tab:minihalo_scalings} (see also Appendix~\ref{app:samplefull} and \citealt{giacintucci2019}). 
    The Pearson and Spearman correlation ranks, with their probabilities, are reported at the top of the plots. The best-fit relations obtained with LinMix are shown as a blue line.
    }
    \label{fig:scalingrel}
\end{figure*}

\subsection{Mini-halo scaling relations}\label{subsec:discscalingrelations}
Scaling relations between mini-halo and host cluster properties have previously been investigated \citep{bravi2016,gitti2018,giacintucci2019,laferriere2020,kolokythas2025}. Weak to moderate correlations between 1.4~GHz radio power $P_{1.4}$ and mass M$_{500}$ have been reported \citep{giacintucci2019,laferriere2020,kolokythas2025}, suggesting that more massive systems may host more powerful non-thermal emission, as in radio halos (e.g., \citealt{cuciti2021}). However, the location of mini-halos in cool cores implies a stronger link with core thermodynamics than with global properties \citep{bravi2016,giacintucci2019}, supported by the tight correlation between $P_{1.4}$ and the X-ray luminosity within 70~kpc, $L_{\rm X,70}$. Moderate correlations with the BCG radio power, $P_{\mathrm{BCG},1.4}$, have also been reported (e.g., \citealt{giacintucci2019,laferriere2020}), suggesting a link with the central AGN. 

In this subsection, we add our results on Abell~496 and ZwCl~235 to the sample presented in \citet{giacintucci2019}, for which not only M$_{500}$, but also $L_{\rm X,70}$ and P$_{\rm BCG,1.4}$ were reported. The values adopted for the full sample of systems are reported in Appendix~\ref{app:samplefull}, where we also detail how we obtained the quantities $P_{1.4}$, M$_{500}$, $L_{\rm X,70}$, and P$_{\rm BCG,1.4}$ for the two low-mass clusters. Here, we simply note that, for the diffuse radio power at 1.4~GHz of Abell~496 and ZwCl~235, we considered the total mini-halo flux density (\say{Diffuse (total)} in Tab.~\ref{tab:fluxtable}). This choice is motivated by two considerations: as discussed in Sect.~\ref{subsec:diffuseemission}, the \say{Diffuse (AGN-masked)} flux densities underestimate the total ones; moreover, adopting the total flux density facilitates comparison with the literature, where the region containing the central AGN is generally not masked, even in maps from which discrete sources have been subtracted. We show in the top panels of Fig. \ref{fig:scalingrel} the comparison between $P_{1.4}$ and $L_{\rm X,70}$, M$_{500}$, and $P_{\rm BCG,1.4}$. We fitted the data with power-laws using LinMix, and we also computed Spearman and Pearson ranks, reported at the top of each plot. The best-fit parameters and correlation coefficients are reported in Tab.~\ref{tab:mh_scaling}, which also presents the results obtained using the \say{Diffuse (AGN-masked)} flux density to compute the radio power. Since the total and AGN-masked flux densities differ by factors of $\leq 2$, whereas the relations we investigate span orders of magnitude, the differences between the two sets of results are minimal and remain within the uncertainties. 

Based on these results, it is clear that, as previously found by \citet{giacintucci2019}, the 1.4 GHz radio power of mini-halos is more strongly correlated to the core X-ray luminosity rather than to the total mass of the cluster or the BCG radio power. We note that there are several factors that might wash out correlations between diffuse radio power and global cluster properties; these include the timings and strength of sloshing motions (i.e., how long ago and how \say{minor} was the merger that triggered sloshing), and of AGN activity (i.e., the energy and duration of recent outbursts).

\begin{table*}[htp!]
\centering
\caption{Best-fit parameters and correlation ranks for the mini-halo scaling relations in Fig.~\ref{fig:scalingrel}.}
\label{tab:mh_scaling}
\small
\setlength{\tabcolsep}{4pt}
\begin{tabular}{lcccccc|cccccc}
\toprule
& \multicolumn{6}{c}{Diffuse (total)}
& \multicolumn{6}{c}{Diffuse (AGN-masked)} \\
\cmidrule(lr){2-7} \cmidrule(lr){8-13}
Relation
& Slope & Intercept & $r_{\rm S}$ & $p_{\rm S}$
& $r_{\rm P}$ & $p_{\rm P}$
& Slope & Intercept & $r_{\rm S}$ & $p_{\rm S}$
& $r_{\rm P}$ & $p_{\rm P}$ \\
(1) & (2) & (3) & (4) & (5) & (6) & (7) & (8) & (9) & (10) & (11) & (12) & (13) \\
\midrule
$\log P_{1.4}$ -- $\log L_{\mathrm{X},70}$
& $1.5\pm0.2$ & $45.1\pm11.1$ & $0.73$ & $<0.01\%$ & $0.83$ & $<0.01\%$
& $1.6\pm0.3$ & $49.4\pm11.5$ & $0.73$ & $<0.01\%$ & $0.83$ & $<0.01\%$ \\

$\log P_{1.4}$ -- $\log M_{500}$
& $2.5\pm0.8$ & $12.8\pm12.5$ & $0.26$ & $20.6\%$ & $0.55$ & $0.4\%$
& $2.8\pm0.9$ & $17.2\pm13.0$ & $0.26$ & $20.6\%$ & $0.57$ & $0.3\%$ \\

$\log P_{1.4}$ -- $\log P_{\mathrm{BCG}}$
& $0.6\pm0.2$ & $9.2\pm4.4$ & $0.48$ & $1.6\%$ & $0.62$ & $0.1\%$
& $0.6\pm0.2$ & $8.6\pm5.0$ & $0.48$ & $1.6\%$ & $0.60$ & $0.2\%$ \\

$\log$J$_{R_{\rm MH}}^{R}$ -- $\log$J$_{R_{\rm MH}}^{X}$
& $1.2\pm0.2$ & $7.9\pm5.2$ & $0.77$ & $<0.01\%$ & $0.82$ & $0.01\%$
& -- & -- & -- & -- & -- & -- \\

$\alpha$ -- $\log L_{\mathrm{X},70}$
& $0.3\pm0.1$ & $-15.9\pm4.6$ & $0.67$ & $0.2\%$ & $0.69$ & $0.2\%$
& -- & -- & -- & -- & -- & -- \\

$\alpha$ -- $\log P_{1.4}$
& $0.2\pm0.1$ & $-5.9\pm1.2$ & $0.64$ & $0.4\%$ & $0.78$ & $0.01\%$
& $0.2\pm0.1$ & $-5.8\pm1.0$ & $0.64$ & $0.44\%$ & $0.81$ & $<0.01\%$ \\
\bottomrule
\end{tabular}
\tablefoot{Results are reported using the ``Diffuse (total)''  and the ``Diffuse (AGN masked)''  flux densities to compute the total radio power of Abell~496 and ZwCl~235. Columns: (1) Relation; (2-7) best-fit parameters and correlation coefficients (slope, intercept, Spearman rank and probability, Pearson rank and probability) for the plots shown in Fig.~\ref{fig:scalingrel}, where the total radio power is computed from the \say{Diffuse (total) flux density}; (8-13) are the same quantities as in columns (2-7), but using the \say{Diffuse (AGN masked)} flux density to compute the total radio power. For the \say{Diffuse (AGN masked)} case, we do not report results for the $\log J_{R_{\rm MH}}^{R}$--$\log J_{R_{\rm MH}}^{X}$ relation, as it would require subtracting the volume occupied by the AGN-related region from the total mini-halo volume within $R_{\rm MH}$. We also do not report results for the $\alpha$ -- $\log L_{\mathrm{X},70}$ since the radio power is not involved.}
\end{table*}

\begin{figure}[ht!]
    \centering
    \includegraphics[width=\linewidth]{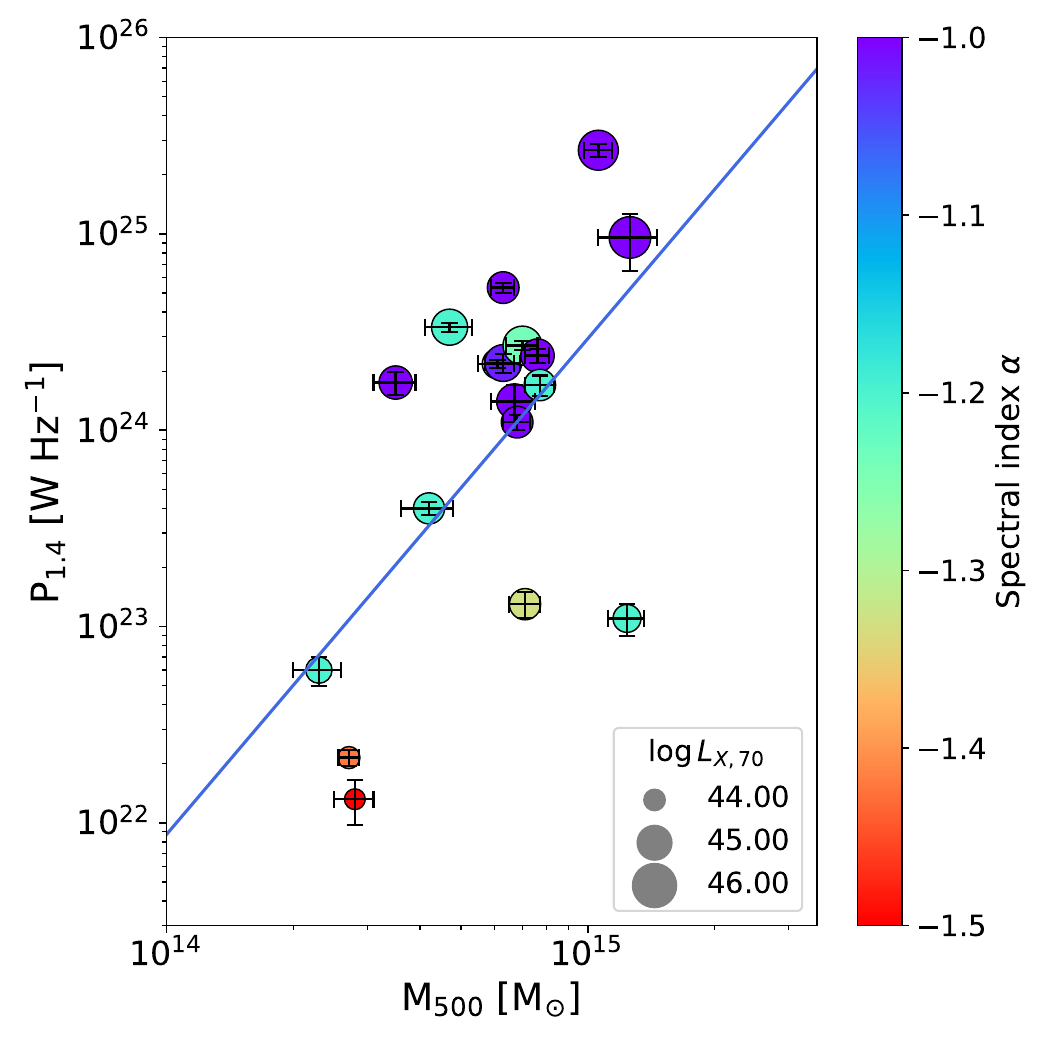}
    \caption{Radio power of mini-halos with available measurements of spectral index versus M$_{500}$ (see Tab.~\ref{tab:mimics}). Points are color-coded by the spectral index of the diffuse radio emission (see Tab.~\ref{tab:minihalo_scalings}; mostly computed in the literature between a sub-GHz frequency and 1.4 GHz); the size of the points scales with the logarithm of the core X-ray luminosity.
    The best fit from Fig.~\ref{fig:scalingrel} (top center) is shown with a blue line.}
    \label{fig:lrm500-colorspix}
\end{figure}

A limitation of the above analysis is that the X-ray luminosity is measured within 70~kpc (the typical size of cool cores), while the radio power of mini-halos is computed over their full extent. Since these two regions do not necessarily coincide (some mini-halos in \citealt{giacintucci2019} extend to $\sim$300~kpc), the resulting scaling relations may be difficult to interpret. To address this issue, we translated radio and X-ray powers to emissivities averaged over the same emitting volumes. First, we recompute the X-ray luminosity within the mini-halo radius, $L_{X,\mathrm{MH}}$  (see Tab.~\ref{tab:minihalo_scalings}), starting from the values measured within 70~kpc. This rescaling is performed by integrating the surface-brightness profile described by a double-$\beta$ model, using parameters available in the literature for each system (see  Appendix~\ref{app:sub:scalingxraylum}). We then estimate the radio and X-ray average emissivities, J$^{\rm R}_{R_{\rm MH}}$ and J$^{\rm X}_{R_{\rm MH}}$, by dividing the mini-halo radio power and $L_{X,\mathrm{MH}}$ by the spherical volume enclosed within $R_{\mathrm{MH}}$. The average emissivities are reported in Tab.~\ref{tab:minihalo_scalings} and compared in Fig.~\ref{fig:scalingrel} (\textit{bottom left}), along with the best-fit relation. The radio and X-ray average emissivities within $R_{\mathrm{MH}}$ appear correlated, with: ${\log J^{\rm R}_{R_{\rm MH}} = (1.2\pm0.2)\log J^{\rm X}_{R_{\rm MH}} - (7.9\pm5.2)}$.
This result can be used to provide some first-order constraints on the mechanisms that sustain the radio emission of mini-halos. The X-ray emissivity emitted through Bremsstrahlung from an optically thin plasma of constant electron density $n_{e}$ and temperature $T$ is: 
\begin{equation}
    J^{X}\propto n_e^2\times\Lambda (T),
\end{equation}
where $\Lambda(T)$ is the cooling function. For the clusters in Fig.~\ref{fig:scalingrel} and \ref{fig:lrm500-colorspix}, the core temperature is $\geq2$~keV (see Tab.~\ref{tab:minihalo_scalings}), a regime where $\Lambda(T)\propto T^{1/2}$ (e.g., \citealt{1993ApJS...88..253S}). Our observed correlation of $J^{\rm R}_{R_{\rm MH}} \propto \left(J^{\rm X}_{R_{\rm MH}}\right)^{1.2\pm0.2}$ implies that:
\begin{equation}
    \langle J^{\rm R}_{R_{\rm MH}}\rangle\appropto \langle n_{\rm e, \leq R_{\rm MH}}^{2.4\pm0.4}\rangle\times \langle T_{\rm \leq R_{\rm MH}}^{0.6\pm0.1} \rangle
\end{equation}
meaning that the density of thermal particles may be the major driver of the observed non-thermal radio emission. If thermal density plays a role, one might naively assume that hadronic interactions with the thermal target within the emitting volume are the primary drivers. However, the two mini-halos in our sample that follow this correlation (Fig.~\ref{fig:scalingrel}) are clearly not powered by hadronic processes. In principle, other key factors such as the turbulent energy flux ($\propto ndV^3$), the magnetic field strength and confinement of relativistic particles, and even the work done by the AGN in the inner regions also depend on the thermal density and energy density. Consequently, interpreting this correlation is more complex and requires more robust observational constraints. For instance, a significant step forward for future studies will be to derive correlations based on emission-weighted quantities rather than volume-weighted ones.

In Fig.~\ref{fig:scalingrel} we also plot the spectral indices of the confirmed mini-halos compiled by \citet{giacintucci2019}, for which spectral information is available (see Tab.~\ref{tab:minihalo_scalings}), together with the values measured for Abell~496 and ZwCl~235, versus $L_{X,70}$, and $P_{1.4}$. Although the scatter is large, we observe for the first time tentative indications that mini-halos hosted by less X-ray luminous cool cores or characterized by lower radio power may tend to exhibit steeper synchrotron spectra. 
We note that these trends should be interpreted with caution, as the spectral indices have been derived over varying physical regions, frequency ranges (between 0.1--1.4~GHz), and with different instruments; these factors likely contribute to the scatter of the points in the plots of Fig.~\ref{fig:scalingrel}.

In Fig.~\ref{fig:lrm500-colorspix} we show again the P$_{1.4}$ -- M$_{500}$ relation, color-coded by the spectral index of the mini-halos and with the size of points scaled by $\log L_{\rm X,70}$. 
This potential trend of steeper spectra in smaller cool cores with less powerful mini-halos is reminiscent of the behavior of ultra-steep-spectrum radio halos in unrelaxed clusters (e.g., \citealt{cassano2009,cuciti2021,pasini2024}), which are predicted to be more common in lower-mass systems and/or to originate from minor mergers.  
In unrelaxed clusters, the higher occurrence of steep-spectrum radio halos in lower masses supports turbulent re-acceleration scenarios (e.g., \citealt{cuciti2021,cassano2023}). For the mini-halos of relaxed clusters, instead, it is less straightforward to associate a possible $\alpha-L_{X,70}$ trend with a turbulent reacceleration scenario, since the AGN contribution to both particle injection and re-acceleration is unclear (but not negligible), and hadronic processes cannot be excluded (e.g., \citealt{ignesti2020,riseley2022,vanWeeren2026}).

While these results await confirmation from a larger, uniformly analyzed sample, the available evidence may suggest that the origin of mini-halos could involve a combination of hadronic and leptonic mechanisms (see also \citealt{riseley2022,hoang2025,vanWeeren2026}) whose relative importance varies with the global and cool core properties of the host cluster.
We note that the densest cool cores might host a large target population of protons for hadronic interactions (e.g., \citealt{pfrommer2004}). In this regime, the continuous injection of secondary electrons could partially sustain the diffuse synchrotron emission (on top of turbulent re-acceleration processes, \citealt{brunettilazarian2011}) and result in flatter spectral indices ($\mathbf{-1.2 \lesssim \alpha \lesssim -0.9}$). In less dense cool cores, turbulent re-acceleration of fossil electrons might instead dominate the radio emission, producing in some cases relatively steeper $\mathbf{\alpha\leq-1.2}$. Some (moderate) scaling with the BCG radio activity would be expected in both cases, as the central AGN could be both the source of cosmic-ray protons in the hadronic case (e.g., \citealt{pfrommer2004}), and the source of seed electrons (e.g., \citealt{2021ApJ...914...73Z,vazza2024,dominguezfernandez2024}) and turbulence (e.g., \citealt{sotira2025}) in the leptonic case. We also note that this scenario does not uniquely discriminate between sloshing- and AGN-driven turbulence: in lower-density cool cores, AGN jets are relatively less powerful (e.g., \citealt{rafferty2008}), possibly injecting less turbulence -- in a leptonic scenario, this would lead to fainter, steeper-spectrum mini-halos.

Such a mixed hadronic/leptonic scenario would explain why $P_{1.4}$ correlates more strongly with the core X-ray luminosity than with $M_{500}$:  
as the core X-ray average emissivity increases -- reflecting higher gas densities rather than higher total cluster masses -- the hadronic contribution can become progressively more substantial. Conversely, in relatively-less X-ray luminous cool cores, turbulence induced by gas sloshing and the mechanical interaction of the AGN with the ambient medium may dominate the acceleration of relativistic electrons, resulting in fainter emission and steeper spectra. This is consistent with the Ophiuchus and PSZ1~G139.61+24.20 clusters being outliers in the $P_{1.4}$–$M_{500}$ plane of Fig.~\ref{fig:scalingrel} and Fig.~\ref{fig:lrm500-colorspix}: following the scenario outlined above, the relatively lower thermal gas density of these cool cores compared to equally-massive clusters leads to a reduced efficiency of hadronic processes and, consequently, a lower radio power. From Fig.~\ref{fig:lrm500-colorspix} (yellow and cyan points in the bottom right side), Ophiuchus and PSZ1~G139.61+24.20 also seem to have relatively steeper spectra compared to clusters of similar mass. Instead, the steeper-spectrum mini-halos at lower core X-ray luminosities, if confirmed with a larger sample, may be consistent with a turbulence-dominated regime. Finally, this mixed scenario may also be consistent with the $I_r/I_x$ slopes. In clusters where the contribution from hadronic processes is more substantial, one expects a super-linear $I_r$–$I_x$ relation, as observed by \citet{ignesti2020} for several mini-halos in X-ray bright cool cores. In contrast, in relatively less luminous cool cores at a given halo mass -- such as Abell~496 and ZwCl~235 (this work), or PSZ1~G139.61+24.20 \citep{biava2024} -- the increasing role of turbulence may explain why sub-linear or nearly-linear $I_r$–$I_x$ relations are observed. 

To conclude, the steep spectra and other observed properties of Abell~496 and ZwCl~235 favor a turbulence-dominated interpretation, with a dominant hadronic origin strongly disfavored in both systems. At the population level, a mixed hadronic and leptonic scenario remains a plausible hypothesis, although this interpretation remains to be tested with a larger and more uniformly analyzed sample.

\section{Conclusions}

In this pilot study of the MIMIC project, we investigated the presence and nature of diffuse radio emission in two low-mass, relaxed galaxy clusters, Abell~496 and ZwCl~235. Our main conclusions can be summarized as follows:

\begin{itemize}[noitemsep, leftmargin=*,topsep=0pt, nosep]
    \item We detect faint, centrally located diffuse radio emission in both clusters, confined within sloshing-induced cold fronts and satisfying the classification criteria for radio mini-halos. These systems represent the least powerful confirmed mini-halos known to date, extending the mini-halo population into the low-mass, low-radio-power regime.
    \item The steep synchrotron spectra ($\alpha\leq-1.4$), the sub-linear radio--X-ray correlations, the spatial association with cold fronts, and, for Abell~496 only, the outward spectral steepening favor turbulent re-acceleration as a dominant mechanism in these systems, although some advection of fossil plasma from the BCG cannot be excluded.
    \item By placing Abell~496 and ZwCl~235 in the context of known mini-halos, we find possible evidence that steeper mini-halo spectra are preferentially associated with lower cool-core X-ray luminosities and lower radio powers, although confirmation from a larger sample with uniform radio coverage and consistent spectral-index measurements is needed before firm conclusions can be drawn.
    \item We discuss a scenario in which mini-halos originate from a combination of hadronic and leptonic processes, whose relative importance varies with cluster properties: in dense, X-ray luminous cool cores, a contribution from hadronic processes could sustain flatter-spectrum diffuse radio emission, while in less X-ray luminous systems turbulent re-acceleration would dominate, leading to fainter, steeper-spectrum mini-halos.
\end{itemize}

Overall, while the properties of Abell~496 and ZwCl~235 favor a turbulence-dominated interpretation, the broader mixed hadronic/leptonic scenario is a population-level hypothesis that remains to be tested with future work. In this context, we note that this scenario makes testable predictions. In particular, one may expect turbulent properties of the ICM (e.g., velocity dispersion $\sigma_{v}$) to correlate with the non-thermal properties of mini-halos (e.g., radio power or spectral index) more strongly in low-mass or steep-spectrum mini-halo galaxy clusters. The forthcoming full statistical analysis of the MIMIC sample will allow us to further test these scenarios.

\begin{acknowledgements}
      We sincerely thank the referee for their very useful comments, which have improved the analysis presented in this manuscript. FU and MG acknowledge the financial contribution from contract PRIN 2022 - CUP J53D23001610006. AB and MB acknowledge support from the ERC through the ERC Co Grant $\vec B ELOVED$ GA n 101169773. TP and FdG acknowledge support from the ERC Consolidator Grant ULU 101086378. 
      MB acknowledge the financial contribution from the INAF GO grant 1.05.24.02.10 {\it Extended Radio Emission in Galaxy Clusters at deep focus with MeerKAT}. PDF acknowledges the support provided by PAPIIT-DGAPA-UNAM through Project No. IN108526 and the Future Faculty Leaders Fellowship at the Center for Astrophysics, Harvard-Smithsonian. The MeerKAT telescope is operated by the South African Radio Astronomy Observatory, which is a facility of the National Research Foundation, an agency of the Department of Science and Innovation. We thank the staff of the GMRT that made these observations possible. GMRT is run by the National Centre for Radio Astrophysics of the Tata Institute of Fundamental Research. Basic research in radio astronomy at the Naval Research Laboratory is supported by 6.1 Base funding. LOFAR is the Low Frequency Array designed and constructed by ASTRON. It has observing, data processing, and data storage facilities in several countries, which are owned by various parties (each with their own funding sources), and which are collectively operated by the LOFAR ERIC under a joint scientific policy. The LOFAR resources have benefited from the following recent major funding sources: CNRS-INSU, Observatoire de Paris and Université d'Orléans, France; BMFTR, MKW-NRW, MPG, Germany; Science Foundation Ireland (SFI), Department of Business, Enterprise and Innovation (DBEI), Ireland; NWO, The Netherlands; The Science and Technology Facilities Council, UK; Ministry of Science and Higher Education, Poland; The Istituto Nazionale di Astrofisica (INAF), Italy. This research made use of the Dutch national e-infrastructure with support of the SURF Cooperative (e-infra 180169) and the LOFAR e-infra group. The Jülich LOFAR Long Term Archive and the German LOFAR network are both coordinated and operated by the Jülich Supercomputing Centre (JSC), and computing resources on the supercomputer JUWELS at JSC were provided by the Gauss Centre for Supercomputing e.V. (grant CHTB00) through the John von Neumann Institute for Computing (NIC). This research made use of the University of Hertfordshire high-performance computing facility and the LOFAR-UK computing facility located at the University of Hertfordshire and supported by STFC [ST/P000096/1], and of the Italian LOFAR-IT computing infrastructure supported and operated by INAF, including the resources within the PLEIADI special ``LOFAR'' project by USC-C of INAF, and by the Physics Department of Turin university (under an agreement with Consorzio Interuniversitario per la Fisica Spaziale) at the C3S Supercomputing Centre, Italy. This research is part of the project LOFAR Data Valorization (LDV) [project numbers 2020.031, 2022.033, and 2024.047] of the research program Computing Time on National Computer Facilities using SPIDER that is (co-)funded by the Dutch Research Council (NWO), hosted by SURF through the call for proposals of Computing Time on National Computer Facilities. 
\end{acknowledgements}

%
   \bibliographystyle{aa} 
   \bibliography{aa60664-26.bib} 

@ARTICLE{vanHaarlem2013,
       author = {{van Haarlem}, M.~P. and {Wise}, M.~W. and {Gunst}, A.~W. and {Heald}, G. and {McKean}, J.~P. and {Hessels}, J.~W.~T. and {de Bruyn}, A.~G. and {Nijboer}, R. and {Swinbank}, J. and {Fallows}, R. and {Brentjens}, M. and {Nelles}, A. and {Beck}, R. and {Falcke}, H. and {Fender}, R. and {H{\"o}randel}, J. and {Koopmans}, L.~V.~E. and {Mann}, G. and {Miley}, G. and {R{\"o}ttgering}, H. and {Stappers}, B.~W. and {Wijers}, R.~A.~M.~J. and {Zaroubi}, S. and {van den Akker}, M. and {Alexov}, A. and {Anderson}, J. and {Anderson}, K. and {van Ardenne}, A. and {Arts}, M. and {Asgekar}, A. and {Avruch}, I.~M. and {Batejat}, F. and {B{\"a}hren}, L. and {Bell}, M.~E. and {Bell}, M.~R. and {van Bemmel}, I. and {Bennema}, P. and {Bentum}, M.~J. and {Bernardi}, G. and {Best}, P. and {B{\^\i}rzan}, L. and {Bonafede}, A. and {Boonstra}, A. -J. and {Braun}, R. and {Bregman}, J. and {Breitling}, F. and {van de Brink}, R.~H. and {Broderick}, J. and {Broekema}, P.~C. and {Brouw}, W.~N. and {Br{\"u}ggen}, M. and {Butcher}, H.~R. and {van Cappellen}, W. and {Ciardi}, B. and {Coenen}, T. and {Conway}, J. and {Coolen}, A. and {Corstanje}, A. and {Damstra}, S. and {Davies}, O. and {Deller}, A.~T. and {Dettmar}, R. -J. and {van Diepen}, G. and {Dijkstra}, K. and {Donker}, P. and {Doorduin}, A. and {Dromer}, J. and {Drost}, M. and {van Duin}, A. and {Eisl{\"o}ffel}, J. and {van Enst}, J. and {Ferrari}, C. and {Frieswijk}, W. and {Gankema}, H. and {Garrett}, M.~A. and {de Gasperin}, F. and {Gerbers}, M. and {de Geus}, E. and {Grie{\ss}meier}, J. -M. and {Grit}, T. and {Gruppen}, P. and {Hamaker}, J.~P. and {Hassall}, T. and {Hoeft}, M. and {Holties}, H.~A. and {Horneffer}, A. and {van der Horst}, A. and {van Houwelingen}, A. and {Huijgen}, A. and {Iacobelli}, M. and {Intema}, H. and {Jackson}, N. and {Jelic}, V. and {de Jong}, A. and {Juette}, E. and {Kant}, D. and {Karastergiou}, A. and {Koers}, A. and {Kollen}, H. and {Kondratiev}, V.~I. and {Kooistra}, E. and {Koopman}, Y. and {Koster}, A. and {Kuniyoshi}, M. and {Kramer}, M. and {Kuper}, G. and {Lambropoulos}, P. and {Law}, C. and {van Leeuwen}, J. and {Lemaitre}, J. and {Loose}, M. and {Maat}, P. and {Macario}, G. and {Markoff}, S. and {Masters}, J. and {McFadden}, R.~A. and {McKay-Bukowski}, D. and {Meijering}, H. and {Meulman}, H. and {Mevius}, M. and {Middelberg}, E. and {Millenaar}, R. and {Miller-Jones}, J.~C.~A. and {Mohan}, R.~N. and {Mol}, J.~D. and {Morawietz}, J. and {Morganti}, R. and {Mulcahy}, D.~D. and {Mulder}, E. and {Munk}, H. and {Nieuwenhuis}, L. and {van Nieuwpoort}, R. and {Noordam}, J.~E. and {Norden}, M. and {Noutsos}, A. and {Offringa}, A.~R. and {Olofsson}, H. and {Omar}, A. and {Orr{\'u}}, E. and {Overeem}, R. and {Paas}, H. and {Pandey-Pommier}, M. and {Pandey}, V.~N. and {Pizzo}, R. and {Polatidis}, A. and {Rafferty}, D. and {Rawlings}, S. and {Reich}, W. and {de Reijer}, J. -P. and {Reitsma}, J. and {Renting}, G.~A. and {Riemers}, P. and {Rol}, E. and {Romein}, J.~W. and {Roosjen}, J. and {Ruiter}, M. and {Scaife}, A. and {van der Schaaf}, K. and {Scheers}, B. and {Schellart}, P. and {Schoenmakers}, A. and {Schoonderbeek}, G. and {Serylak}, M. and {Shulevski}, A. and {Sluman}, J. and {Smirnov}, O. and {Sobey}, C. and {Spreeuw}, H. and {Steinmetz}, M. and {Sterks}, C.~G.~M. and {Stiepel}, H. -J. and {Stuurwold}, K. and {Tagger}, M. and {Tang}, Y. and {Tasse}, C. and {Thomas}, I. and {Thoudam}, S. and {Toribio}, M.~C. and {van der Tol}, B. and {Usov}, O. and {van Veelen}, M. and {van der Veen}, A. -J. and {ter Veen}, S. and {Verbiest}, J.~P.~W. and {Vermeulen}, R. and {Vermaas}, N. and {Vocks}, C. and {Vogt}, C. and {de Vos}, M. and {van der Wal}, E. and {van Weeren}, R. and {Weggemans}, H. and {Weltevrede}, P. and {White}, S. and {Wijnholds}, S.~J. and {Wilhelmsson}, T. and {Wucknitz}, O. and {Yatawatta}, S. and {Zarka}, P. and {Zensus}, A. and {van Zwieten}, J.},
        title = "{LOFAR: The LOw-Frequency ARray}",
      journal = {\aap},
         year = 2013,
        month = aug,
       volume = {556},
          eid = {A2},
        pages = {A2},
          doi = {10.1051/0004-6361/201220873},
archivePrefix = {arXiv},
       eprint = {1305.3550},
 primaryClass = {astro-ph.IM},
       adsurl = {https://ui.adsabs.harvard.edu/abs/2013A&A...556A...2V}
}

@ARTICLE{vanWeeren2026,
       author = {{van Weeren}, R.~J. and {Osinga}, E. and {Brunetti}, G. and {Riseley}, C.~J. and {Botteon}, A. and {Timmerman}, R. and {Bonafede}, A. and {Br{\"u}ggen}, M. and {Cassano}, R. and {Cuciti}, V. and {Dallacasa}, D. and {de Gasperin}, F. and {de Jong}, J.~M.~G.~H.~J. and {Gastaldello}, F. and {Knowles}, K. and {Zhang}, X.},
        title = "{MeerKAT observations of Abell 1775 and Abell 1795: the discovery of a hadronic radio halo?}",
      journal = {\mnras},
         year = 2026,
        month = feb,
       volume = {546},
       number = {2},
          eid = {stag054},
        pages = {stag054},
          doi = {10.1093/mnras/stag054},
archivePrefix = {arXiv},
       eprint = {2601.16288},
 primaryClass = {astro-ph.CO},
       adsurl = {https://ui.adsabs.harvard.edu/abs/2026MNRAS.546ag054V}
}

@ARTICLE{kolokythas2025,
       author = {{Kolokythas}, Konstantinos and {Venturi}, Tiziana and {Knowles}, Kenda and {Br{\"u}ggen}, Marcus and {de Gasperin}, Francesco and {Sikhosana}, S. Precious and {Clarke}, Tracy E. and {Smirnov}, Oleg and {Loubser}, S. Ilani and {Moodley}, Kavilian},
        title = "{The MeerKAT Galaxy Cluster Legacy Survey {\textendash} II. Catalogue of the diffuse radio emission in MeerKAT-GCLS clusters}",
      journal = {\mnras},
         year = 2025,
        month = oct,
       volume = {543},
       number = {2},
        pages = {1638-1704},
          doi = {10.1093/mnras/staf1499},
archivePrefix = {arXiv},
       eprint = {2509.05442},
 primaryClass = {astro-ph.GA},
       adsurl = {https://ui.adsabs.harvard.edu/abs/2025MNRAS.543.1638K}
}

@ARTICLE{planck_2016_IIDatarelease,
       author = {{Planck Collaboration} and {Ade}, P.~A.~R. and {Aghanim}, N. and {Arnaud}, M. and {Ashdown}, M. and {Aumont}, J. and {Baccigalupi}, C. and {Banday}, A.~J. and {Barreiro}, R.~B. and {Barrena}, R. and {Bartlett}, J.~G. and {Bartolo}, N. and {Battaner}, E. and {Battye}, R. and {Benabed}, K. and {Beno{\^\i}t}, A. and {Benoit-L{\'e}vy}, A. and {Bernard}, J.-P. and {Bersanelli}, M. and {Bielewicz}, P. and {Bikmaev}, I. and {B{\"o}hringer}, H. and {Bonaldi}, A. and {Bonavera}, L. and {Bond}, J.~R. and {Borrill}, J. and {Bouchet}, F.~R. and {Bucher}, M. and {Burenin}, R. and {Burigana}, C. and {Butler}, R.~C. and {Calabrese}, E. and {Cardoso}, J.-F. and {Carvalho}, P. and {Catalano}, A. and {Challinor}, A. and {Chamballu}, A. and {Chary}, R.-R. and {Chiang}, H.~C. and {Chon}, G. and {Christensen}, P.~R. and {Clements}, D.~L. and {Colombi}, S. and {Colombo}, L.~P.~L. and {Combet}, C. and {Comis}, B. and {Couchot}, F. and {Coulais}, A. and {Crill}, B.~P. and {Curto}, A. and {Cuttaia}, F. and {Dahle}, H. and {Danese}, L. and {Davies}, R.~D. and {Davis}, R.~J. and {de Bernardis}, P. and {de Rosa}, A. and {de Zotti}, G. and {Delabrouille}, J. and {D{\'e}sert}, F.-X. and {Dickinson}, C. and {Diego}, J.~M. and {Dolag}, K. and {Dole}, H. and {Donzelli}, S. and {Dor{\'e}}, O. and {Douspis}, M. and {Ducout}, A. and {Dupac}, X. and {Efstathiou}, G. and {Eisenhardt}, P.~R.~M. and {Elsner}, F. and {En{\ss}lin}, T.~A. and {Eriksen}, H.~K. and {Falgarone}, E. and {Fergusson}, J. and {Feroz}, F. and {Ferragamo}, A. and {Finelli}, F. and {Forni}, O. and {Frailis}, M. and {Fraisse}, A.~A. and {Franceschi}, E. and {Frejsel}, A. and {Galeotta}, S. and {Galli}, S. and {Ganga}, K. and {G{\'e}nova-Santos}, R.~T. and {Giard}, M. and {Giraud-H{\'e}raud}, Y. and {Gjerl{\o}w}, E. and {Gonz{\'a}lez-Nuevo}, J. and {G{\'o}rski}, K.~M. and {Grainge}, K.~J.~B. and {Gratton}, S. and {Gregorio}, A. and {Gruppuso}, A. and {Gudmundsson}, J.~E. and {Hansen}, F.~K. and {Hanson}, D. and {Harrison}, D.~L. and {Hempel}, A. and {Henrot-Versill{\'e}}, S. and {Hern{\'a}ndez-Monteagudo}, C. and {Herranz}, D. and {Hildebrandt}, S.~R. and {Hivon}, E. and {Hobson}, M. and {Holmes}, W.~A. and {Hornstrup}, A. and {Hovest}, W. and {Huffenberger}, K.~M. and {Hurier}, G. and {Jaffe}, A.~H. and {Jaffe}, T.~R. and {Jin}, T. and {Jones}, W.~C. and {Juvela}, M. and {Keih{\"a}nen}, E. and {Keskitalo}, R. and {Khamitov}, I. and {Kisner}, T.~S. and {Kneissl}, R. and {Knoche}, J. and {Kunz}, M. and {Kurki-Suonio}, H. and {Lagache}, G. and {Lamarre}, J.-M. and {Lasenby}, A. and {Lattanzi}, M. and {Lawrence}, C.~R. and {Leonardi}, R. and {Lesgourgues}, J. and {Levrier}, F. and {Liguori}, M. and {Lilje}, P.~B. and {Linden-V{\o}rnle}, M. and {L{\'o}pez-Caniego}, M. and {Lubin}, P.~M. and {Mac{\'\i}as-P{\'e}rez}, J.~F. and {Maggio}, G. and {Maino}, D. and {Mak}, D.~S.~Y. and {Mandolesi}, N. and {Mangilli}, A. and {Martin}, P.~G. and {Mart{\'\i}nez-Gonz{\'a}lez}, E. and {Masi}, S. and {Matarrese}, S. and {Mazzotta}, P. and {McGehee}, P. and {Mei}, S. and {Melchiorri}, A. and {Melin}, J.-B. and {Mendes}, L. and {Mennella}, A. and {Migliaccio}, M. and {Mitra}, S. and {Miville-Desch{\^e}nes}, M.-A. and {Moneti}, A. and {Montier}, L. and {Morgante}, G. and {Mortlock}, D. and {Moss}, A. and {Munshi}, D. and {Murphy}, J.~A. and {Naselsky}, P. and {Nastasi}, A. and {Nati}, F. and {Natoli}, P. and {Netterfield}, C.~B. and {N{\o}rgaard-Nielsen}, H.~U. and {Noviello}, F. and {Novikov}, D. and {Novikov}, I. and {Olamaie}, M. and {Oxborrow}, C.~A. and {Paci}, F. and {Pagano}, L. and {Pajot}, F. and {Paoletti}, D. and {Pasian}, F. and {Patanchon}, G. and {Pearson}, T.~J. and {Perdereau}, O. and {Perotto}, L. and {Perrott}, Y.~C. and {Perrotta}, F. and {Pettorino}, V. and {Piacentini}, F. and {Piat}, M. and {Pierpaoli}, E. and {Pietrobon}, D. and {Plaszczynski}, S. and {Pointecouteau}, E. and {Polenta}, G. and {Pratt}, G.~W. and {Pr{\'e}zeau}, G. and {Prunet}, S. and {Puget}, J.-L.},
        title = "{Planck 2015 results. XXVII. The second Planck catalogue of Sunyaev-Zeldovich sources}",
      journal = {\aap},
         year = 2016,
        month = sep,
       volume = {594},
          eid = {A27},
        pages = {A27},
          doi = {10.1051/0004-6361/201525823},
archivePrefix = {arXiv},
       eprint = {1502.01598},
 primaryClass = {astro-ph.CO},
       adsurl = {https://ui.adsabs.harvard.edu/abs/2016A&A...594A..27P}
}

@ARTICLE{bellomi2025,
       author = {{Bellomi}, Elena and {ZuHone}, John A. and {Truong}, Nhut and {Zhuravleva}, Irina and {Weinberger}, Rainer and {Pfrommer}, Christoph and {Zhang}, Congyao and {Heinrich}, Annie and {Ruszkowski}, Mateusz and {McNamara}, Brian and {Hlavacek-Larrondo}, Julie and {Gendron-Marsolais}, Marie-Lou and {Vigneron}, Benjamin},
        title = "{Disentangling AGN Feedback and Sloshing in the Perseus Cluster with XRISM: Insights from Simulations}",
      journal = {arXiv e-prints},
         year = 2025,
        month = dec,
          eid = {arXiv:2512.12754},
        pages = {arXiv:2512.12754},
          doi = {10.48550/arXiv.2512.12754},
archivePrefix = {arXiv},
       eprint = {2512.12754},
 primaryClass = {astro-ph.HE},
       adsurl = {https://ui.adsabs.harvard.edu/abs/2025arXiv251212754B},
      note    = {submitted to \apj},
}

@ARTICLE{riseley2023,
       author = {{Riseley}, C.~J. and {Biava}, N. and {Lusetti}, G. and {Bonafede}, A. and {Bonnassieux}, E. and {Botteon}, A. and {Loi}, F. and {Brunetti}, G. and {Cassano}, R. and {Osinga}, E. and {Rajpurohit}, K. and {R{\"o}ttgering}, H.~J.~A. and {Shimwell}, T. and {Timmerman}, R. and {van Weeren}, R.~J.},
        title = "{A MeerKAT-meets-LOFAR study of Abell 1413: a moderately disturbed non-cool-core cluster hosting a 500 kpc 'mini'-halo}",
      journal = {\mnras},
         year = 2023,
        month = oct,
       volume = {524},
       number = {4},
        pages = {6052-6070},
          doi = {10.1093/mnras/stad2218},
archivePrefix = {arXiv},
       eprint = {2308.01464},
 primaryClass = {astro-ph.GA},
       adsurl = {https://ui.adsabs.harvard.edu/abs/2023MNRAS.524.6052R}
}

@ARTICLE{scaife2012,
       author = {{Scaife}, Anna M.~M. and {Heald}, George H.},
        title = "{A broad-band flux scale for low-frequency radio telescopes}",
      journal = {\mnras},
         year = 2012,
        month = jun,
       volume = {423},
       number = {1},
        pages = {L30-L34},
          doi = {10.1111/j.1745-3933.2012.01251.x},
archivePrefix = {arXiv},
       eprint = {1203.0977},
 primaryClass = {astro-ph.IM},
       adsurl = {https://ui.adsabs.harvard.edu/abs/2012MNRAS.423L..30S}
}

@ARTICLE{shimwell19,
       author = {{Shimwell}, T.~W. and {Tasse}, C. and {Hardcastle}, M.~J. and {Mechev}, A.~P. and {Williams}, W.~L. and {Best}, P.~N. and {R{\"o}ttgering}, H.~J.~A. and {Callingham}, J.~R. and {Dijkema}, T.~J. and {de Gasperin}, F. and {Hoang}, D.~N. and {Hugo}, B. and {Mirmont}, M. and {Oonk}, J.~B.~R. and {Prandoni}, I. and {Rafferty}, D. and {Sabater}, J. and {Smirnov}, O. and {van Weeren}, R.~J. and {White}, G.~J. and {Atemkeng}, M. and {Bester}, L. and {Bonnassieux}, E. and {Br{\"u}ggen}, M. and {Brunetti}, G. and {Chy{\.z}y}, K.~T. and {Cochrane}, R. and {Conway}, J.~E. and {Croston}, J.~H. and {Danezi}, A. and {Duncan}, K. and {Haverkorn}, M. and {Heald}, G.~H. and {Iacobelli}, M. and {Intema}, H.~T. and {Jackson}, N. and {Jamrozy}, M. and {Jarvis}, M.~J. and {Lakhoo}, R. and {Mevius}, M. and {Miley}, G.~K. and {Morabito}, L. and {Morganti}, R. and {Nisbet}, D. and {Orr{\'u}}, E. and {Perkins}, S. and {Pizzo}, R.~F. and {Schrijvers}, C. and {Smith}, D.~J.~B. and {Vermeulen}, R. and {Wise}, M.~W. and {Alegre}, L. and {Bacon}, D.~J. and {van Bemmel}, I.~M. and {Beswick}, R.~J. and {Bonafede}, A. and {Botteon}, A. and {Bourke}, S. and {Brienza}, M. and {Calistro Rivera}, G. and {Cassano}, R. and {Clarke}, A.~O. and {Conselice}, C.~J. and {Dettmar}, R.~J. and {Drabent}, A. and {Dumba}, C. and {Emig}, K.~L. and {En{\ss}lin}, T.~A. and {Ferrari}, C. and {Garrett}, M.~A. and {G{\'e}nova-Santos}, R.~T. and {Goyal}, A. and {G{\"u}rkan}, G. and {Hale}, C. and {Harwood}, J.~J. and {Heesen}, V. and {Hoeft}, M. and {Horellou}, C. and {Jackson}, C. and {Kokotanekov}, G. and {Kondapally}, R. and {Kunert-Bajraszewska}, M. and {Mahatma}, V. and {Mahony}, E.~K. and {Mandal}, S. and {McKean}, J.~P. and {Merloni}, A. and {Mingo}, B. and {Miskolczi}, A. and {Mooney}, S. and {Nikiel-Wroczy{\'n}ski}, B. and {O'Sullivan}, S.~P. and {Quinn}, J. and {Reich}, W. and {Roskowi{\'n}ski}, C. and {Rowlinson}, A. and {Savini}, F. and {Saxena}, A. and {Schwarz}, D.~J. and {Shulevski}, A. and {Sridhar}, S.~S. and {Stacey}, H.~R. and {Urquhart}, S. and {van der Wiel}, M.~H.~D. and {Varenius}, E. and {Webster}, B. and {Wilber}, A.},
        title = "{The LOFAR Two-metre Sky Survey. II. First data release}",
      journal = {\aap},
         year = 2019,
        month = feb,
       volume = {622},
          eid = {A1},
        pages = {A1},
          doi = {10.1051/0004-6361/201833559},
archivePrefix = {arXiv},
       eprint = {1811.07926},
 primaryClass = {astro-ph.GA},
       adsurl = {https://ui.adsabs.harvard.edu/abs/2019A&A...622A...1S}
}

@ARTICLE{Shimwell22,
       author = {{Shimwell}, T.~W. and {Hardcastle}, M.~J. and {Tasse}, C. and {Best}, P.~N. and {R{\"o}ttgering}, H.~J.~A. and {Williams}, W.~L. and {Botteon}, A. and {Drabent}, A. and {Mechev}, A. and {Shulevski}, A. and {van Weeren}, R.~J. and {Bester}, L. and {Br{\"u}ggen}, M. and {Brunetti}, G. and {Callingham}, J.~R. and {Chy{\.z}y}, K.~T. and {Conway}, J.~E. and {Dijkema}, T.~J. and {Duncan}, K. and {de Gasperin}, F. and {Hale}, C.~L. and {Haverkorn}, M. and {Hugo}, B. and {Jackson}, N. and {Mevius}, M. and {Miley}, G.~K. and {Morabito}, L.~K. and {Morganti}, R. and {Offringa}, A. and {Oonk}, J.~B.~R. and {Rafferty}, D. and {Sabater}, J. and {Smith}, D.~J.~B. and {Schwarz}, D.~J. and {Smirnov}, O. and {O'Sullivan}, S.~P. and {Vedantham}, H. and {White}, G.~J. and {Albert}, J.~G. and {Alegre}, L. and {Asabere}, B. and {Bacon}, D.~J. and {Bonafede}, A. and {Bonnassieux}, E. and {Brienza}, M. and {Bilicki}, M. and {Bonato}, M. and {Calistro Rivera}, G. and {Cassano}, R. and {Cochrane}, R. and {Croston}, J.~H. and {Cuciti}, V. and {Dallacasa}, D. and {Danezi}, A. and {Dettmar}, R.~J. and {Di Gennaro}, G. and {Edler}, H.~W. and {En{\ss}lin}, T.~A. and {Emig}, K.~L. and {Franzen}, T.~M.~O. and {Garc{\'\i}a-Vergara}, C. and {Grange}, Y.~G. and {G{\"u}rkan}, G. and {Hajduk}, M. and {Heald}, G. and {Heesen}, V. and {Hoang}, D.~N. and {Hoeft}, M. and {Horellou}, C. and {Iacobelli}, M. and {Jamrozy}, M. and {Jeli{\'c}}, V. and {Kondapally}, R. and {Kukreti}, P. and {Kunert-Bajraszewska}, M. and {Magliocchetti}, M. and {Mahatma}, V. and {Ma{\l}ek}, K. and {Mandal}, S. and {Massaro}, F. and {Meyer-Zhao}, Z. and {Mingo}, B. and {Mostert}, R.~I.~J. and {Nair}, D.~G. and {Nakoneczny}, S.~J. and {Nikiel-Wroczy{\'n}ski}, B. and {Orr{\'u}}, E. and {Pajdosz-{\'S}mierciak}, U. and {Pasini}, T. and {Prandoni}, I. and {van Piggelen}, H.~E. and {Rajpurohit}, K. and {Retana-Montenegro}, E. and {Riseley}, C.~J. and {Rowlinson}, A. and {Saxena}, A. and {Schrijvers}, C. and {Sweijen}, F. and {Siewert}, T.~M. and {Timmerman}, R. and {Vaccari}, M. and {Vink}, J. and {West}, J.~L. and {Wo{\l}owska}, A. and {Zhang}, X. and {Zheng}, J.},
        title = "{The LOFAR Two-metre Sky Survey. V. Second data release}",
      journal = {\aap},
         year = 2022,
        month = mar,
       volume = {659},
          eid = {A1},
        pages = {A1},
          doi = {10.1051/0004-6361/202142484},
archivePrefix = {arXiv},
       eprint = {2202.11733},
 primaryClass = {astro-ph.GA},
       adsurl = {https://ui.adsabs.harvard.edu/abs/2022A&A...659A...1S}
}

@MISC{vanDiepen18,
       author = {{van Diepen}, Ger and {Dijkema}, Tammo Jan and {Offringa}, Andr{\'e}},
        title = "{DPPP: Default Pre-Processing Pipeline}",
 howpublished = {Astrophysics Source Code Library, record ascl:1804.003},
         year = 2018,
        month = apr,
          eid = {ascl:1804.003},
        pages = {ascl:1804.003},
archivePrefix = {ascl},
       eprint = {1804.003},
       adsurl = {https://ui.adsabs.harvard.edu/abs/2018ascl.soft04003V}
}

@ARTICLE{Ubertosi2024c,
       author = {{Ubertosi}, F. and {Giacintucci}, S. and {Clarke}, T. and {Markevitch}, M. and {Venturi}, T. and {O'Sullivan}, E. and {Gitti}, M.},
        title = "{Multi-epoch jet outbursts in Abell 496: Synchrotron ageing and buoyant X-ray cavities draped by warm gas filaments}",
      journal = {\aap},
         year = 2024,
        month = nov,
       volume = {691},
          eid = {A294},
        pages = {A294},
          doi = {10.1051/0004-6361/202451766},
archivePrefix = {arXiv},
       eprint = {2409.15440},
 primaryClass = {astro-ph.GA},
       adsurl = {https://ui.adsabs.harvard.edu/abs/2024A&A...691A.294U}
}

@ARTICLE{pfrommerensslin2003,
       author = {{Pfrommer}, C. and {En{\ss}lin}, T.~A.},
        title = "{Probing the cosmic ray population of the giant elliptical galaxy M 87 with observed TeV gamma -rays}",
      journal = {\aap},
         year = 2003,
        month = aug,
       volume = {407},
        pages = {L73-L77},
          doi = {10.1051/0004-6361:20031088},
archivePrefix = {arXiv},
       eprint = {astro-ph/0306258},
 primaryClass = {astro-ph},
       adsurl = {https://ui.adsabs.harvard.edu/abs/2003A&A...407L..73P}
}

@ARTICLE{brunetti2008,
       author = {{Brunetti}, G. and {Giacintucci}, S. and {Cassano}, R. and {Lane}, W. and {Dallacasa}, D. and {Venturi}, T. and {Kassim}, N.~E. and {Setti}, G. and {Cotton}, W.~D. and {Markevitch}, M.},
        title = "{A low-frequency radio halo associated with a cluster of galaxies}",
      journal = {\nat},
         year = 2008,
        month = oct,
       volume = {455},
       number = {7215},
        pages = {944-947},
          doi = {10.1038/nature07379},
archivePrefix = {arXiv},
       eprint = {0810.4288},
 primaryClass = {astro-ph},
       adsurl = {https://ui.adsabs.harvard.edu/abs/2008Natur.455..944B}
}

@ARTICLE{Harwood13,
       author = {{Harwood}, Jeremy J. and {Hardcastle}, Martin J. and {Croston}, Judith H. and {Goodger}, Joanna L.},
        title = "{Spectral ageing in the lobes of FR-II radio galaxies: new methods of analysis for broad-band radio data}",
      journal = {\mnras},
         year = 2013,
        month = nov,
       volume = {435},
       number = {4},
        pages = {3353-3375},
          doi = {10.1093/mnras/stt1526},
archivePrefix = {arXiv},
       eprint = {1308.4137},
 primaryClass = {astro-ph.CO},
       adsurl = {https://ui.adsabs.harvard.edu/abs/2013MNRAS.435.3353H}
}

@ARTICLE{chandra2004,
       author = {{Chandra}, P. and {Ray}, A. and {Bhatnagar}, S.},
        title = "{The Late-Time Radio Emission from SN 1993J at Meter Wavelengths}",
      journal = {\apj},
         year = 2004,
        month = sep,
       volume = {612},
       number = {2},
        pages = {974-987},
          doi = {10.1086/422675},
archivePrefix = {arXiv},
       eprint = {astro-ph/0405448},
 primaryClass = {astro-ph},
       adsurl = {https://ui.adsabs.harvard.edu/abs/2004ApJ...612..974C}
}

@ARTICLE{deGasperin2023,
       author = {{de Gasperin}, F. and {Edler}, H.~W. and {Williams}, W.~L. and {Callingham}, J.~R. and {Asabere}, B. and {Br{\"u}ggen}, M. and {Brunetti}, G. and {Dijkema}, T.~J. and {Hardcastle}, M.~J. and {Iacobelli}, M. and {Offringa}, A. and {Norden}, M.~J. and {R{\"o}ttgering}, H.~J.~A. and {Shimwell}, T. and {van Weeren}, R.~J. and {Tasse}, C. and {Bomans}, D.~J. and {Bonafede}, A. and {Botteon}, A. and {Cassano}, R. and {Chy{\.z}y}, K.~T. and {Cuciti}, V. and {Emig}, K.~L. and {Kadler}, M. and {Miley}, G. and {Mingo}, B. and {Oei}, M.~S.~S.~L. and {Prandoni}, I. and {Schwarz}, D.~J. and {Zarka}, P.},
        title = "{The LOFAR LBA Sky Survey. II. First data release}",
      journal = {\aap},
         year = 2023,
        month = may,
       volume = {673},
          eid = {A165},
        pages = {A165},
          doi = {10.1051/0004-6361/202245389},
archivePrefix = {arXiv},
       eprint = {2301.12724},
 primaryClass = {astro-ph.IM},
       adsurl = {https://ui.adsabs.harvard.edu/abs/2023A&A...673A.165D}
}

@ARTICLE{hoang2025,
       author = {{Hoang}, D.~N. and {Br{\"u}ggen}, M. and {Bonafede}, A. and {Koch}, P.~M. and {Brunetti}, G. and {Bulbul}, E. and {Di Gennaro}, G. and {Liu}, A. and {Riseley}, C.~J. and {R{\"o}ttgering}, H.~J.~A. and {van Weeren}, R.~J.},
        title = "{Discovery of large-scale radio emission enveloping the mini-halo in the most X-ray luminous galaxy cluster RX J1347.5{\textendash}1145}",
      journal = {\aap},
         year = 2025,
        month = mar,
       volume = {695},
          eid = {A240},
        pages = {A240},
          doi = {10.1051/0004-6361/202452052},
archivePrefix = {arXiv},
       eprint = {2502.09472},
 primaryClass = {astro-ph.HE},
       adsurl = {https://ui.adsabs.harvard.edu/abs/2025A&A...695A.240H}
}

@ARTICLE{Hogan15b,
       author = {{Hogan}, M.~T. and {Edge}, A.~C. and {Hlavacek-Larrondo}, J. and {Grainge}, K.~J.~B. and {Hamer}, S.~L. and {Mahony}, E.~K. and {Russell}, H.~R. and {Fabian}, A.~C. and {McNamara}, B.~R. and {Wilman}, R.~J.},
        title = "{A comprehensive study of the radio properties of brightest cluster galaxies}",
      journal = {\mnras},
         year = 2015,
        month = oct,
       volume = {453},
       number = {2},
        pages = {1201-1222},
          doi = {10.1093/mnras/stv1517},
archivePrefix = {arXiv},
       eprint = {1507.03019},
 primaryClass = {astro-ph.GA},
       adsurl = {https://ui.adsabs.harvard.edu/abs/2015MNRAS.453.1201H}
}

@ARTICLE{rafferty2008,
       author = {{Rafferty}, D.~A. and {McNamara}, B.~R. and {Nulsen}, P.~E.~J.},
        title = "{The Regulation of Cooling and Star Formation in Luminous Galaxies by Active Galactic Nucleus Feedback and the Cooling-Time/Entropy Threshold for the Onset of Star Formation}",
      journal = {\apj},
         year = 2008,
        month = nov,
       volume = {687},
       number = {2},
        pages = {899-918},
          doi = {10.1086/591240},
archivePrefix = {arXiv},
       eprint = {0802.1864},
 primaryClass = {astro-ph},
       adsurl = {https://ui.adsabs.harvard.edu/abs/2008ApJ...687..899R}
}

@ARTICLE{macario2010,
       author = {{Macario}, G. and {Venturi}, T. and {Brunetti}, G. and {Dallacasa}, D. and {Giacintucci}, S. and {Cassano}, R. and {Bardelli}, S. and {Athreya}, R.},
        title = "{The very steep spectrum radio halo in Abell 697}",
      journal = {\aap},
         year = 2010,
        month = jul,
       volume = {517},
          eid = {A43},
        pages = {A43},
          doi = {10.1051/0004-6361/201014109},
archivePrefix = {arXiv},
       eprint = {1004.1515},
 primaryClass = {astro-ph.CO},
       adsurl = {https://ui.adsabs.harvard.edu/abs/2010A&A...517A..43M}
}

@ARTICLE{Doria12,
       author = {{Doria}, Alberto and {Gitti}, Myriam and {Ettori}, Stefano and {Brighenti}, Fabrizio and {Nulsen}, Paul E.~J. and {McNamara}, Brian R.},
        title = "{A Chandra-VLA Investigation of the X-Ray Cavity System and Radio Mini-Halo in the Galaxy Cluster RBS 797}",
      journal = {\apj},
         year = 2012,
        month = jul,
       volume = {753},
       number = {1},
          eid = {47},
        pages = {47},
          doi = {10.1088/0004-637X/753/1/47},
archivePrefix = {arXiv},
       eprint = {1204.6191},
 primaryClass = {astro-ph.CO},
       adsurl = {https://ui.adsabs.harvard.edu/abs/2012ApJ...753...47D}
}

@PREAMBLE{ "\newcommand{\noopsort}[1]{} "
# "\newcommand{\printfirst}[2]{#1} "
# "\newcommand{\singleletter}[1]{#1} "
# "\newcommand{\switchargs}[2]{#2#1} " }

@ARTICLE{ettori2013,
       author = {{Ettori}, S. and {Gastaldello}, F. and {Gitti}, M. and {O'Sullivan}, E. and {Gaspari}, M. and {Brighenti}, F. and {David}, L. and {Edge}, A.~C.},
        title = "{Cold fronts and metal anisotropies in the X-ray cool core of the galaxy cluster Zw 1742+3306}",
      journal = {\aap},
         year = 2013,
        month = jul,
       volume = {555},
          eid = {A93},
        pages = {A93},
          doi = {10.1051/0004-6361/201321107},
archivePrefix = {arXiv},
       eprint = {1305.3926},
 primaryClass = {astro-ph.CO},
       adsurl = {https://ui.adsabs.harvard.edu/abs/2013A&A...555A..93E}
}

@ARTICLE{giacintucci2014,
       author = {{Giacintucci}, Simona and {Markevitch}, Maxim and {Venturi}, Tiziana and {Clarke}, Tracy E. and {Cassano}, Rossella and {Mazzotta}, Pasquale},
        title = "{New Detections of Radio Minihalos in Cool Cores of Galaxy Clusters}",
      journal = {\apj},
         year = 2014,
        month = jan,
       volume = {781},
       number = {1},
          eid = {9},
        pages = {9},
          doi = {10.1088/0004-637X/781/1/9},
archivePrefix = {arXiv},
       eprint = {1311.5248},
 primaryClass = {astro-ph.CO},
       adsurl = {https://ui.adsabs.harvard.edu/abs/2014ApJ...781....9G}
}

@ARTICLE{giacintucci2019,
       author = {{Giacintucci}, Simona and {Markevitch}, Maxim and {Cassano}, Rossella and {Venturi}, Tiziana and {Clarke}, Tracy E. and {Kale}, Ruta and {Cuciti}, Virginia},
        title = "{Expanding the Sample of Radio Minihalos in Galaxy Clusters}",
      journal = {\apj},
         year = 2019,
        month = aug,
       volume = {880},
       number = {2},
          eid = {70},
        pages = {70},
          doi = {10.3847/1538-4357/ab29f1},
archivePrefix = {arXiv},
       eprint = {1906.07087},
 primaryClass = {astro-ph.HE},
       adsurl = {https://ui.adsabs.harvard.edu/abs/2019ApJ...880...70G}
}

@ARTICLE{rotella2025,
       author = {{Rotella}, N. and {Ubertosi}, F. and {Gitti}, M. and {Rossetti}, M. and {Gastaldello}, F. and {Pratt}, G.~W. and {Brighenti}, F. and {Torresi}, E. and {Grandi}, P.},
        title = "{A JVLA, GMRT, and XMM study of Abell 795: Large-scale sloshing and a candidate radio phoenix}",
      journal = {\aap},
         year = 2025,
        month = may,
       volume = {697},
          eid = {A232},
        pages = {A232},
          doi = {10.1051/0004-6361/202553841},
archivePrefix = {arXiv},
       eprint = {2504.16178},
 primaryClass = {astro-ph.GA},
       adsurl = {https://ui.adsabs.harvard.edu/abs/2025A&A...697A.232R}
}

@ARTICLE{giacintucci2024,
       author = {{Giacintucci}, S. and {Venturi}, T. and {Markevitch}, M. and {Brunetti}, G. and {Clarke}, T.~E. and {Kale}, R.},
        title = "{Are Radio Minihalos Confined by Cold Fronts in Galaxy Clusters? Minihalos and Large-scale Sloshing in A3444 and MS 1455.0+2232}",
      journal = {\apj},
         year = 2024,
        month = jan,
       volume = {961},
       number = {1},
          eid = {133},
        pages = {133},
          doi = {10.3847/1538-4357/ad12bc},
archivePrefix = {arXiv},
       eprint = {2312.02824},
 primaryClass = {astro-ph.HE},
       adsurl = {https://ui.adsabs.harvard.edu/abs/2024ApJ...961..133G}
}

@ARTICLE{cuciti2021,
       author = {{Cuciti}, V. and {Cassano}, R. and {Brunetti}, G. and {Dallacasa}, D. and {de Gasperin}, F. and {Ettori}, S. and {Giacintucci}, S. and {Kale}, R. and {Pratt}, G.~W. and {van Weeren}, R.~J. and {Venturi}, T.},
        title = "{Radio halos in a mass-selected sample of 75 galaxy clusters. II. Statistical analysis}",
      journal = {\aap},
         year = 2021,
        month = mar,
       volume = {647},
          eid = {A51},
        pages = {A51},
          doi = {10.1051/0004-6361/202039208},
archivePrefix = {arXiv},
       eprint = {2101.01641},
 primaryClass = {astro-ph.CO},
       adsurl = {https://ui.adsabs.harvard.edu/abs/2021A&A...647A..51C}
}

@ARTICLE{brunetti2014,
       author = {{Brunetti}, Gianfranco and {Jones}, Thomas W.},
        title = "{Cosmic Rays in Galaxy Clusters and Their Nonthermal Emission}",
      journal = {Int. J. Mod. Phys D},
         year = 2014,
        month = mar,
       volume = {23},
       number = {4},
          eid = {1430007-98},
        pages = {1430007-98},
          doi = {10.1142/S0218271814300079},
archivePrefix = {arXiv},
       eprint = {1401.7519},
 primaryClass = {astro-ph.CO},
       adsurl = {https://ui.adsabs.harvard.edu/abs/2014IJMPD..2330007B}
}

@ARTICLE{knowles2019,
       author = {{Knowles}, Kenda and {Baker}, Andrew J. and {Bond}, J. Richard and {Gallardo}, Patricio A. and {Gupta}, Neeraj and {Hilton}, Matt and {Hughes}, John P. and {Intema}, Huib and {L{\'o}pez-Caraballo}, Carlos H. and {Moodley}, Kavilan and {Schmitt}, Benjamin L. and {Sievers}, Jonathan and {Sif{\'o}n}, Crist{\'o}bal and {Wollack}, Edward},
        title = "{GMRT 610 MHz observations of galaxy clusters in the ACT equatorial sample}",
      journal = {\mnras},
         year = 2019,
        month = jun,
       volume = {486},
       number = {1},
        pages = {1332-1349},
          doi = {10.1093/mnras/stz823},
archivePrefix = {arXiv},
       eprint = {1806.09579},
 primaryClass = {astro-ph.CO},
       adsurl = {https://ui.adsabs.harvard.edu/abs/2019MNRAS.486.1332K}
}

@ARTICLE{gupta2017,
       author = {{Gupta}, Y. and {Ajithkumar}, B. and {Kale}, H.~S. and {Nayak}, S. and {Sabhapathy}, S. and {Sureshkumar}, S. and {Swami}, R.~V. and {Chengalur}, J.~N. and {Ghosh}, S.~K. and {Ishwara-Chandra}, C.~H. and {Joshi}, B.~C. and {Kanekar}, N. and {Lal}, D.~V. and {Roy}, S.},
        title = "{The upgraded GMRT: opening new windows on the radio Universe}",
      journal = {Current Science},
         year = 2017,
        month = aug,
       volume = {113},
       number = {4},
        pages = {707-714},
          doi = {10.18520/cs/v113/i04/707-714},
       adsurl = {https://ui.adsabs.harvard.edu/abs/2017CSci..113..707G}
}

@ARTICLE{bagchi2002,
       author = {{Bagchi}, Joydeep and {En{\ss}lin}, Torsten A. and {Miniati}, Francesco and {Stalin}, C.~S. and {Singh}, M. and {Raychaudhury}, Somak and {Humeshkar}, N.~B.},
        title = "{Evidence for shock acceleration and intergalactic magnetic fields in a large-scale filament of galaxies ZwCl 2341.1+0000}",
      journal = {\na},
         year = 2002,
        month = jul,
       volume = {7},
       number = {5},
        pages = {249-277},
          doi = {10.1016/S1384-1076(02)00137-9},
archivePrefix = {arXiv},
       eprint = {astro-ph/0204389},
 primaryClass = {astro-ph},
       adsurl = {https://ui.adsabs.harvard.edu/abs/2002NewA....7..249B}
}

@ARTICLE{knowles2022,
       author = {{Knowles}, K. and {Cotton}, W.~D. and {Rudnick}, L. and {Camilo}, F. and {Goedhart}, S. and {Deane}, R. and {Ramatsoku}, M. and {Bietenholz}, M.~F. and {Br{\"u}ggen}, M. and {Button}, C. and {Chen}, H. and {Chibueze}, J.~O. and {Clarke}, T.~E. and {de Gasperin}, F. and {Ianjamasimanana}, R. and {J{\'o}zsa}, G.~I.~G. and {Hilton}, M. and {Kesebonye}, K.~C. and {Kolokythas}, K. and {Kraan-Korteweg}, R.~C. and {Lawrie}, G. and {Lochner}, M. and {Loubser}, S.~I. and {Marchegiani}, P. and {Mhlahlo}, N. and {Moodley}, K. and {Murphy}, E. and {Namumba}, B. and {Oozeer}, N. and {Parekh}, V. and {Pillay}, D.~S. and {Passmoor}, S.~S. and {Ramaila}, A.~J.~T. and {Ranchod}, S. and {Retana-Montenegro}, E. and {Sebokolodi}, L. and {Sikhosana}, S.~P. and {Smirnov}, O. and {Thorat}, K. and {Venturi}, T. and {Abbott}, T.~D. and {Adam}, R.~M. and {Adams}, G. and {Aldera}, M.~A. and {Bauermeister}, E.~F. and {Bennett}, T.~G.~H. and {Bode}, W.~A. and {Botha}, D.~H. and {Botha}, A.~G. and {Brederode}, L.~R.~S. and {Buchner}, S. and {Burger}, J.~P. and {Cheetham}, T. and {de Villiers}, D.~I.~L. and {Dikgale-Mahlakoana}, M.~A. and {du Toit}, L.~J. and {Esterhuyse}, S.~W.~P. and {Fadana}, G. and {Fanaroff}, B.~L. and {Fataar}, S. and {Foley}, A.~R. and {Fourie}, D.~J. and {Frank}, B.~S. and {Gamatham}, R.~R.~G. and {Gatsi}, T.~G. and {Geyer}, M. and {Gouws}, M. and {Gumede}, S.~C. and {Heywood}, I. and {Hlakola}, M.~J. and {Hokwana}, A. and {Hoosen}, S.~W. and {Horn}, D.~M. and {Horrell}, J.~M.~G. and {Hugo}, B.~V. and {Isaacson}, A.~R. and {Jonas}, J.~L. and {Jordaan}, J.~D.~B. and {Joubert}, A.~F. and {Julie}, R.~P.~M. and {Kapp}, F.~B. and {Kasper}, V.~A. and {Kenyon}, J.~S. and {Kotz{\'e}}, P.~P.~A. and {Kotze}, A.~G. and {Kriek}, N. and {Kriel}, H. and {Krishnan}, V.~K. and {Kusel}, T.~W. and {Legodi}, L.~S. and {Lehmensiek}, R. and {Liebenberg}, D. and {Lord}, R.~T. and {Lunsky}, B.~M. and {Madisa}, K. and {Magnus}, L.~G. and {Main}, J.~P.~L. and {Makhaba}, A. and {Makhathini}, S. and {Malan}, J.~A. and {Manley}, J.~R. and {Marais}, S.~J. and {Maree}, M.~D.~J. and {Martens}, A. and {Mauch}, T. and {McAlpine}, K. and {Merry}, B.~C. and {Millenaar}, R.~P. and {Mokone}, O.~J. and {Monama}, T.~E. and {Mphego}, M.~C. and {New}, W.~S. and {Ngcebetsha}, B. and {Ngoasheng}, K.~J. and {Ockards}, M.~T. and {Otto}, A.~J. and {Patel}, A.~A. and {Peens-Hough}, A. and {Perkins}, S.~J. and {Ramanujam}, N.~M. and {Ramudzuli}, Z.~R. and {Ratcliffe}, S.~M. and {Renil}, R. and {Robyntjies}, A. and {Rust}, A.~N. and {Salie}, S. and {Sambu}, N. and {Schollar}, C.~T.~G. and {Schwardt}, L.~C. and {Schwartz}, R.~L. and {Serylak}, M. and {Siebrits}, R. and {Sirothia}, S.~K. and {Slabber}, M. and {Sofeya}, L. and {Taljaard}, B. and {Tasse}, C. and {Tiplady}, A.~J. and {Toruvanda}, O. and {Twum}, S.~N. and {van Balla}, T.~J. and {van der Byl}, A. and {van der Merwe}, C. and {van Dyk}, C.~L. and {Van Tonder}, V. and {Van Wyk}, R. and {Venter}, A.~J. and {Venter}, M. and {Welz}, M.~G. and {Williams}, L.~P. and {Xaia}, B.},
        title = "{The MeerKAT Galaxy Cluster Legacy Survey. I. Survey Overview and Highlights}",
      journal = {\aap},
         year = 2022,
        month = jan,
       volume = {657},
          eid = {A56},
        pages = {A56},
          doi = {10.1051/0004-6361/202141488},
archivePrefix = {arXiv},
       eprint = {2111.05673},
 primaryClass = {astro-ph.GA},
       adsurl = {https://ui.adsabs.harvard.edu/abs/2022A&A...657A..56K}
}

@ARTICLE{botteon2022,
       author = {{Botteon}, A. and {Shimwell}, T.~W. and {Cassano}, R. and {Cuciti}, V. and {Zhang}, X. and {Bruno}, L. and {Camillini}, L. and {Natale}, R. and {Jones}, A. and {Gastaldello}, F. and {Simionescu}, A. and {Rossetti}, M. and {Akamatsu}, H. and {van Weeren}, R.~J. and {Brunetti}, G. and {Br{\"u}ggen}, M. and {Groeneveld}, C. and {Hoang}, D.~N. and {Hardcastle}, M.~J. and {Ignesti}, A. and {Di Gennaro}, G. and {Bonafede}, A. and {Drabent}, A. and {R{\"o}ttgering}, H.~J.~A. and {Hoeft}, M. and {de Gasperin}, F.},
        title = "{The Planck clusters in the LOFAR sky. I. LoTSS-DR2: New detections and sample overview}",
      journal = {\aap},
         year = 2022,
        month = apr,
       volume = {660},
          eid = {A78},
        pages = {A78},
          doi = {10.1051/0004-6361/202143020},
archivePrefix = {arXiv},
       eprint = {2202.11720},
 primaryClass = {astro-ph.CO},
       adsurl = {https://ui.adsabs.harvard.edu/abs/2022A&A...660A..78B}
}

@ARTICLE{pignataro2024c,
       author = {{Pignataro}, G.~V. and {Bonafede}, A. and {Bernardi}, G. and {Balboni}, M. and {Vazza}, F. and {van Weeren}, R.~J. and {Ubertosi}, F. and {Cassano}, R. and {Brunetti}, G. and {Botteon}, A. and {Venturi}, T. and {Akamatsu}, H. and {Drabent}, A. and {Hoeft}, M.},
        title = "{Mind the gap between A2061 and A2067: Unveiling new diffuse, large-scale radio emission}",
      journal = {\aap},
         year = 2024,
        month = nov,
       volume = {691},
          eid = {A99},
        pages = {A99},
          doi = {10.1051/0004-6361/202451529},
archivePrefix = {arXiv},
       eprint = {2409.15412},
 primaryClass = {astro-ph.CO},
       adsurl = {https://ui.adsabs.harvard.edu/abs/2024A&A...691A..99P}
}

@ARTICLE{giacintucci2025,
       author = {{Giacintucci}, S. and {Markevitch}, M. and {Clarke}, T. and {Wik}, Daniel R.},
        title = "{The Structure of the Giant Radio Fossil in the Ophiuchus Galaxy Cluster}",
      journal = {\apj},
         year = 2025,
        month = sep,
       volume = {990},
       number = {1},
          eid = {71},
        pages = {71},
          doi = {10.3847/1538-4357/adef4a},
archivePrefix = {arXiv},
       eprint = {2508.20190},
 primaryClass = {astro-ph.CO},
       adsurl = {https://ui.adsabs.harvard.edu/abs/2025ApJ...990...71G}
}

@ARTICLE{giacintucci2017,
       author = {{Giacintucci}, Simona and {Markevitch}, Maxim and {Cassano}, Rossella and {Venturi}, Tiziana and {Clarke}, Tracy E. and {Brunetti}, Gianfranco},
        title = "{Occurrence of Radio Minihalos in a Mass-limited Sample of Galaxy Clusters}",
      journal = {\apj},
         year = 2017,
        month = jun,
       volume = {841},
       number = {2},
          eid = {71},
        pages = {71},
          doi = {10.3847/1538-4357/aa7069},
archivePrefix = {arXiv},
       eprint = {1701.01364},
 primaryClass = {astro-ph.HE},
       adsurl = {https://ui.adsabs.harvard.edu/abs/2017ApJ...841...71G}
}

@ARTICLE{shimwell2017,
       author = {{Shimwell}, T.~W. and {R{\"o}ttgering}, H.~J.~A. and {Best}, P.~N. and {Williams}, W.~L. and {Dijkema}, T.~J. and {de Gasperin}, F. and {Hardcastle}, M.~J. and {Heald}, G.~H. and {Hoang}, D.~N. and {Horneffer}, A. and {Intema}, H. and {Mahony}, E.~K. and {Mandal}, S. and {Mechev}, A.~P. and {Morabito}, L. and {Oonk}, J.~B.~R. and {Rafferty}, D. and {Retana-Montenegro}, E. and {Sabater}, J. and {Tasse}, C. and {van Weeren}, R.~J. and {Br{\"u}ggen}, M. and {Brunetti}, G. and {Chy{\.z}y}, K.~T. and {Conway}, J.~E. and {Haverkorn}, M. and {Jackson}, N. and {Jarvis}, M.~J. and {McKean}, J.~P. and {Miley}, G.~K. and {Morganti}, R. and {White}, G.~J. and {Wise}, M.~W. and {van Bemmel}, I.~M. and {Beck}, R. and {Brienza}, M. and {Bonafede}, A. and {Calistro Rivera}, G. and {Cassano}, R. and {Clarke}, A.~O. and {Cseh}, D. and {Deller}, A. and {Drabent}, A. and {van Driel}, W. and {Engels}, D. and {Falcke}, H. and {Ferrari}, C. and {Fr{\"o}hlich}, S. and {Garrett}, M.~A. and {Harwood}, J.~J. and {Heesen}, V. and {Hoeft}, M. and {Horellou}, C. and {Israel}, F.~P. and {Kapi{\'n}ska}, A.~D. and {Kunert-Bajraszewska}, M. and {McKay}, D.~J. and {Mohan}, N.~R. and {Orr{\'u}}, E. and {Pizzo}, R.~F. and {Prandoni}, I. and {Schwarz}, D.~J. and {Shulevski}, A. and {Sipior}, M. and {Smith}, D.~J.~B. and {Sridhar}, S.~S. and {Steinmetz}, M. and {Stroe}, A. and {Varenius}, E. and {van der Werf}, P.~P. and {Zensus}, J.~A. and {Zwart}, J.~T.~L.},
        title = "{The LOFAR Two-metre Sky Survey. I. Survey description and preliminary data release}",
      journal = {\aap},
         year = 2017,
        month = feb,
       volume = {598},
          eid = {A104},
        pages = {A104},
          doi = {10.1051/0004-6361/201629313},
archivePrefix = {arXiv},
       eprint = {1611.02700},
 primaryClass = {astro-ph.IM},
       adsurl = {https://ui.adsabs.harvard.edu/abs/2017A&A...598A.104S}
}

@ARTICLE{paul2021,
       author = {{Paul}, Surajit and {Gupta}, Prateek and {Salunkhe}, Sameer and {Bhagat}, Shubham and {Sonkamble}, Satish and {Hiray}, Manish and {Dabhade}, Pratik and {Raychaudhury}, Somak},
        title = "{uGMRT detection of cluster radio emission in low-mass Planck Sunyaev-Zel'dovich clusters}",
      journal = {\mnras},
         year = 2021,
        month = oct,
       volume = {506},
       number = {4},
        pages = {5389-5397},
          doi = {10.1093/mnras/stab1965},
archivePrefix = {arXiv},
       eprint = {2107.03408},
 primaryClass = {astro-ph.GA},
       adsurl = {https://ui.adsabs.harvard.edu/abs/2021MNRAS.506.5389P}
}

@ARTICLE{knowles2016,
       author = {{Knowles}, K. and {Intema}, H.~T. and {Baker}, A.~J. and {Bharadwaj}, V. and {Bond}, J.~R. and {Cress}, C. and {Gupta}, N. and {Hajian}, A. and {Hilton}, M. and {Hincks}, A.~D. and {Hlozek}, R. and {Hughes}, J.~P. and {Lindner}, R.~R. and {Marriage}, T.~A. and {Menanteau}, F. and {Moodley}, K. and {Niemack}, M.~D. and {Reese}, E.~D. and {Sievers}, J. and {Sif{\'o}n}, C. and {Srianand}, R. and {Wollack}, E.~J.},
        title = "{A giant radio halo in a low-mass SZ-selected galaxy cluster: ACT-CL J0256.5+0006}",
      journal = {\mnras},
         year = 2016,
        month = jul,
       volume = {459},
       number = {4},
        pages = {4240-4258},
          doi = {10.1093/mnras/stw795},
archivePrefix = {arXiv},
       eprint = {1506.01547},
 primaryClass = {astro-ph.GA},
       adsurl = {https://ui.adsabs.harvard.edu/abs/2016MNRAS.459.4240K}
}

@INPROCEEDINGS{cassano2009,
       author = {{Cassano}, R.},
        title = "{Large-scale Diffuse Radio Emission from Clusters of Galaxies and the Importance of Low Frequency Radio Observations}",
    booktitle = {The Low-Frequency Radio Universe},
         year = 2009,
       editor = {{Saikia}, D.~J. and {Green}, D.~A. and {Gupta}, Y. and {Venturi}, T.},
       series = {Astronomical Society of the Pacific Conference Series},
       volume = {407},
        month = sep,
        pages = {223},
          doi = {10.48550/arXiv.0902.2971},
archivePrefix = {arXiv},
       eprint = {0902.2971},
 primaryClass = {astro-ph.CO},
       adsurl = {https://ui.adsabs.harvard.edu/abs/2009ASPC..407..223C}
}

@ARTICLE{pfrommer2004,
       author = {{Pfrommer}, C. and {En{\ss}lin}, T.~A.},
        title = "{Constraining the population of cosmic ray protons in cooling flow clusters with {\ensuremath{\gamma}}-ray and radio observations:  Are radio mini-halos of hadronic origin?}",
      journal = {\aap},
         year = 2004,
        month = jan,
       volume = {413},
        pages = {17-36},
          doi = {10.1051/0004-6361:20031464},
       adsurl = {https://ui.adsabs.harvard.edu/abs/2004A&A...413...17P}
}

@ARTICLE{bruno2023,
       author = {{Bruno}, L. and {Botteon}, A. and {Shimwell}, T. and {Cuciti}, V. and {de Gasperin}, F. and {Brunetti}, G. and {Dallacasa}, D. and {Gastaldello}, F. and {Rossetti}, M. and {van Weeren}, R.~J. and {Venturi}, T. and {Russo}, S.~A. and {Taffoni}, G. and {Cassano}, R. and {Biava}, N. and {Lusetti}, G. and {Bonafede}, A. and {Ghizzardi}, S. and {De Grandi}, S.},
        title = "{A three-component giant radio halo: The puzzling case of the galaxy cluster Abell 2142}",
      journal = {\aap},
         year = 2023,
        month = oct,
       volume = {678},
          eid = {A133},
        pages = {A133},
          doi = {10.1051/0004-6361/202347245},
archivePrefix = {arXiv},
       eprint = {2308.07603},
 primaryClass = {astro-ph.CO},
       adsurl = {https://ui.adsabs.harvard.edu/abs/2023A&A...678A.133B}
}

@ARTICLE{pasini2024,
       author = {{Pasini}, T. and {De Gasperin}, F. and {Br{\"u}ggen}, M. and {Cassano}, R. and {Botteon}, A. and {Brunetti}, G. and {Edler}, H.~W. and {van Weeren}, R.~J. and {Cuciti}, V. and {Shimwell}, T. and {Di Gennaro}, G. and {Gaspari}, M. and {Hardcastle}, M. and {Rottgering}, H.~J.~A. and {Tasse}, C.},
        title = "{Ultra-low-frequency LOFAR spectral indices of cluster radio halos}",
      journal = {\aap},
         year = 2024,
        month = sep,
       volume = {689},
          eid = {A218},
        pages = {A218},
          doi = {10.1051/0004-6361/202450697},
archivePrefix = {arXiv},
       eprint = {2406.12005},
 primaryClass = {astro-ph.CO},
       adsurl = {https://ui.adsabs.harvard.edu/abs/2024A&A...689A.218P}
}

@ARTICLE{brunetti2001,
       author = {{Brunetti}, G. and {Setti}, G. and {Feretti}, L. and {Giovannini}, G.},
        title = "{Particle reacceleration in the Coma cluster: radio properties and hard X-ray emission}",
      journal = {\mnras},
         year = 2001,
        month = jan,
       volume = {320},
       number = {3},
        pages = {365-378},
          doi = {10.1046/j.1365-8711.2001.03978.x},
archivePrefix = {arXiv},
       eprint = {astro-ph/0008518},
 primaryClass = {astro-ph},
       adsurl = {https://ui.adsabs.harvard.edu/abs/2001MNRAS.320..365B}
}

@ARTICLE{bonafede2022,
       author = {{Bonafede}, A. and {Brunetti}, G. and {Rudnick}, L. and {Vazza}, F. and {Bourdin}, H. and {Giovannini}, G. and {Shimwell}, T.~W. and {Zhang}, X. and {Mazzotta}, P. and {Simionescu}, A. and {Biava}, N. and {Bonnassieux}, E. and {Brienza}, M. and {Br{\"u}ggen}, M. and {Rajpurohit}, K. and {Riseley}, C.~J. and {Stuardi}, C. and {Feretti}, L. and {Tasse}, C. and {Botteon}, A. and {Carretti}, E. and {Cassano}, R. and {Cuciti}, V. and {de Gasperin}, F. and {Gastaldello}, F. and {Rossetti}, M. and {Rottgering}, H.~J.~A. and {Venturi}, T. and {van Weeren}, R.~J.},
        title = "{The Coma Cluster at LOFAR Frequencies. II. The Halo, Relic, and a New Accretion Relic}",
      journal = {\apj},
         year = 2022,
        month = jul,
       volume = {933},
       number = {2},
          eid = {218},
        pages = {218},
          doi = {10.3847/1538-4357/ac721d},
archivePrefix = {arXiv},
       eprint = {2203.01958},
 primaryClass = {astro-ph.HE},
       adsurl = {https://ui.adsabs.harvard.edu/abs/2022ApJ...933..218B}
}

@ARTICLE{rajpurohit2021,
       author = {{Rajpurohit}, K. and {Brunetti}, G. and {Bonafede}, A. and {van Weeren}, R.~J. and {Botteon}, A. and {Vazza}, F. and {Hoeft}, M. and {Riseley}, C.~J. and {Bonnassieux}, E. and {Brienza}, M. and {Forman}, W.~R. and {R{\"o}ttgering}, H.~J.~A. and {Rajpurohit}, A.~S. and {Locatelli}, N. and {Shimwell}, T.~W. and {Cassano}, R. and {Di Gennaro}, G. and {Br{\"u}ggen}, M. and {Wittor}, D. and {Drabent}, A. and {Ignesti}, A.},
        title = "{Physical insights from the spectrum of the radio halo in MACS J0717.5+3745}",
      journal = {\aap},
         year = 2021,
        month = feb,
       volume = {646},
          eid = {A135},
        pages = {A135},
          doi = {10.1051/0004-6361/202039591},
archivePrefix = {arXiv},
       eprint = {2012.14373},
 primaryClass = {astro-ph.GA},
       adsurl = {https://ui.adsabs.harvard.edu/abs/2021A&A...646A.135R}
}

@ARTICLE{rajpurohit2024,
       author = {{Rajpurohit}, K. and {O'Sullivan}, E. and {Schellenberger}, G. and {Brienza}, M. and {Vrtilek}, J.~M. and {Forman}, W. and {David}, L.~P. and {Clarke}, T. and {Botteon}, A. and {Vazza}, F. and {Giacintucci}, S. and {Jones}, C. and {Br{\"u}ggen}, M. and {Shimwell}, T.~W. and {Drabent}, A. and {Loi}, F. and {Loubser}, S.~I. and {Kolokythas}, K. and {Babyk}, I. and {R{\"o}ttgering}, H.~J.~A.},
        title = "{A Deep Dive into the NGC 741 Galaxy Group: Insights into a Spectacular Head-tail Radio Galaxy from VLA, MeerKAT, uGMRT, and LOFAR}",
      journal = {\apj},
         year = 2024,
        month = nov,
       volume = {976},
       number = {1},
          eid = {64},
        pages = {64},
          doi = {10.3847/1538-4357/ad8136},
archivePrefix = {arXiv},
       eprint = {2408.15197},
 primaryClass = {astro-ph.GA},
       adsurl = {https://ui.adsabs.harvard.edu/abs/2024ApJ...976...64R}
}

@ARTICLE{dominguezfernandez2024,
       author = {{Dom{\'\i}nguez-Fern{\'a}ndez}, P. and {ZuHone}, J. and {Weinberger}, R. and {Bellomi}, E. and {Hernquist}, L. and {Nulsen}, P. and {Brunetti}, G.},
        title = "{Jet Interaction with Galaxy Cluster Mergers}",
      journal = {\apj},
         year = 2024,
        month = dec,
       volume = {977},
       number = {2},
          eid = {221},
        pages = {221},
          doi = {10.3847/1538-4357/ad9028},
archivePrefix = {arXiv},
       eprint = {2406.19681},
 primaryClass = {astro-ph.HE},
       adsurl = {https://ui.adsabs.harvard.edu/abs/2024ApJ...977..221D}
}

@ARTICLE{timmerman2021,
       author = {{Timmerman}, R. and {van Weeren}, R.~J. and {McDonald}, M. and {Ignesti}, A. and {McNamara}, B.~R. and {Hlavacek-Larrondo}, J. and {R{\"o}ttgering}, H.~J.~A.},
        title = "{Very Large Array observations of the mini-halo and AGN feedback in the Phoenix cluster}",
      journal = {\aap},
         year = 2021,
        month = feb,
       volume = {646},
          eid = {A38},
        pages = {A38},
          doi = {10.1051/0004-6361/202039075},
archivePrefix = {arXiv},
       eprint = {2009.13238},
 primaryClass = {astro-ph.CO},
       adsurl = {https://ui.adsabs.harvard.edu/abs/2021A&A...646A..38T}
}

@ARTICLE{bruno2024tail,
       author = {{Bruno}, L. and {Venturi}, T. and {Dallacasa}, D. and {Brienza}, M. and {Ignesti}, A. and {Brunetti}, G. and {Riseley}, C.~J. and {Rossetti}, M. and {Gastaldello}, F. and {Botteon}, A. and {Rudnick}, L. and {van Weeren}, R.~J. and {Shulevski}, A. and {Lal}, D.~V.},
        title = "{Ageing and dynamics of the tailed radio galaxies in Abell 2142}",
      journal = {\aap},
         year = 2024,
        month = oct,
       volume = {690},
          eid = {A329},
        pages = {A329},
          doi = {10.1051/0004-6361/202451515},
archivePrefix = {arXiv},
       eprint = {2409.03453},
 primaryClass = {astro-ph.GA},
       adsurl = {https://ui.adsabs.harvard.edu/abs/2024A&A...690A.329B}
}

@ARTICLE{trehaeven2023,
       author = {{Trehaeven}, K.~S. and {Parekh}, V. and {Oozeer}, N. and {Hugo}, B. and {Smirnov}, O. and {Bernardi}, G. and {Knowles}, K. and {Tasse}, C. and {Asad}, K.~M.~B. and {Giacintucci}, S.},
        title = "{Mining mini-halos with MeerKAT I. Calibration and imaging}",
      journal = {\mnras},
         year = 2023,
        month = apr,
       volume = {520},
       number = {3},
        pages = {4410-4426},
          doi = {10.1093/mnras/stad391},
archivePrefix = {arXiv},
       eprint = {2303.08427},
 primaryClass = {astro-ph.GA},
       adsurl = {https://ui.adsabs.harvard.edu/abs/2023MNRAS.520.4410T}
}

@ARTICLE{ignesti2020,
       author = {{Ignesti}, A. and {Brunetti}, G. and {Gitti}, M. and {Giacintucci}, S.},
        title = "{Radio and X-ray connection in radio mini-halos: Implications for hadronic models}",
      journal = {\aap},
         year = 2020,
        month = aug,
       volume = {640},
          eid = {A37},
        pages = {A37},
          doi = {10.1051/0004-6361/201937207},
archivePrefix = {arXiv},
       eprint = {2006.09254},
 primaryClass = {astro-ph.HE},
       adsurl = {https://ui.adsabs.harvard.edu/abs/2020A&A...640A..37I}
}

@ARTICLE{cassano2008,
       author = {{Cassano}, R. and {Gitti}, M. and {Brunetti}, G.},
        title = "{A morphological comparison between giant radio halos and radio mini-halos in galaxy clusters}",
      journal = {\aap},
         year = 2008,
        month = aug,
       volume = {486},
       number = {3},
        pages = {L31-L34},
          doi = {10.1051/0004-6361:200810179},
archivePrefix = {arXiv},
       eprint = {0806.1817},
 primaryClass = {astro-ph},
       adsurl = {https://ui.adsabs.harvard.edu/abs/2008A&A...486L..31C}
}

@ARTICLE{2025A&A...695A.180B,
       author = {{Balboni}, M. and {Ettori}, S. and {Gastaldello}, F. and {Cassano}, R. and {Bonafede}, A. and {Cuciti}, V. and {Botteon}, A. and {Brunetti}, G. and {Bartalucci}, I. and {Gaspari}, M. and {Gavazzi}, R. and {Ghizzardi}, S. and {Gitti}, M. and {Lovisari}, L. and {Maughan}, B.~J. and {Molendi}, S. and {Pointecouteau}, E. and {Pratt}, G.~W. and {Rasia}, E. and {Riva}, G. and {Rossetti}, M. and {Rottgering}, H. and {Sayers}, J. and {van Weeren}, R.~J.},
        title = "{CHEX-MATE: Scaling relations of radio halo profiles for clusters in the LoTSS DR2 area}",
      journal = {\aap},
         year = 2025,
        month = mar,
       volume = {695},
          eid = {A180},
        pages = {A180},
          doi = {10.1051/0004-6361/202453183},
archivePrefix = {arXiv},
       eprint = {2502.18568},
 primaryClass = {astro-ph.CO},
       adsurl = {https://ui.adsabs.harvard.edu/abs/2025A&A...695A.180B}
}

@ARTICLE{cassano2013,
       author = {{Cassano}, R. and {Ettori}, S. and {Brunetti}, G. and {Giacintucci}, S. and {Pratt}, G.~W. and {Venturi}, T. and {Kale}, R. and {Dolag}, K. and {Markevitch}, M.},
        title = "{Revisiting Scaling Relations for Giant Radio Halos in Galaxy Clusters}",
      journal = {\apj},
         year = 2013,
        month = nov,
       volume = {777},
       number = {2},
          eid = {141},
        pages = {141},
          doi = {10.1088/0004-637X/777/2/141},
archivePrefix = {arXiv},
       eprint = {1306.4379},
 primaryClass = {astro-ph.CO},
       adsurl = {https://ui.adsabs.harvard.edu/abs/2013ApJ...777..141C}
}

@ARTICLE{savini2018,
       author = {{Savini}, F. and {Bonafede}, A. and {Br{\"u}ggen}, M. and {van Weeren}, R. and {Brunetti}, G. and {Intema}, H. and {Botteon}, A. and {Shimwell}, T. and {Wilber}, A. and {Rafferty}, D. and {Giacintucci}, S. and {Cassano}, R. and {Cuciti}, V. and {de Gasperin}, F. and {R{\"o}ttgering}, H. and {Hoeft}, M. and {White}, G.},
        title = "{First evidence of diffuse ultra-steep-spectrum radio emission surrounding the cool core of a cluster}",
      journal = {\mnras},
         year = 2018,
        month = aug,
       volume = {478},
       number = {2},
        pages = {2234-2242},
          doi = {10.1093/mnras/sty1125},
archivePrefix = {arXiv},
       eprint = {1805.01900},
 primaryClass = {astro-ph.GA},
       adsurl = {https://ui.adsabs.harvard.edu/abs/2018MNRAS.478.2234S}
}

@ARTICLE{biava2021_minihalo,
       author = {{Biava}, N. and {de Gasperin}, F. and {Bonafede}, A. and {Edler}, H.~W. and {Giacintucci}, S. and {Mazzotta}, P. and {Brunetti}, G. and {Botteon}, A. and {Br{\"u}ggen}, M. and {Cassano}, R. and {Drabent}, A. and {Edge}, A.~C. and {En{\ss}lin}, T. and {Gastaldello}, F. and {Riseley}, C.~J. and {Rossetti}, M. and {Rottgering}, H.~J.~A. and {Shimwell}, T.~W. and {Tasse}, C. and {van Weeren}, R.~J.},
        title = "{The ultra-steep diffuse radio emission observed in the cool-core cluster RX J1720.1+2638 with LOFAR at 54 MHz}",
      journal = {\mnras},
         year = 2021,
        month = dec,
       volume = {508},
       number = {3},
        pages = {3995-4007},
          doi = {10.1093/mnras/stab2840},
archivePrefix = {arXiv},
       eprint = {2110.01629},
 primaryClass = {astro-ph.GA},
       adsurl = {https://ui.adsabs.harvard.edu/abs/2021MNRAS.508.3995B}
}

@ARTICLE{giacintucci2011,
       author = {{Giacintucci}, S. and {Markevitch}, M. and {Brunetti}, G. and {Cassano}, R. and {Venturi}, T.},
        title = "{A radio minihalo in the extreme cool-core galaxy cluster RXC J1504.1-0248}",
      journal = {\aap},
         year = 2011,
        month = jan,
       volume = {525},
          eid = {L10},
        pages = {L10},
          doi = {10.1051/0004-6361/201015882},
archivePrefix = {arXiv},
       eprint = {1011.3141},
 primaryClass = {astro-ph.HE},
       adsurl = {https://ui.adsabs.harvard.edu/abs/2011A&A...525L..10G}
}

@ARTICLE{ignesti2022,
       author = {{Ignesti}, A. and {Brunetti}, G. and {Shimwell}, T. and {Gitti}, M. and {Birzan}, L. and {Botteon}, A. and {Br{\"u}ggen}, M. and {de Gasperin}, F. and {Di Gennaro}, G. and {Edge}, A.~C. and {Riseley}, C.~J. and {R{\"o}ttgering}, H.~J.~A. and {van Weeren}, R.~J.},
        title = "{A LOFAR view into the stormy environment of the galaxy cluster 2A0335+096}",
      journal = {\aap},
         year = 2022,
        month = mar,
       volume = {659},
          eid = {A20},
        pages = {A20},
          doi = {10.1051/0004-6361/202142549},
archivePrefix = {arXiv},
       eprint = {2111.15517},
 primaryClass = {astro-ph.CO},
       adsurl = {https://ui.adsabs.harvard.edu/abs/2022A&A...659A..20I}
}

@ARTICLE{botteon2024,
       author = {{Botteon}, A. and {van Weeren}, R.~J. and {Eckert}, D. and {Gastaldello}, F. and {Markevitch}, M. and {Giacintucci}, S. and {Brunetti}, G. and {Kale}, R. and {Venturi}, T.},
        title = "{The prototypical major cluster merger Abell 754: I. Calibration of MeerKAT data and radio/X-ray spectral mapping of the cluster}",
      journal = {\aap},
         year = 2024,
        month = oct,
       volume = {690},
          eid = {A222},
        pages = {A222},
          doi = {10.1051/0004-6361/202451293},
archivePrefix = {arXiv},
       eprint = {2406.18983},
 primaryClass = {astro-ph.CO},
       adsurl = {https://ui.adsabs.harvard.edu/abs/2024A&A...690A.222B}
}

@ARTICLE{feretti1992,
       author = {{Feretti}, L. and {Perola}, G.~C. and {Fanti}, R.},
        title = "{Tailed radio sources as probes of the intergalactic medium pressure.}",
      journal = {\aap},
         year = 1992,
        month = nov,
       volume = {265},
        pages = {9-18},
       adsurl = {https://ui.adsabs.harvard.edu/abs/1992A&A...265....9F}
}

@ARTICLE{derubeis2025,
       author = {{De Rubeis}, E. and {Bondi}, M. and {Botteon}, A. and {van Weeren}, R.~J. and {de Jong}, J.~M.~G.~H.~J. and {Rudnick}, L. and {Brunetti}, G. and {Rajpurohit}, K. and {Gheller}, C. and {R{\"o}ttgering}, H.~J.~A.},
        title = "{Revealing the intricacies of radio galaxies and filaments in the merging galaxy cluster Abell 2255: I. Insights from deep LOFAR-VLBI sub-arcsecond resolution images}",
      journal = {\aap},
         year = 2025,
        month = jul,
       volume = {699},
          eid = {A229},
        pages = {A229},
          doi = {10.1051/0004-6361/202555268},
archivePrefix = {arXiv},
       eprint = {2505.13595},
 primaryClass = {astro-ph.GA},
       adsurl = {https://ui.adsabs.harvard.edu/abs/2025A&A...699A.229D}
}

@ARTICLE{girardi1993,
       author = {{Girardi}, M. and {Biviano}, A. and {Giuricin}, G. and {Mardirossian}, F. and {Mezzetti}, M.},
        title = "{Velocity Dispersions in Galaxy Clusters}",
      journal = {\apj},
         year = 1993,
        month = feb,
       volume = {404},
        pages = {38},
          doi = {10.1086/172256},
       adsurl = {https://ui.adsabs.harvard.edu/abs/1993ApJ...404...38G}
}

@ARTICLE{riseley2022,
       author = {{Riseley}, C.~J. and {Rajpurohit}, K. and {Loi}, F. and {Botteon}, A. and {Timmerman}, R. and {Biava}, N. and {Bonafede}, A. and {Bonnassieux}, E. and {Brunetti}, G. and {En{\ss}lin}, T. and {Di Gennaro}, G. and {Ignesti}, A. and {Shimwell}, T. and {Stuardi}, C. and {Vernstrom}, T. and {van Weeren}, R.~J.},
        title = "{A MeerKAT-meets-LOFAR study of MS 1455.0 + 2232: a 590 kiloparsec 'mini'-halo in a sloshing cool-core cluster}",
      journal = {\mnras},
         year = 2022,
        month = may,
       volume = {512},
       number = {3},
        pages = {4210-4230},
          doi = {10.1093/mnras/stac672},
archivePrefix = {arXiv},
       eprint = {2112.07364},
 primaryClass = {astro-ph.HE},
       adsurl = {https://ui.adsabs.harvard.edu/abs/2022MNRAS.512.4210R}
}

@ARTICLE{rykoff2016,
       author = {{Rykoff}, E.~S. and {Rozo}, E. and {Hollowood}, D. and {Bermeo-Hernandez}, A. and {Jeltema}, T. and {Mayers}, J. and {Romer}, A.~K. and {Rooney}, P. and {Saro}, A. and {Vergara Cervantes}, C. and {Wechsler}, R.~H. and {Wilcox}, H. and {Abbott}, T.~M.~C. and {Abdalla}, F.~B. and {Allam}, S. and {Annis}, J. and {Benoit-L{\'e}vy}, A. and {Bernstein}, G.~M. and {Bertin}, E. and {Brooks}, D. and {Burke}, D.~L. and {Capozzi}, D. and {Carnero Rosell}, A. and {Carrasco Kind}, M. and {Castander}, F.~J. and {Childress}, M. and {Collins}, C.~A. and {Cunha}, C.~E. and {D'Andrea}, C.~B. and {da Costa}, L.~N. and {Davis}, T.~M. and {Desai}, S. and {Diehl}, H.~T. and {Dietrich}, J.~P. and {Doel}, P. and {Evrard}, A.~E. and {Finley}, D.~A. and {Flaugher}, B. and {Fosalba}, P. and {Frieman}, J. and {Glazebrook}, K. and {Goldstein}, D.~A. and {Gruen}, D. and {Gruendl}, R.~A. and {Gutierrez}, G. and {Hilton}, M. and {Honscheid}, K. and {Hoyle}, B. and {James}, D.~J. and {Kay}, S.~T. and {Kuehn}, K. and {Kuropatkin}, N. and {Lahav}, O. and {Lewis}, G.~F. and {Lidman}, C. and {Lima}, M. and {Maia}, M.~A.~G. and {Mann}, R.~G. and {Marshall}, J.~L. and {Martini}, P. and {Melchior}, P. and {Miller}, C.~J. and {Miquel}, R. and {Mohr}, J.~J. and {Nichol}, R.~C. and {Nord}, B. and {Ogando}, R. and {Plazas}, A.~A. and {Reil}, K. and {Sahl{\'e}n}, M. and {Sanchez}, E. and {Santiago}, B. and {Scarpine}, V. and {Schubnell}, M. and {Sevilla-Noarbe}, I. and {Smith}, R.~C. and {Soares-Santos}, M. and {Sobreira}, F. and {Stott}, J.~P. and {Suchyta}, E. and {Swanson}, M.~E.~C. and {Tarle}, G. and {Thomas}, D. and {Tucker}, D. and {Uddin}, S. and {Viana}, P.~T.~P. and {Vikram}, V. and {Walker}, A.~R. and {Zhang}, Y. and {DES Collaboration}},
        title = "{The RedMaPPer Galaxy Cluster Catalog From DES Science Verification Data}",
      journal = {\apjs},
         year = 2016,
        month = may,
       volume = {224},
       number = {1},
          eid = {1},
        pages = {1},
          doi = {10.3847/0067-0049/224/1/1},
archivePrefix = {arXiv},
       eprint = {1601.00621},
 primaryClass = {astro-ph.CO},
       adsurl = {https://ui.adsabs.harvard.edu/abs/2016ApJS..224....1R}
}

@ARTICLE{desi2019,
       author = {{Dey}, Arjun and {Schlegel}, David J. and {Lang}, Dustin and {Blum}, Robert and {Burleigh}, Kaylan and {Fan}, Xiaohui and {Findlay}, Joseph R. and {Finkbeiner}, Doug and {Herrera}, David and {Juneau}, St{\'e}phanie and {Landriau}, Martin and {Levi}, Michael and {McGreer}, Ian and {Meisner}, Aaron and {Myers}, Adam D. and {Moustakas}, John and {Nugent}, Peter and {Patej}, Anna and {Schlafly}, Edward F. and {Walker}, Alistair R. and {Valdes}, Francisco and {Weaver}, Benjamin A. and {Y{\`e}che}, Christophe and {Zou}, Hu and {Zhou}, Xu and {Abareshi}, Behzad and {Abbott}, T.~M.~C. and {Abolfathi}, Bela and {Aguilera}, C. and {Alam}, Shadab and {Allen}, Lori and {Alvarez}, A. and {Annis}, James and {Ansarinejad}, Behzad and {Aubert}, Marie and {Beechert}, Jacqueline and {Bell}, Eric F. and {BenZvi}, Segev Y. and {Beutler}, Florian and {Bielby}, Richard M. and {Bolton}, Adam S. and {Brice{\~n}o}, C{\'e}sar and {Buckley-Geer}, Elizabeth J. and {Butler}, Karen and {Calamida}, Annalisa and {Carlberg}, Raymond G. and {Carter}, Paul and {Casas}, Ricard and {Castander}, Francisco J. and {Choi}, Yumi and {Comparat}, Johan and {Cukanovaite}, Elena and {Delubac}, Timoth{\'e}e and {DeVries}, Kaitlin and {Dey}, Sharmila and {Dhungana}, Govinda and {Dickinson}, Mark and {Ding}, Zhejie and {Donaldson}, John B. and {Duan}, Yutong and {Duckworth}, Christopher J. and {Eftekharzadeh}, Sarah and {Eisenstein}, Daniel J. and {Etourneau}, Thomas and {Fagrelius}, Parker A. and {Farihi}, Jay and {Fitzpatrick}, Mike and {Font-Ribera}, Andreu and {Fulmer}, Leah and {G{\"a}nsicke}, Boris T. and {Gaztanaga}, Enrique and {George}, Koshy and {Gerdes}, David W. and {Gontcho}, Satya Gontcho A. and {Gorgoni}, Claudio and {Green}, Gregory and {Guy}, Julien and {Harmer}, Diane and {Hernandez}, M. and {Honscheid}, Klaus and {Huang}, Lijuan Wendy and {James}, David J. and {Jannuzi}, Buell T. and {Jiang}, Linhua and {Joyce}, Richard and {Karcher}, Armin and {Karkar}, Sonia and {Kehoe}, Robert and {Kneib}, Jean-Paul and {Kueter-Young}, Andrea and {Lan}, Ting-Wen and {Lauer}, Tod R. and {Le Guillou}, Laurent and {Le Van Suu}, Auguste and {Lee}, Jae Hyeon and {Lesser}, Michael and {Perreault Levasseur}, Laurence and {Li}, Ting S. and {Mann}, Justin L. and {Marshall}, Robert and {Mart{\'\i}nez-V{\'a}zquez}, C.~E. and {Martini}, Paul and {du Mas des Bourboux}, H{\'e}lion and {McManus}, Sean and {Meier}, Tobias Gabriel and {M{\'e}nard}, Brice and {Metcalfe}, Nigel and {Mu{\~n}oz-Guti{\'e}rrez}, Andrea and {Najita}, Joan and {Napier}, Kevin and {Narayan}, Gautham and {Newman}, Jeffrey A. and {Nie}, Jundan and {Nord}, Brian and {Norman}, Dara J. and {Olsen}, Knut A.~G. and {Paat}, Anthony and {Palanque-Delabrouille}, Nathalie and {Peng}, Xiyan and {Poppett}, Claire L. and {Poremba}, Megan R. and {Prakash}, Abhishek and {Rabinowitz}, David and {Raichoor}, Anand and {Rezaie}, Mehdi and {Robertson}, A.~N. and {Roe}, Natalie A. and {Ross}, Ashley J. and {Ross}, Nicholas P. and {Rudnick}, Gregory and {Safonova}, Sasha and {Saha}, Abhijit and {S{\'a}nchez}, F. Javier and {Savary}, Elodie and {Schweiker}, Heidi and {Scott}, Adam and {Seo}, Hee-Jong and {Shan}, Huanyuan and {Silva}, David R. and {Slepian}, Zachary and {Soto}, Christian and {Sprayberry}, David and {Staten}, Ryan and {Stillman}, Coley M. and {Stupak}, Robert J. and {Summers}, David L. and {Sien Tie}, Suk and {Tirado}, H. and {Vargas-Maga{\~n}a}, Mariana and {Vivas}, A. Katherina and {Wechsler}, Risa H. and {Williams}, Doug and {Yang}, Jinyi and {Yang}, Qian and {Yapici}, Tolga and {Zaritsky}, Dennis and {Zenteno}, A. and {Zhang}, Kai and {Zhang}, Tianmeng and {Zhou}, Rongpu and {Zhou}, Zhimin},
        title = "{Overview of the DESI Legacy Imaging Surveys}",
      journal = {\aj},
         year = 2019,
        month = may,
       volume = {157},
       number = {5},
          eid = {168},
        pages = {168},
          doi = {10.3847/1538-3881/ab089d},
archivePrefix = {arXiv},
       eprint = {1804.08657},
 primaryClass = {astro-ph.IM},
       adsurl = {https://ui.adsabs.harvard.edu/abs/2019AJ....157..168D}
}

@ARTICLE{offringa2016,
       author = {{Offringa}, A.~R.},
        title = "{Compression of interferometric radio-astronomical data}",
      journal = {\aap},
         year = 2016,
        month = nov,
       volume = {595},
          eid = {A99},
        pages = {A99},
          doi = {10.1051/0004-6361/201629565},
archivePrefix = {arXiv},
       eprint = {1609.02019},
 primaryClass = {astro-ph.IM},
       adsurl = {https://ui.adsabs.harvard.edu/abs/2016A&A...595A..99O}
}

@ARTICLE{balboni2025,
       author = {{Balboni}, M. and {Gastaldello}, F. and {Bonafede}, A. and {Botteon}, A. and {Bartalucci}, I. and {Cassano}, R. and {De Grandi}, S. and {Ettori}, S. and {Gaspari}, M. and {Ghizzardi}, S. and {Gitti}, M. and {Johnston-Hollitt}, M. and {Lovisari}, L. and {Molendi}, S. and {Pointecouteau}, E. and {Pratt}, G.~W. and {Riva}, G. and {Rossetti}, M. and {Sayers}, J. and {Sereno}, M. and {van Weeren}, R.~J.},
        title = "{CHEX-MATE: New detections and properties of the radio diffuse emission in massive clusters with MeerKAT}",
      journal = {\aap},
         year = 2026,
        month = mar,
       volume = {707},
          eid = {A143},
        pages = {A143},
          doi = {10.1051/0004-6361/202556148},
archivePrefix = {arXiv},
       eprint = {2507.00133},
 primaryClass = {astro-ph.CO},
       adsurl = {https://ui.adsabs.harvard.edu/abs/2026A&A...707A.143B}
}

@ARTICLE{2008ApJ...687..899R,
       author = {{Rafferty}, D.~A. and {McNamara}, B.~R. and {Nulsen}, P.~E.~J.},
        title = "{The Regulation of Cooling and Star Formation in Luminous Galaxies by Active Galactic Nucleus Feedback and the Cooling-Time/Entropy Threshold for the Onset of Star Formation}",
      journal = {\apj},
         year = 2008,
        month = nov,
       volume = {687},
       number = {2},
        pages = {899-918},
          doi = {10.1086/591240},
archivePrefix = {arXiv},
       eprint = {0802.1864},
 primaryClass = {astro-ph},
       adsurl = {https://ui.adsabs.harvard.edu/abs/2008ApJ...687..899R}
}

@ARTICLE{bohringer2004,
       author = {{B{\"o}hringer}, H. and {Schuecker}, P. and {Guzzo}, L. and {Collins}, C.~A. and {Voges}, W. and {Cruddace}, R.~G. and {Ortiz-Gil}, A. and {Chincarini}, G. and {De Grandi}, S. and {Edge}, A.~C. and {MacGillivray}, H.~T. and {Neumann}, D.~M. and {Schindler}, S. and {Shaver}, P.},
        title = "{The ROSAT-ESO Flux Limited X-ray (REFLEX) Galaxy cluster survey. V. The cluster catalogue}",
      journal = {\aap},
         year = 2004,
        month = oct,
       volume = {425},
        pages = {367-383},
          doi = {10.1051/0004-6361:20034484},
archivePrefix = {arXiv},
       eprint = {astro-ph/0405546},
 primaryClass = {astro-ph},
       adsurl = {https://ui.adsabs.harvard.edu/abs/2004A&A...425..367B}
}

@ARTICLE{mcxc_ii,
       author = {{Sadibekova}, T. and {Arnaud}, M. and {Pratt}, G.~W. and {Tarr{\'\i}o}, P. and {Melin}, J.-B.},
        title = "{MCXC-II: Second release of the Meta-Catalogue of X-ray detected Clusters of galaxies}",
      journal = {\aap},
         year = 2024,
        month = aug,
       volume = {688},
          eid = {A187},
        pages = {A187},
          doi = {10.1051/0004-6361/202449427},
archivePrefix = {arXiv},
       eprint = {2402.01538},
 primaryClass = {astro-ph.CO},
       adsurl = {https://ui.adsabs.harvard.edu/abs/2024A&A...688A.187S}
}

@ARTICLE{ebeling1998,
       author = {{Ebeling}, H. and {Edge}, A.~C. and {Bohringer}, H. and {Allen}, S.~W. and {Crawford}, C.~S. and {Fabian}, A.~C. and {Voges}, W. and {Huchra}, J.~P.},
        title = "{The ROSAT Brightest Cluster Sample - I. The compilation of the sample and the cluster log N-log S distribution}",
      journal = {\mnras},
         year = 1998,
        month = dec,
       volume = {301},
       number = {4},
        pages = {881-914},
          doi = {10.1046/j.1365-8711.1998.01949.x},
archivePrefix = {arXiv},
       eprint = {astro-ph/9812394},
 primaryClass = {astro-ph},
       adsurl = {https://ui.adsabs.harvard.edu/abs/1998MNRAS.301..881E}
}

@ARTICLE{1999ApJ...517..627M,
       author = {{Mohr}, Joseph J. and {Mathiesen}, Benjamin and {Evrard}, August E.},
        title = "{Properties of the Intracluster Medium in an Ensemble of Nearby Galaxy Clusters}",
      journal = {\apj},
         year = 1999,
        month = jun,
       volume = {517},
       number = {2},
        pages = {627-649},
          doi = {10.1086/307227},
archivePrefix = {arXiv},
       eprint = {astro-ph/9901281},
 primaryClass = {astro-ph},
       adsurl = {https://ui.adsabs.harvard.edu/abs/1999ApJ...517..627M}
}

@ARTICLE{1993ApJS...88..253S,
       author = {{Sutherland}, Ralph S. and {Dopita}, M.~A.},
        title = "{Cooling Functions for Low-Density Astrophysical Plasmas}",
      journal = {The Astrophysical Journals},
         year = 1993,
        month = sep,
       volume = {88},
        pages = {253},
          doi = {10.1086/191823},
       adsurl = {https://ui.adsabs.harvard.edu/abs/1993ApJS...88..253S}
}

@ARTICLE{biava2024,
       author = {{Biava}, N. and {Bonafede}, A. and {Gastaldello}, F. and {Botteon}, A. and {Brienza}, M. and {Shimwell}, T.~W. and {Brunetti}, G. and {Bruno}, L. and {Rajpurohit}, K. and {Riseley}, C.~J. and {van Weeren}, R.~J. and {Rossetti}, M. and {Cassano}, R. and {De Gasperin}, F. and {Drabent}, A. and {Rottgering}, H.~J.~A. and {Edge}, A.~C. and {Tasse}, C.},
        title = "{First evidence of a connection between cluster-scale diffuse radio emission in cool-core galaxy clusters and sloshing features}",
      journal = {\aap},
         year = 2024,
        month = jun,
       volume = {686},
          eid = {A82},
        pages = {A82},
          doi = {10.1051/0004-6361/202348045},
archivePrefix = {arXiv},
       eprint = {2403.09802},
 primaryClass = {astro-ph.CO},
       adsurl = {https://ui.adsabs.harvard.edu/abs/2024A&A...686A..82B}
}

@ARTICLE{ubertosi2024,
       author = {{Ubertosi}, Francesco and {Schellenberger}, Gerrit and {O'Sullivan}, Ewan and {Vrtilek}, Jan and {Giacintucci}, Simona and {David}, Laurence P. and {Forman}, William and {Gitti}, Myriam and {Venturi}, Tiziana and {Jones}, Christine and {Brighenti}, Fabrizio},
        title = "{Jet Reorientation in Central Galaxies of Clusters and Groups: Insights from VLBA and Chandra Data}",
      journal = {\apj},
         year = 2024,
        month = jan,
       volume = {961},
       number = {1},
          eid = {134},
        pages = {134},
          doi = {10.3847/1538-4357/ad11d8},
archivePrefix = {arXiv},
       eprint = {2312.02283},
 primaryClass = {astro-ph.GA},
       adsurl = {https://ui.adsabs.harvard.edu/abs/2024ApJ...961..134U}
}

@ARTICLE{govoni2001,
       author = {{Govoni}, F. and {En{\ss}lin}, T.~A. and {Feretti}, L. and {Giovannini}, G.},
        title = "{A comparison of radio and X-ray morphologies of four clusters of galaxies containing radio halos}",
      journal = {\aap},
         year = 2001,
        month = apr,
       volume = {369},
        pages = {441-449},
          doi = {10.1051/0004-6361:20010115},
archivePrefix = {arXiv},
       eprint = {astro-ph/0101418},
 primaryClass = {astro-ph},
       adsurl = {https://ui.adsabs.harvard.edu/abs/2001A&A...369..441G}
}

@ARTICLE{2009ApJS..182...12C,
       author = {{Cavagnolo}, Kenneth W. and {Donahue}, Megan and {Voit}, G. Mark and {Sun}, Ming},
        title = "{Intracluster Medium Entropy Profiles for a Chandra Archival Sample of Galaxy Clusters}",
      journal = {The Astrophysical Journals},
         year = 2009,
        month = may,
       volume = {182},
       number = {1},
        pages = {12-32},
          doi = {10.1088/0067-0049/182/1/12},
archivePrefix = {arXiv},
       eprint = {0902.1802},
 primaryClass = {astro-ph.CO},
       adsurl = {https://ui.adsabs.harvard.edu/abs/2009ApJS..182...12C}
}

@ARTICLE{vanweeren2019,
       author = {{van Weeren}, R.~J. and {de Gasperin}, F. and {Akamatsu}, H. and {Br{\"u}ggen}, M. and {Feretti}, L. and {Kang}, H. and {Stroe}, A. and {Zandanel}, F.},
        title = "{Diffuse Radio Emission from Galaxy Clusters}",
      journal = {\ssr},
         year = 2019,
        month = feb,
       volume = {215},
       number = {1},
          eid = {16},
        pages = {16},
          doi = {10.1007/s11214-019-0584-z},
archivePrefix = {arXiv},
       eprint = {1901.04496},
 primaryClass = {astro-ph.HE},
       adsurl = {https://ui.adsabs.harvard.edu/abs/2019SSRv..215...16V}
}

@ARTICLE{vanweeren2024,
       author = {{van Weeren}, R.~J. and {Timmerman}, R. and {Vaidya}, V. and {Gendron-Marsolais}, M.-L. and {Botteon}, A. and {Roberts}, I.~D. and {Hlavacek-Larrondo}, J. and {Bonafede}, A. and {Br{\"u}ggen}, M. and {Brunetti}, G. and {Cassano}, R. and {Cuciti}, V. and {Edge}, A.~C. and {Gastaldello}, F. and {Groeneveld}, C. and {Shimwell}, T.~W.},
        title = "{LOFAR high-band antenna observations of the Perseus cluster: The discovery of a giant radio halo}",
      journal = {\aap},
         year = 2024,
        month = dec,
       volume = {692},
          eid = {A12},
        pages = {A12},
          doi = {10.1051/0004-6361/202451618},
archivePrefix = {arXiv},
       eprint = {2410.02863},
 primaryClass = {astro-ph.CO},
       adsurl = {https://ui.adsabs.harvard.edu/abs/2024A&A...692A..12V}
}

@ARTICLE{bonafede2023,
       author = {{Bonafede}, A. and {Gitti}, M. and {La Bella}, N. and {Biava}, N. and {Ubertosi}, F. and {Brunetti}, G. and {Lusetti}, G. and {Brienza}, M. and {Riseley}, C.~J. and {Stuardi}, C. and {Botteon}, A. and {Ignesti}, A. and {R{\"o}ttgering}, H. and {van Weeren}, R.~J.},
        title = "{Shock imprints on the radio mini halo in RBS 797}",
      journal = {\aap},
         year = 2023,
        month = dec,
       volume = {680},
          eid = {A5},
        pages = {A5},
          doi = {10.1051/0004-6361/202347567},
archivePrefix = {arXiv},
       eprint = {2310.07773},
 primaryClass = {astro-ph.CO},
       adsurl = {https://ui.adsabs.harvard.edu/abs/2023A&A...680A...5B}
}

@ARTICLE{bravi2016,
       author = {{Bravi}, L. and {Gitti}, M. and {Brunetti}, G.},
        title = "{Do radio mini-halos and gas heating in cool-core clusters have a common origin?}",
      journal = {\mnras},
         year = 2016,
        month = jan,
       volume = {455},
       number = {1},
        pages = {L41-L45},
          doi = {10.1093/mnrasl/slv137},
archivePrefix = {arXiv},
       eprint = {1509.06911},
 primaryClass = {astro-ph.CO},
       adsurl = {https://ui.adsabs.harvard.edu/abs/2016MNRAS.455L..41B}
}

@ARTICLE{brunettilazarian2011,
       author = {{Brunetti}, G. and {Lazarian}, A.},
        title = "{Particle reacceleration by compressible turbulence in galaxy clusters: effects of a reduced mean free path}",
      journal = {\mnras},
         year = 2011,
        month = apr,
       volume = {412},
       number = {2},
        pages = {817-824},
          doi = {10.1111/j.1365-2966.2010.17937.x},
archivePrefix = {arXiv},
       eprint = {1011.1198},
 primaryClass = {astro-ph.CO},
       adsurl = {https://ui.adsabs.harvard.edu/abs/2011MNRAS.412..817B}
}

@ARTICLE{Botteon2018,
       author = {{Botteon}, A. and {Shimwell}, T.~W. and {Bonafede}, A. and {Dallacasa}, D. and {Brunetti}, G. and {Mandal}, S. and {van Weeren}, R.~J. and {Br{\"u}ggen}, M. and {Cassano}, R. and {de Gasperin}, F. and {Hoang}, D.~N. and {Hoeft}, M. and {R{\"o}ttgering}, H.~J.~A. and {Savini}, F. and {White}, G.~J. and {Wilber}, A. and {Venturi}, T.},
        title = "{LOFAR discovery of a double radio halo system in Abell 1758 and radio/X-ray study of the cluster pair}",
      journal = {\mnras},
         year = 2018,
        month = jul,
       volume = {478},
       number = {1},
        pages = {885-898},
          doi = {10.1093/mnras/sty1102},
archivePrefix = {arXiv},
       eprint = {1804.09187},
 primaryClass = {astro-ph.CO},
       adsurl = {https://ui.adsabs.harvard.edu/abs/2018MNRAS.478..885B}
}

@ARTICLE{Veronica2026,
       author = {{Veronica}, Angie and {Reiprich}, Thomas H. and {Ota}, Naomi and {Dietl}, Jakob and {Groth}, Frederick and {Dolag}, Klaus and {Gattuzz}, Efrain and {Pacaud}, Florian and {Roediger}, Elke and {Sanders}, Jeremy S. and {Seidel}, Benjamin and {Zhao}, Yuanyuan},
        title = "{The Quiescent Sloshing Core of Abell 496 with XRISM}",
      journal = {arXiv e-prints},
         year = 2026,
        month = jun,
          eid = {arXiv:2607.00114},
        pages = {arXiv:2607.00114},
          doi = {10.48550/arXiv.2607.00114},
archivePrefix = {arXiv},
       eprint = {2607.00114},
 primaryClass = {astro-ph.CO},
       adsurl = {https://ui.adsabs.harvard.edu/abs/2026arXiv260700114V},
    note    = {submitted to \aap}
}

@ARTICLE{gitti2002,
       author = {{Gitti}, M. and {Brunetti}, G. and {Setti}, G.},
        title = "{Modeling the interaction between ICM and relativistic plasma in cooling flows: The case of the Perseus cluster}",
      journal = {\aap},
         year = 2002,
        month = may,
       volume = {386},
        pages = {456-463},
          doi = {10.1051/0004-6361:20020284},
archivePrefix = {arXiv},
       eprint = {astro-ph/0202279},
 primaryClass = {astro-ph},
       adsurl = {https://ui.adsabs.harvard.edu/abs/2002A&A...386..456G}
}

@ARTICLE{laferriere2020,
       author = {{Richard-Laferri{\`e}re}, A. and {Hlavacek-Larrondo}, J. and {Nemmen}, R.~S. and {Rhea}, C.~L. and {Taylor}, G.~B. and {Prasow-{\'E}mond}, M. and {Gendron-Marsolais}, M. and {Latulippe}, M. and {Edge}, A.~C. and {Fabian}, A.~C. and {Sanders}, J.~S. and {Hogan}, M.~T. and {Demontigny}, G.},
        title = "{On the relation between mini-halos and AGN feedback in clusters of galaxies}",
      journal = {\mnras},
         year = 2020,
        month = oct,
       volume = {499},
       number = {2},
        pages = {2934-2958},
          doi = {10.1093/mnras/staa2877},
archivePrefix = {arXiv},
       eprint = {2007.01306},
 primaryClass = {astro-ph.GA},
       adsurl = {https://ui.adsabs.harvard.edu/abs/2020MNRAS.499.2934R}
}

@ARTICLE{ubertosi23AA,
       author = {{Ubertosi}, F. and {Gitti}, M. and {Brighenti}, F.},
        title = "{Chasing ICM cooling and AGN feedback from the macro to the meso scales in the galaxy cluster ZwCl 235}",
      journal = {\aap},
         year = 2023,
        month = feb,
       volume = {670},
          eid = {A23},
        pages = {A23},
          doi = {10.1051/0004-6361/202244023},
archivePrefix = {arXiv},
       eprint = {2211.09141},
 primaryClass = {astro-ph.GA},
       adsurl = {https://ui.adsabs.harvard.edu/abs/2023A&A...670A..23U}
}

@ARTICLE{2012MNRAS.420.3632R,
       author = {{Roediger}, E. and {Lovisari}, L. and {Dupke}, R. and {Ghizzardi}, S. and {Br{\"u}ggen}, M. and {Kraft}, R.~P. and {Machacek}, M.~E.},
        title = "{Gas sloshing, cold fronts, Kelvin-Helmholtz instabilities and the merger history of the cluster of galaxies Abell 496}",
      journal = {\mnras},
         year = 2012,
        month = mar,
       volume = {420},
       number = {4},
        pages = {3632-3648},
          doi = {10.1111/j.1365-2966.2011.20287.x},
archivePrefix = {arXiv},
       eprint = {1112.1407},
 primaryClass = {astro-ph.CO},
       adsurl = {https://ui.adsabs.harvard.edu/abs/2012MNRAS.420.3632R}
}

@ARTICLE{2021ApJ...914...73Z,
       author = {{ZuHone}, John A. and {Markevitch}, Maxim and {Weinberger}, Rainer and {Nulsen}, Paul and {Ehlert}, Kristian},
        title = "{How Merger-driven Gas Motions in Galaxy Clusters Can Turn AGN Bubbles into Radio Relics}",
      journal = {\apj},
         year = 2021,
        month = jun,
       volume = {914},
       number = {1},
          eid = {73},
        pages = {73},
          doi = {10.3847/1538-4357/abf7bc},
archivePrefix = {arXiv},
       eprint = {2012.02001},
 primaryClass = {astro-ph.HE},
       adsurl = {https://ui.adsabs.harvard.edu/abs/2021ApJ...914...73Z}
}

@ARTICLE{piffaretti2011,
       author = {{Piffaretti}, R. and {Arnaud}, M. and {Pratt}, G.~W. and {Pointecouteau}, E. and {Melin}, J. -B.},
        title = "{The MCXC: a meta-catalogue of x-ray detected clusters of galaxies}",
      journal = {\aap},
         year = 2011,
        month = oct,
       volume = {534},
          eid = {A109},
        pages = {A109},
          doi = {10.1051/0004-6361/201015377},
archivePrefix = {arXiv},
       eprint = {1007.1916},
 primaryClass = {astro-ph.CO},
       adsurl = {https://ui.adsabs.harvard.edu/abs/2011A&A...534A.109P}
}

@ARTICLE{osullivan2024,
       author = {{O'Sullivan}, Ewan and {Rajpurohit}, Kamlesh and {Schellenberger}, Gerrit and {Vrtilek}, Jan and {David}, Laurence P. and {Babul}, Arif and {Olivares}, Valeria and {Ubertosi}, Francesco and {Kolokythas}, Konstantinos and {Babyk}, Iurii and {Loubser}, Ilani},
        title = "{A Hot Core in the Group-dominant Elliptical Galaxy NGC 777}",
      journal = {\apj},
         year = 2024,
        month = jul,
       volume = {970},
       number = {1},
          eid = {65},
        pages = {65},
          doi = {10.3847/1538-4357/ad4ed6},
archivePrefix = {arXiv},
       eprint = {2405.13667},
 primaryClass = {astro-ph.GA},
       adsurl = {https://ui.adsabs.harvard.edu/abs/2024ApJ...970...65O}
}

@article{offringa-wsclean-2017,
  author = {Offringa, A. R. and Smirnov, O.},
  title = {{An optimized algorithm for multiscale wideband deconvolution of radio astronomical images}},
  volume = {471},
  number = {1},
  pages = {301-316},
  year = {2017},
  doi = {10.1093/mnras/stx1547},
  journal = {MNRAS}
}

@article{offringa-wsclean-2014,
  author = {Offringa, A. R. and McKinley, B. and Hurley-Walker and others},
  title = {{WSClean: an implementation of a fast, generic wide-field imager for radio astronomy}},
  volume = {444},
  number = {1},
  pages = {606-619},
  year = {2014},
  doi = {10.1093/mnras/stu1368},
  journal = {MNRAS}
}

@INPROCEEDINGS{jonas2016,
       author = {{Jonas}, J. and {MeerKAT Team}},
        title = "{The MeerKAT Radio Telescope}",
    booktitle = {MeerKAT Science: On the Pathway to the SKA},
         year = 2016,
        month = jan,
          eid = {1},
        pages = {1},
          doi = {10.22323/1.277.0001},
       adsurl = {https://ui.adsabs.harvard.edu/abs/2016mks..confE...1J}
}

@ARTICLE{gitti2004,
       author = {{Gitti}, M. and {Brunetti}, G. and {Feretti}, L. and {Setti}, G.},
        title = "{Particle acceleration in cooling flow clusters of galaxies: The case of Abell 2626}",
      journal = {\aap},
         year = 2004,
        month = apr,
       volume = {417},
        pages = {1-11},
          doi = {10.1051/0004-6361:20031750},
archivePrefix = {arXiv},
       eprint = {astro-ph/0311275},
 primaryClass = {astro-ph},
       adsurl = {https://ui.adsabs.harvard.edu/abs/2004A&A...417....1G}
}

@ARTICLE{vazza2024,
       author = {{Vazza}, Franco and {Botteon}, Andrea},
        title = "{The Seeding of Cosmic Ray Electrons by Cluster Radio Galaxies: A Review}",
      journal = {Galaxies},
         year = 2024,
        month = apr,
       volume = {12},
       number = {2},
          eid = {19},
        pages = {19},
          doi = {10.3390/galaxies12020019},
archivePrefix = {arXiv},
       eprint = {2403.16068},
 primaryClass = {astro-ph.GA},
       adsurl = {https://ui.adsabs.harvard.edu/abs/2024Galax..12...19V}
}

@ARTICLE{cassano2023,
       author = {{Cassano}, R. and {Cuciti}, V. and {Brunetti}, G. and {Botteon}, A. and {Rossetti}, M. and {Bruno}, L. and {Simionescu}, A. and {Gastaldello}, F. and {van Weeren}, R.~J. and {Br{\"u}ggen}, M. and {Dallacasa}, D. and {Zhang}, X. and {Akamatsu}, H. and {Bonafede}, A. and {Di Gennaro}, G. and {Shimwell}, T.~W. and {de Gasperin}, F. and {R{\"o}ttgering}, H.~J.~A. and {Jones}, A.},
        title = "{The Planck clusters in the LOFAR sky. IV. LoTSS-DR2: Statistics of radio haloes and re-acceleration models}",
      journal = {\aap},
         year = 2023,
        month = apr,
       volume = {672},
          eid = {A43},
        pages = {A43},
          doi = {10.1051/0004-6361/202244876},
archivePrefix = {arXiv},
       eprint = {2301.08052},
 primaryClass = {astro-ph.CO},
       adsurl = {https://ui.adsabs.harvard.edu/abs/2023A&A...672A..43C}
}

@ARTICLE{mcdonald2012nature,
       author = {{McDonald}, M. and {Bayliss}, M. and {Benson}, B.~A. and {Foley}, R.~J. and {Ruel}, J. and {Sullivan}, P. and {Veilleux}, S. and {Aird}, K.~A. and {Ashby}, M.~L.~N. and {Bautz}, M. and {Bazin}, G. and {Bleem}, L.~E. and {Brodwin}, M. and {Carlstrom}, J.~E. and {Chang}, C.~L. and {Cho}, H.~M. and {Clocchiatti}, A. and {Crawford}, T.~M. and {Crites}, A.~T. and {de Haan}, T. and {Desai}, S. and {Dobbs}, M.~A. and {Dudley}, J.~P. and {Egami}, E. and {Forman}, W.~R. and {Garmire}, G.~P. and {George}, E.~M. and {Gladders}, M.~D. and {Gonzalez}, A.~H. and {Halverson}, N.~W. and {Harrington}, N.~L. and {High}, F.~W. and {Holder}, G.~P. and {Holzapfel}, W.~L. and {Hoover}, S. and {Hrubes}, J.~D. and {Jones}, C. and {Joy}, M. and {Keisler}, R. and {Knox}, L. and {Lee}, A.~T. and {Leitch}, E.~M. and {Liu}, J. and {Lueker}, M. and {Luong-van}, D. and {Mantz}, A. and {Marrone}, D.~P. and {McMahon}, J.~J. and {Mehl}, J. and {Meyer}, S.~S. and {Miller}, E.~D. and {Mocanu}, L. and {Mohr}, J.~J. and {Montroy}, T.~E. and {Murray}, S.~S. and {Natoli}, T. and {Padin}, S. and {Plagge}, T. and {Pryke}, C. and {Rawle}, T.~D. and {Reichardt}, C.~L. and {Rest}, A. and {Rex}, M. and {Ruhl}, J.~E. and {Saliwanchik}, B.~R. and {Saro}, A. and {Sayre}, J.~T. and {Schaffer}, K.~K. and {Shaw}, L. and {Shirokoff}, E. and {Simcoe}, R. and {Song}, J. and {Spieler}, H.~G. and {Stalder}, B. and {Staniszewski}, Z. and {Stark}, A.~A. and {Story}, K. and {Stubbs}, C.~W. and {{\v{S}}uhada}, R. and {van Engelen}, A. and {Vanderlinde}, K. and {Vieira}, J.~D. and {Vikhlinin}, A. and {Williamson}, R. and {Zahn}, O. and {Zenteno}, A.},
        title = "{A massive, cooling-flow-induced starburst in the core of a luminous cluster of galaxies}",
      journal = {\nat},
         year = 2012,
        month = aug,
       volume = {488},
       number = {7411},
        pages = {349-352},
          doi = {10.1038/nature11379},
archivePrefix = {arXiv},
       eprint = {1208.2962},
 primaryClass = {astro-ph.CO},
       adsurl = {https://ui.adsabs.harvard.edu/abs/2012Natur.488..349M}
}

@ARTICLE{sotira2025,
       author = {{Sotira}, S. and {Vazza}, F. and {Brighenti}, F.},
        title = "{Impact of active galactic nucleus feedback modes on the turbulent properties of the multiphase intracluster medium}",
      journal = {\aap},
         year = 2025,
        month = may,
       volume = {697},
          eid = {A138},
        pages = {A138},
          doi = {10.1051/0004-6361/202452549},
archivePrefix = {arXiv},
       eprint = {2410.07314},
 primaryClass = {astro-ph.GA},
       adsurl = {https://ui.adsabs.harvard.edu/abs/2025A&A...697A.138S}
}

@ARTICLE{intema2009,
       author = {{Intema}, H.~T. and {van der Tol}, S. and {Cotton}, W.~D. and {Cohen}, A.~S. and {van Bemmel}, I.~M. and {R{\"o}ttgering}, H.~J.~A.},
        title = "{Ionospheric calibration of low frequency radio interferometric observations using the peeling scheme. I. Method description and first results}",
      journal = {\aap},
         year = 2009,
        month = jul,
       volume = {501},
       number = {3},
        pages = {1185-1205},
          doi = {10.1051/0004-6361/200811094},
archivePrefix = {arXiv},
       eprint = {0904.3975},
 primaryClass = {astro-ph.IM},
       adsurl = {https://ui.adsabs.harvard.edu/abs/2009A&A...501.1185I}
}

@ARTICLE{Sanderson2006,
       author = {{Sanderson}, Alastair J.~R. and {Ponman}, Trevor J. and {O'Sullivan}, Ewan},
        title = "{A statistically selected Chandra sample of 20 galaxy clusters - I. Temperature and cooling time profiles}",
      journal = {\mnras},
         year = 2006,
        month = nov,
       volume = {372},
       number = {4},
        pages = {1496-1508},
          doi = {10.1111/j.1365-2966.2006.10956.x},
archivePrefix = {arXiv},
       eprint = {astro-ph/0608423},
 primaryClass = {astro-ph},
       adsurl = {https://ui.adsabs.harvard.edu/abs/2006MNRAS.372.1496S}
}

@INCOLLECTION{zuhone2022,
       author = {{ZuHone}, John and {Su}, Yuanyuan},
        title = "{The Merger Dynamics of the X-Ray-Emitting Plasma in Clusters of Galaxies}",
    booktitle = {Handbook of X-ray and Gamma-ray Astrophysics},
         year = 2022,
       editor = {{Bambi}, Cosimo and {Sangangelo}, Andrea},
          eid = {93},
        pages = {93},
          doi = {10.1007/978-981-16-4544-0_124-1},
       adsurl = {https://ui.adsabs.harvard.edu/abs/2022hxga.book...93Z}
}

@ARTICLE{shitanishi2018,
       author = {{Shitanishi}, Jennifer A. and {Pierpaoli}, Elena and {Sayers}, Jack and {Golwala}, Sunil R. and {Ameglio}, Silvia and {Mantz}, Adam B. and {Mroczkowski}, Tony K. and {Rasia}, Elena and {Siegel}, Seth},
        title = "{Thermodynamic profiles of galaxy clusters from a joint X-ray/SZ analysis}",
      journal = {\mnras},
         year = 2018,
        month = nov,
       volume = {481},
       number = {1},
        pages = {749-792},
          doi = {10.1093/mnras/sty2195},
archivePrefix = {arXiv},
       eprint = {1712.05464},
 primaryClass = {astro-ph.CO},
       adsurl = {https://ui.adsabs.harvard.edu/abs/2018MNRAS.481..749S}
}

@ARTICLE{vikhlinin06,
       author = {{Vikhlinin}, A. and {Kravtsov}, A. and {Forman}, W. and {Jones}, C. and {Markevitch}, M. and {Murray}, S.~S. and {Van Speybroeck}, L.},
        title = "{Chandra Sample of Nearby Relaxed Galaxy Clusters: Mass, Gas Fraction, and Mass-Temperature Relation}",
      journal = {\apj},
         year = 2006,
        month = apr,
       volume = {640},
       number = {2},
        pages = {691-709},
          doi = {10.1086/500288},
archivePrefix = {arXiv},
       eprint = {astro-ph/0507092},
 primaryClass = {astro-ph},
       adsurl = {https://ui.adsabs.harvard.edu/abs/2006ApJ...640..691V}
}

@ARTICLE{fornasiero2025,
       author = {{Fornasiero}, I. and {Ubertosi}, F. and {Gitti}, M.},
        title = "{Investigating AGN feedback in H{\ensuremath{\alpha}}-luminous galaxy clusters: First Chandra X-ray analysis of Abell 2009}",
      journal = {\aap},
         year = 2025,
        month = mar,
       volume = {695},
          eid = {A265},
        pages = {A265},
          doi = {10.1051/0004-6361/202453315},
archivePrefix = {arXiv},
       eprint = {2503.07781},
 primaryClass = {astro-ph.GA},
       adsurl = {https://ui.adsabs.harvard.edu/abs/2025A&A...695A.265F}
}

@ARTICLE{gitti2018,
       author = {{Gitti}, M. and {Brunetti}, G. and {Cassano}, R. and {Ettori}, S.},
        title = "{Radio-continuum surveys with SKA and LOFAR: a first look at the perspectives for radio mini-halos}",
      journal = {\aap},
         year = 2018,
        month = sep,
       volume = {617},
          eid = {A11},
        pages = {A11},
          doi = {10.1051/0004-6361/201832749},
archivePrefix = {arXiv},
       eprint = {1806.05751},
 primaryClass = {astro-ph.CO},
       adsurl = {https://ui.adsabs.harvard.edu/abs/2018A&A...617A..11G}
}

@ARTICLE{kotov2005,
       author = {{Kotov}, O. and {Vikhlinin}, A.},
        title = "{XMM-Newton Observations of Evolution of Cluster X-Ray Scaling Relations at z=0.4-0.7}",
      journal = {\apj},
         year = 2005,
        month = nov,
       volume = {633},
       number = {2},
        pages = {781-790},
          doi = {10.1086/433170},
archivePrefix = {arXiv},
       eprint = {astro-ph/0504233},
 primaryClass = {astro-ph},
       adsurl = {https://ui.adsabs.harvard.edu/abs/2005ApJ...633..781K}
}

@ARTICLE{Jaffe1977,
       author = {{Jaffe}, W.~J.},
        title = "{Origin and transport of electrons in the halo radio source in the Coma cluster.}",
      journal = {\apj},
         year = 1977,
        month = feb,
       volume = {212},
        pages = {1-7},
          doi = {10.1086/155011},
       adsurl = {https://ui.adsabs.harvard.edu/abs/1977ApJ...212....1J}
}

@ARTICLE{yu2011,
       author = {{Yu}, Heng and {Zhu}, Zong-Hong},
        title = "{Combining optical and X-ray observations of galaxy clusters to constrain cosmological parameters}",
      journal = {Research in Astronomy and Astrophysics},
         year = 2011,
        month = jul,
       volume = {11},
       number = {7},
        pages = {776-786},
          doi = {10.1088/1674-4527/11/7/004},
archivePrefix = {arXiv},
       eprint = {1011.6060},
 primaryClass = {astro-ph.CO},
       adsurl = {https://ui.adsabs.harvard.edu/abs/2011RAA....11..776Y}
}

@ARTICLE{lancaster2011,
       author = {{Lancaster}, Katy and {Birkinshaw}, Mark and {Gawro{\'n}ski}, Marcin P. and {Battye}, Richard and {Browne}, Ian and {Davis}, Richard and {Giles}, Paul and {Feiler}, Roman and {Kus}, Andrzej and {Lew}, Bartosz and {Lowe}, Stuart and {Maughan}, Ben and {Alareedh}, Abdulaziz and {Pazderska}, Bogna and {Pazderski}, Eugeniusz and {Peel}, Mike and {Roukema}, Boudewijn F. and {Wilkinson}, Peter},
        title = "{Sunyaev-Zel'dovich observations of a statistically complete sample of galaxy clusters with OCRA-p}",
      journal = {\mnras},
         year = 2011,
        month = dec,
       volume = {418},
       number = {3},
        pages = {1441-1451},
          doi = {10.1111/j.1365-2966.2011.19241.x},
archivePrefix = {arXiv},
       eprint = {1106.3766},
 primaryClass = {astro-ph.CO},
       adsurl = {https://ui.adsabs.harvard.edu/abs/2011MNRAS.418.1441L}
}

@ARTICLE{mazzotta2003,
       author = {{Mazzotta}, P. and {Edge}, A.~C. and {Markevitch}, M.},
        title = "{A Chandra Study of the Complex Structure in the Core of 2A 0335+096}",
      journal = {\apj},
         year = 2003,
        month = oct,
       volume = {596},
       number = {1},
        pages = {190-203},
          doi = {10.1086/377633},
archivePrefix = {arXiv},
       eprint = {astro-ph/0303314},
 primaryClass = {astro-ph},
       adsurl = {https://ui.adsabs.harvard.edu/abs/2003ApJ...596..190M}
}

@ARTICLE{landry2013,
       author = {{Landry}, D. and {Bonamente}, M. and {Giles}, P. and {Maughan}, B. and {Joy}, M. and {Murray}, S.},
        title = "{Chandra measurements of a complete sample of X-ray luminous galaxy clusters: the gas mass fraction}",
      journal = {\mnras},
         year = 2013,
        month = aug,
       volume = {433},
       number = {4},
        pages = {2790-2811},
          doi = {10.1093/mnras/stt901},
archivePrefix = {arXiv},
       eprint = {1211.4626},
 primaryClass = {astro-ph.CO},
       adsurl = {https://ui.adsabs.harvard.edu/abs/2013MNRAS.433.2790L}
}

@ARTICLE{vanweeren2021,
       author = {{van Weeren}, R.~J. and {Shimwell}, T.~W. and {Botteon}, A. and {Brunetti}, G. and {Br{\"u}ggen}, M. and {Boxelaar}, J.~M. and {Cassano}, R. and {Di Gennaro}, G. and {Andrade-Santos}, F. and {Bonnassieux}, E. and {Bonafede}, A. and {Cuciti}, V. and {Dallacasa}, D. and {de Gasperin}, F. and {Gastaldello}, F. and {Hardcastle}, M.~J. and {Hoeft}, M. and {Kraft}, R.~P. and {Mandal}, S. and {Rossetti}, M. and {R{\"o}ttgering}, H.~J.~A. and {Tasse}, C. and {Wilber}, A.~G.},
        title = "{LOFAR observations of galaxy clusters in HETDEX. Extraction and self-calibration of individual LOFAR targets}",
      journal = {\aap},
         year = 2021,
        month = jul,
       volume = {651},
          eid = {A115},
        pages = {A115},
          doi = {10.1051/0004-6361/202039826},
archivePrefix = {arXiv},
       eprint = {2011.02387},
 primaryClass = {astro-ph.CO},
       adsurl = {https://ui.adsabs.harvard.edu/abs/2021A&A...651A.115V}
}

@ARTICLE{birzan2020,
       author = {{B{\^\i}rzan}, L. and {Rafferty}, D.~A. and {Br{\"u}ggen}, M. and {Botteon}, A. and {Brunetti}, G. and {Cuciti}, V. and {Edge}, A.~C. and {Morganti}, R. and {R{\"o}ttgering}, H.~J.~A. and {Shimwell}, T.~W.},
        title = "{LOFAR observations of X-ray cavity systems}",
      journal = {\mnras},
         year = 2020,
        month = aug,
       volume = {496},
       number = {3},
        pages = {2613-2635},
          doi = {10.1093/mnras/staa1594},
archivePrefix = {arXiv},
       eprint = {2006.09708},
 primaryClass = {astro-ph.GA},
       adsurl = {https://ui.adsabs.harvard.edu/abs/2020MNRAS.496.2613B}
}

@ARTICLE{deGasperin2021,
       author = {{de Gasperin}, F. and {Williams}, W.~L. and {Best}, P. and {Br{\"u}ggen}, M. and {Brunetti}, G. and {Cuciti}, V. and {Dijkema}, T.~J. and {Hardcastle}, M.~J. and {Norden}, M.~J. and {Offringa}, A. and {Shimwell}, T. and {van Weeren}, R. and {Bomans}, D. and {Bonafede}, A. and {Botteon}, A. and {Callingham}, J.~R. and {Cassano}, R. and {Chy{\.z}y}, K.~T. and {Emig}, K.~L. and {Edler}, H. and {Haverkorn}, M. and {Heald}, G. and {Heesen}, V. and {Iacobelli}, M. and {Intema}, H.~T. and {Kadler}, M. and {Ma{\l}ek}, K. and {Mevius}, M. and {Miley}, G. and {Mingo}, B. and {Morabito}, L.~K. and {Sabater}, J. and {Morganti}, R. and {Orr{\'u}}, E. and {Pizzo}, R. and {Prandoni}, I. and {Shulevski}, A. and {Tasse}, C. and {Vaccari}, M. and {Zarka}, P. and {R{\"o}ttgering}, H.},
        title = "{The LOFAR LBA Sky Survey. I. Survey description and preliminary data release}",
      journal = {\aap},
         year = 2021,
        month = apr,
       volume = {648},
          eid = {A104},
        pages = {A104},
          doi = {10.1051/0004-6361/202140316},
archivePrefix = {arXiv},
       eprint = {2102.09238},
 primaryClass = {astro-ph.IM},
       adsurl = {https://ui.adsabs.harvard.edu/abs/2021A&A...648A.104D}
}

@ARTICLE{vanWeeren2016,
       author = {{van Weeren}, R.~J. and {Williams}, W.~L. and {Hardcastle}, M.~J. and {Shimwell}, T.~W. and {Rafferty}, D.~A. and {Sabater}, J. and {Heald}, G. and {Sridhar}, S.~S. and {Dijkema}, T.~J. and {Brunetti}, G. and {Br{\"u}ggen}, M. and {Andrade-Santos}, F. and {Ogrean}, G.~A. and {R{\"o}ttgering}, H.~J.~A. and {Dawson}, W.~A. and {Forman}, W.~R. and {de Gasperin}, F. and {Jones}, C. and {Miley}, G.~K. and {Rudnick}, L. and {Sarazin}, C.~L. and {Bonafede}, A. and {Best}, P.~N. and {B{\^\i}rzan}, L. and {Cassano}, R. and {Chy{\.z}y}, K.~T. and {Croston}, J.~H. and {Ensslin}, T. and {Ferrari}, C. and {Hoeft}, M. and {Horellou}, C. and {Jarvis}, M.~J. and {Kraft}, R.~P. and {Mevius}, M. and {Intema}, H.~T. and {Murray}, S.~S. and {Orr{\'u}}, E. and {Pizzo}, R. and {Simionescu}, A. and {Stroe}, A. and {van der Tol}, S. and {White}, G.~J.},
        title = "{LOFAR Facet Calibration}",
      journal = {\apjs},
         year = 2016,
        month = mar,
       volume = {223},
       number = {1},
          eid = {2},
        pages = {2},
          doi = {10.3847/0067-0049/223/1/2},
archivePrefix = {arXiv},
       eprint = {1601.05422},
 primaryClass = {astro-ph.IM},
       adsurl = {https://ui.adsabs.harvard.edu/abs/2016ApJS..223....2V}
}

@ARTICLE{Pasini2022,
       author = {{Pasini}, T. and {Edler}, H.~W. and {Br{\"u}ggen}, M. and {de Gasperin}, F. and {Botteon}, A. and {Rajpurohit}, K. and {van Weeren}, R.~J. and {Gastaldello}, F. and {Gaspari}, M. and {Brunetti}, G. and {Cuciti}, V. and {Nanci}, C. and {di Gennaro}, G. and {Rossetti}, M. and {Dallacasa}, D. and {Hoang}, D.~N. and {Riseley}, C.~J.},
        title = "{Particle re-acceleration and diffuse radio sources in the galaxy cluster Abell 1550}",
      journal = {\aap},
         year = 2022,
        month = jul,
       volume = {663},
          eid = {A105},
        pages = {A105},
          doi = {10.1051/0004-6361/202243833},
archivePrefix = {arXiv},
       eprint = {2205.12281},
 primaryClass = {astro-ph.CO},
       adsurl = {https://ui.adsabs.harvard.edu/abs/2022A&A...663A.105P}
}
%

\begin{appendix}
\section{The MIMIC sample}\label{app:mimics}
In Table \ref{tab:mimics} of this Appendix we present an initial selection of the low-mass galaxy clusters that form the MIMIC (Mini-halo in Mini-Clusters) project. The objects are in decreasing M$_{500}$ order.

\begin{table*}[ht]
    \centering
    {\renewcommand{\arraystretch}{1.3}
    \caption{The MIMIC sample.}\label{tab:mimics}}
    \begin{tabular}{lccccc}
Cluster & RA (J2000) & DEC (J2000) & z & M$_{500}^{X}$ [M$_{\odot}$] & M$_{500}^{SZ}$ [M$_{\odot}$] \\
(1) & (2) & (3) & (4) & (5) & (6) \\
\hline
Abell~2199 & 16 28 34.5 & +39 32 26 & 0.029  & $2.93\times10^{14}$  & $2.87\times10^{14}$ \\
ZwCl~235  & 00 43 52.0 & +24 24 21 & 0.083  & $2.80\times10^{14}  $ & -- \\
Abell~496    & 04 33 37.7 & $-$13 15 41  & 0.033 & $2.80 \times10^{14} $ & $2.71\times10^{14}$ \\
Abell~S1101 & 23 13 58.5 & $-$42 43 39 & 0.056 & $2.80 \times10^{14} $& $1.46\times10^{14}$ \\
Abell~4059    & 23 57 02.2 & $-$34 45 58   & 0.048 & $2.65 \times10^{14} $ & $2.46\times10^{14}$\\
Abell~2495 & 22 50 19.7 & +10 54 11 & 0.077 & $2.54 \times10^{14} $ & $2.23\times10^{14}$ \\
Abell~2052 & 15 16 44.7 & +07 01 21 & 0.035 & $2.47 \times10^{14} $ & -- \\
Abell~2415 & 22 05 38.6 & $-$05 35 34 & 0.058 & $2.09 \times10^{14} $ & $1.78\times10^{14}$\\
Abell~1668 & 13 03 46.5 & +19 16 13 & 0.063 & $1.89 \times10^{14} $ & $1.86\times10^{14}$ \\
Abell~3880 & 22 27 54.4 & $-$30 34 37 & 0.058 & $1.79 \times10^{14} $& -- \\
Abell~1991 & 14 54 31.6 & +18 38 32 & 0.059  & $1.68 \times10^{14} $& -- \\
CAN~010 & 01 20 58.8 & $-$13 51 7 & 0.052 & $1.68 \times10^{14} $ & $1.67\times10^{14}$\\
Abell~3526  & 12 48 48.9 & $-$41 18 45   & 0.011 & $1.63 \times10^{14} $& $1.32\times10^{14}$ \\

\hline
    \end{tabular}
    \tablefoot{Columns: (1) Name (2) Right ascension (3) Declination (4) redshift (5) mass within $R_{500}$ (from \citealt{piffaretti2011,mcxc_ii}) (6) mass within $R_{500}$  (from \citealt{planck_2016_IIDatarelease}).}
\end{table*}

\section{Calibration of interferometric radio observations presented in this work}\label{app:datareduction}
This Appendix summarizes the data reduction steps employed to calibrate the interferometric radio observations and X-ray observations presented in this work. Separate subsections are dedicated to different instruments and frequency ranges. We also provide a table with a summary of the data and images (Tab.~\ref{tab:radioimages}), and a table with a summary of the flux densities of the targets measured from each data set (Tab.~\ref{tab:fluxtable}).
\begin{table*}[ht!]
    \centering
        \caption{Radio observations and images presented in this work.} \label{tab:radioimages}\renewcommand{\arraystretch}{1.0}
    \begin{tabular}{cccccccc}
    Cluster & Map & UV & Briggs & $\nu$ & $\sigma_{rms}$ & $\theta_{\text{FWHM}}$ & Figure \\
     &  & & & [MHz] & [mJy~beam$^{-1}$] & [$\arcsec$] & \\
     (1) & (2) & (3) & (4) & (5) & (6) & (7) & (8) \\
    \hline
        \multirow{6}{*}{Abell 496} & \multirow{2}{*}{Full-uv}  & \multirow{2}{*}{-} & \multirow{2}{*}{$-0.5$} & 650 & 0.05 &$11.4\times9.7$, 153$^{\circ}$ & Fig. \ref{fig:total_A496}, top left \\
        & & & & 1280 & 0.005 &$5.9\times4.8$, -12$^{\circ}$ & Fig. \ref{fig:total_A496}, bottom left \\
        & \multirow{2}{*}{Discrete sources}  & \multirow{2}{*}{$>2$~k$\lambda$} & \multirow{2}{*}{$+0.5$} & 650  & 0.05 &$12.7\times10.0$, 149$^{\circ}$ & Fig. \ref{fig:total_A496}, top center \\
        & & & & 1280  & 0.008 &$11.3\times11.2$, 1.6$^{\circ}$ & Fig. \ref{fig:total_A496}, bottom center \\
        & \multirow{2}{*}{Diffuse emission}  & \multirow{2}{*}{Full-uv, taper$=$20$\arcsec$} & \multirow{2}{*}{$+0.5$} & 650 & 0.09 &$22.0\times22.0$, 0$^{\circ}$ & Fig. \ref{fig:total_A496}, top right \\
        & & & & 1280  & 0.025 &$22.0\times22.0$, 0$^{\circ}$ & Fig. \ref{fig:total_A496}, bottom right \\
        \hline
        \multirow{9}{*}{ZwCl 235} & \multirow{3}{*}{Full-uv}  & \multirow{3}{*}{-} & \multirow{3}{*}{$-0.5$} & 54  & 4.0 &$17.3\times11.2$, 97$^{\circ}$ & Fig. \ref{fig:total_z235}, top left \\
        & & & & 144 & 0.08 &$8.0\times4.9$, 85$^{\circ}$& Fig. \ref{fig:total_z235}, middle left \\
        & & & & 400 & 0.035 &$12.7\times7.8$, 50$^{\circ}$ & Fig. \ref{fig:total_z235}, bottom left \\
        & \multirow{3}{*}{Discrete sources}  & \multirow{3}{*}{$>5$~k$\lambda$} & \multirow{3}{*}{$-0.5$} & 54  & 4.6 &$16.9\times11.1$, 96$^{\circ}$ & Fig. \ref{fig:total_z235}, top center \\
        & & & & 144 & 0.2 &$5.5\times3.6$, 83$^{\circ}$ & Fig. \ref{fig:total_z235}, middle center \\
        & & & & 400 & 0.03 &$10.2\times3.6$, 46$^{\circ}$ & Fig. \ref{fig:total_z235}, bottom center \\
        & \multirow{3}{*}{Diffuse emission}  & \multirow{3}{*}{Full-uv, taper$=$10$\arcsec$} & \multirow{3}{*}{$+0.5$} & 54  & 5.0 &$20.0\times20.0$, 0$^{\circ}$ & Fig. \ref{fig:total_z235}, top right \\
        & & & & 144 & 0.5 &$20.0\times20.0$, 0$^{\circ}$ & Fig. \ref{fig:total_z235}, middle right \\
        & & & & 400 & 0.1 &$20.0\times20.0$, 0$^{\circ}$ & Fig. \ref{fig:total_z235}, bottom right \\
        \hline
    \end{tabular}
   \tablefoot{Columns: (1) Name of the cluster (2) type of map (3) Selection of uv-cut or tapering (4) Briggs weighting (\texttt{robust} parameter) (5) central frequency (6) rms noise (7) angular resolution (8) figure where the image is shown.}
\end{table*}

\subsection{Radio observations of ZwCl 235}\label{sub:z235data}

\subsubsection{LOFAR LBA}\label{subsub:lofarlba}

LOFAR LBA observations of ZwCl~235 have been obtained as part of the LOFAR LBA Sky Survey (LoLSS, \citealt{deGasperin2023})\footnote{This data will be included in the second Data Release (de Gasperin et al. in prep.).}. The target is covered by two survey pointings, P010+26 and P013+26. The frequency range for both is $42-66$~MHz. Observations were calibrated using the automated Pipeline for LOFAR LBA (PiLL), which is described in detail in \citet{deGasperin2021, deGasperin2023}. Below, we briefly summarise the main steps.
\\\indent After an initial round of radio-frequency interference (RFI) flagging and data averaging, bandpass and phase solutions are derived for the calibrator source (3C295) using the calibrator model provided within PiLL (using the \citealt{scaife2012} flux density scale), and subsequently transferred to the target field. A coarse direction-dependent calibration is then performed to correct for first-order ionospheric delays while a small region with bright sources is used to correct for Faraday rotation and second-order beam effects.

A second, more refined, direction-dependent calibration is carried out following the facet-based approach \citep{vanWeeren2016}. All sources are first subtracted from the visibilities, after which the brightest source is reinserted to derive calibration solutions (on top of those derived from the coarse direction-dependent calibration) for that specific direction. This procedure is repeated for each sufficiently bright source in the field of view, which serves as a direction-dependent calibrator (DD-calibrator). The field is subsequently divided into facets defined by the positions of the DD-calibrators, and the corresponding solutions are applied to all sources within each facet. These steps are performed using the facet imaging mode in WSClean \citep{offringa-wsclean-2014,offringa-wsclean-2017}.
\\\indent The correction of direction-dependent ionospheric effects was further improved by employing the extraction procedure originally developed for HBA observations \citep{vanweeren2021} and later adapted for LBA data \citep{Pasini2022}. All sources outside of a $\sim35'$-wide region, centered on the target, were subtracted from the visibilities, shifting then the phase centre to ZwCl~235. The data were then averaged in time and frequency to reduce the data volume and to smear residual emission from imperfectly subtracted off-axis sources. The primary beam effect was then corrected with respect to the new phase centre. Subsequent self-calibration cycles were performed to further reduce the noise and improve the overall image fidelity. The images obtained at the final round of self calibration using the full uv-range reached an rms noise level of $\sim4$~mJy~beam$^{-1}$ with an angular resolutions of $\sim15\arcsec$. This is higher than the typical noise of the LoLSS survey ($\sim1-2$~mJy~beam$^{-1}$, \citealt{deGasperin2023}) but compatible with the position of ZwCl~235 at the edges of the considered pointings and the sub-optimal declination for LOFAR. The LOFAR LBA images reported in the paper have a central frequency of 54 MHz, and we adopt a systematic uncertainty of 6\% on the flux density measurements \citep{deGasperin2023}.

\subsubsection{LOFAR HBA}\label{subsub:lofarhba}
LOFAR HBA observations of ZwCl~235 in the frequency range 120--168 MHz were obtained as part of the LOFAR Two-meter Sky Survey (LoTSS, \citealt{shimwell2017, shimwell19}) and are included in the DR2 release \citep{Shimwell22}. The target falls within the overlapping areas of the pointings P011+23 and P010+26, at $\sim$1 deg and $\sim$1.7 deg from the respective pointing centers. To improve the calibration towards the direction of ZwCl~235, we reprocessed the survey data with the ``extract+recalibration'' method \citep{vanweeren2021} presented above for the LBA data, in this case with a field of view of $15' \times 15'$. 
The images obtained at the final round of self calibration using the full uv-range and no tapering with angular resolutions of $\sim$6$\arcsec$ reach rms noise levels of $\sim$80$\,\mu$Jy~beam$^{-1}$, in very good agreement with the median rms noise of the LoTSS survey (83~$\mu$Jy~beam$^{-1}$, \citealt{Shimwell22}). 
\\ \indent To correct for possible offsets in the absolute flux density scale in the extracted and recalibrated field, we compared the flux density of unresolved sources in the field of our data with that of the LoTSS-DR2 catalog \citep{Shimwell22}, finding a multiplicative correction factor of 1.06 that was applied to the images. The LOFAR HBA images reported in the paper have a central frequency of 144 MHz, and we adopt a systematic uncertainty of 10\% on the flux density measurements \citep{Shimwell22}.

\subsubsection{uGMRT Band 3}\label{subsub:ugmrtband3}
The uGMRT Band 3 (300 - 500 MHz) observations of ZwCl 235 were recently carried out on August 15, 2025, under project code 48\_048 (PI: F. Ubertosi), for a time on source of 7~h, using the GMRT Wideband Backend (GWB). The source 3C295 was included as a primary calibrator. The data were processed using the Source Peeling and Atmospheric Modeling (SPAM; \citealt{intema2009}) pipeline\footnote{See \url{https://www.intema.nl/doku.php?id=huibintema:spam:pipeline}.}. The pipeline calibrates the flux density scale using the primary calibrator (adopting the scale from \citealt{scaife2012}), averages the data and flags radio frequency interferences (RFI), corrects for the bandpass, and performs automated rounds of direction-independent and direction-dependent calibration. These steps are performed after splitting the wideband data into 6 subbands, which are then re-combined during imaging giving a representative frequency of 400 MHz. The images obtained at the final round of self calibration using the full uv-range and no tapering with angular resolutions of $\sim$10$\arcsec$ reach rms noise levels of $\sim$35$\,\mu$Jy~beam$^{-1}$, which is higher than the reference value from the uGMRT exposure calculator for 7h on source and a fudge factor of 5 ($\sim$15$\,\mu$Jy~beam$^{-1}$) but in line with literature uGMRT observations at Band 3 (e.g., \citealt{rajpurohit2024,osullivan2024}). When measuring flux densities from the images, we adopted a systematic uncertainty of 10\% on the flux density scale \citep{chandra2004}.

\subsection{Radio observations of Abell 496}\label{sub:a496data}

\subsubsection{uGMRT Band 4}\label{subsub:ugmrtband4}
uGMRT Band 4 observations of Abell 496 were performed on March 29, 2024, under project code 45\_017 (PI: A. Botteon), for a total time on source of 10~h using the GWB. The source 3C147 was included as a primary calibrator. The data were calibrated again using SPAM (\citealt{intema2009}; see Sect.~\ref{subsub:ugmrtband3}). Although the original frequency range of the data was 550 MHz - 900 MHz, the sensitivity of the instrument is negligible above 750 MHz. Therefore, the effective bandwidth of the data is 550 MHz - 750 MHz and the representative frequency is 650 MHz. The images obtained at the final round of self calibration using the full uv-range and no tapering with angular resolutions of $\sim$5$\arcsec$ reach rms noise levels far from other sources of $\sim$15$\,\mu$Jy~beam$^{-1}$, which is compatible with the reference value from the uGMRT exposure calculator for 10h on source and a fudge factor of 5 ($\sim$12$\,\mu$Jy~beam$^{-1}$). When measuring flux densities from the images, we adopted a systematic uncertainty of 10\% on the flux density measurements \citep{chandra2004}.

\begin{table*}[ht!]
    \centering
    \caption{Flux density measurements for the AGN, total diffuse emission, and AGN-masked diffuse emission in Abell 496 and ZwCl 235.} \label{tab:fluxtable}\renewcommand{\arraystretch}{1.3}
    \begin{tabular}{cccccc}
    Cluster & $\nu$ & Component & $S_{\nu}$ & Area & $N_{beam}$ \\
     & [MHz] & & [mJy] & [$\arcsec^2$] & \\
     (1) & (2) & (3) & (4) & (5) & (6) \\
    \hline
        \multirow{6}{*}{Abell 496} & \multirow{3}{*}{650} & AGN & $225.80\pm22.58$ & 3940.0 & 27.38 \\
        & & Diffuse (total) & $24.13\pm2.49$ & 25992.0 & 47.39 \\
        & & Diffuse (AGN-masked) & $16.07\pm1.71$ & 22052.0 & 40.21 \\
        & \multirow{3}{*}{1280} & AGN & $125.50\pm6.28$ & 5249.0 & 36.60 \\
        & & Diffuse (total) & $10.10\pm0.54$ & 27928.0 & 50.92 \\
        & & Diffuse (AGN-masked) & $6.15\pm0.35$ & 22679.0 & 41.35 \\
        \hline
        \multirow{9}{*}{ZwCl 235} & \multirow{3}{*}{54} & AGN & $280.00\pm20.62$ & 1435.5 & 6.75 \\
        & & Diffuse (total) & $330.00\pm26.33$ & 5465.2 & 12.06 \\
        & & Diffuse (AGN-masked) & $126.00\pm16.72$ & 4029.8 & 8.89 \\
        & \multirow{3}{*}{144} & AGN & $151.00\pm15.12$ & 298.9 & 13.32 \\
        & & Diffuse (total) & $75.00\pm7.68$ & 4909.5 & 10.83 \\
        & & Diffuse (AGN-masked) & $19.50\pm2.52$ & 4610.6 & 10.17 \\
        & \multirow{3}{*}{400} & AGN & $70.10\pm7.01$ & 675.0 & 16.22 \\
        & & Diffuse (total) & $8.80\pm0.95$ & 5823.0 & 12.85 \\
        & & Diffuse (AGN-masked) & $3.10\pm0.46$ & 5148.0 & 11.36 \\
        \hline
    \end{tabular}
   \tablefoot{The flux densities reported in this Table are plotted in Fig.~\ref{fig:spix}. Columns: (1) Name of the cluster; (2) central frequency; (3) source component: the central AGN, the total diffuse emission, or the diffuse emission with the AGN region masked; (4) flux density and associated uncertainty from Eq.~\ref{eq:uncertflux}; (5) area over which the flux density is integrated; (6) number of restoring beams within the area over which the flux density is integrated (as in Eq.~\ref{eq:uncertflux}).}
\end{table*}

\subsubsection{MeerKAT L band}
We recently observed Abell 496 with MeerKAT at L band (856 MHz - 1712 MHz) for a total night-time on source of 7~h (Proposal ID SCI-20241101; PI: F. Ubertosi). The observations were split into two blocks (one rising and one setting) performed in January and August 2025 (Capture block IDs: 1736188157 and 1759012213). The calibration was performed following \citet{botteon2024,balboni2025}. The target data calibrated through the SARAO Science Data Processor (SDP\footnote{See \url{https://skaafrica.atlassian.net/wiki/spaces/ESDKB/pages/338723406/SDP+pipelines+overview}.}) pipeline, which corrects for delay, bandpass, and gains, were downloaded from the MeerKAT archive and then compressed using \texttt{dysco} \citep{offringa2016}. We note that the field of Abell 496 includes three relatively bright (0.1 - 1 Jy flux density at GHz frequencies) radio sources: the BCG of Abell 496 itself, and two more radio sources at less than 10$^{\prime}$ from the target, PKS~0431-133 and PKS~0431-135. Residual sidelobes from these sources were present in the SDP calibrated data, limiting the dynamic range of the images. Therefore, to improve the data quality, we performed additional cycles of self-calibration using \texttt{facetselfcal} \citep{vanweeren2021}. Specifically, we performed direction-independent self-calibration in four cycles (two phase-only and two phase and amplitude), during which solutions are computed using \texttt{DP3} \citep{vanDiepen18} and imaging is performed with \texttt{WSclean} \citep{offringa-wsclean-2014}. No direction-dependent calibration was performed, since the rms noise close to the target at the final direction-independent step was already at the level of the expected one. The images obtained at the final round of self calibration using the full uv-range and no tapering with angular resolutions of $\sim$5$\arcsec$ reach rms noise levels far from other sources of $\sim$5$\,\mu$Jy~beam$^{-1}$, in good agreement with the reference value from the MeerKAT exposure calculator for 7h on source ($\sim$5$\,\mu$Jy~beam$^{-1}$). When measuring flux densities from the images, we adopted a systematic uncertainty of 5\% on the flux density scale\footnote{Our MeerKAT observations used J0408–6545 as the flux density scale calibrator, whose model is based on \href{https://skaafrica.atlassian.net/wiki/spaces/ESDKB/pages/1481408634/Flux+and+bandpass+calibration}{SARAO prescriptions}. J0408–6545 and its model were also used in \citet{knowles2022}, where an absolute flux density scale uncertainty of $\sim$5\% is reported, which we adopt here.} \citep{knowles2022}.

\subsection{X-ray observations of Abell 496 and ZwCl~235}\label{sub:X-raydata}
Both clusters were observed with the \textit{Chandra} X-ray telescope and the data were reduced following standard procedures. Abell~496 was observed with ACIS-S in 2000 (ObsID 931, 18.9 ks), 2001 (ObsID 3361, 10 ks), and 2004 (ObsID 4976, 75 ks). ZwCl~235 was observed with ACIS-S in 2009 (ObsID 11735), with a total uncleaned exposure of 19.8 ks. We reprocessed the event files using CIAO~4.14 and CALDB~4.9.7. Periods affected by background flares were removed, obtaining a cleaned exposure time after filtering of 65.3 ks for Abell~496 and 19.5~ks for ZwCl~235. The background was modeled using the appropriate blank-sky event files from CALDB, reprojected onto the observation using the aspect solution and normalized using the ratio of the observed to blank-sky count rates in the 9--12 keV band. Point sources were identified using \texttt{wavdetect} and masked before the morphological analysis. The event files from the three datasets of Abell~496 were merged into a single mosaic. For both systems, we produced background-subtracted and exposure-corrected images in the 0.5--7 keV band, which were used for the X-ray surface brightness analysis and for the point-to-point comparison with the radio emission (Sect.~\ref{subsec:xray}).

\section{Testing different choices of minimum uv-cut to subtract the AGN-related emission}\label{app:testuv}

\begin{figure*}[htp!]
    \centering
    \includegraphics[width=0.24\linewidth]{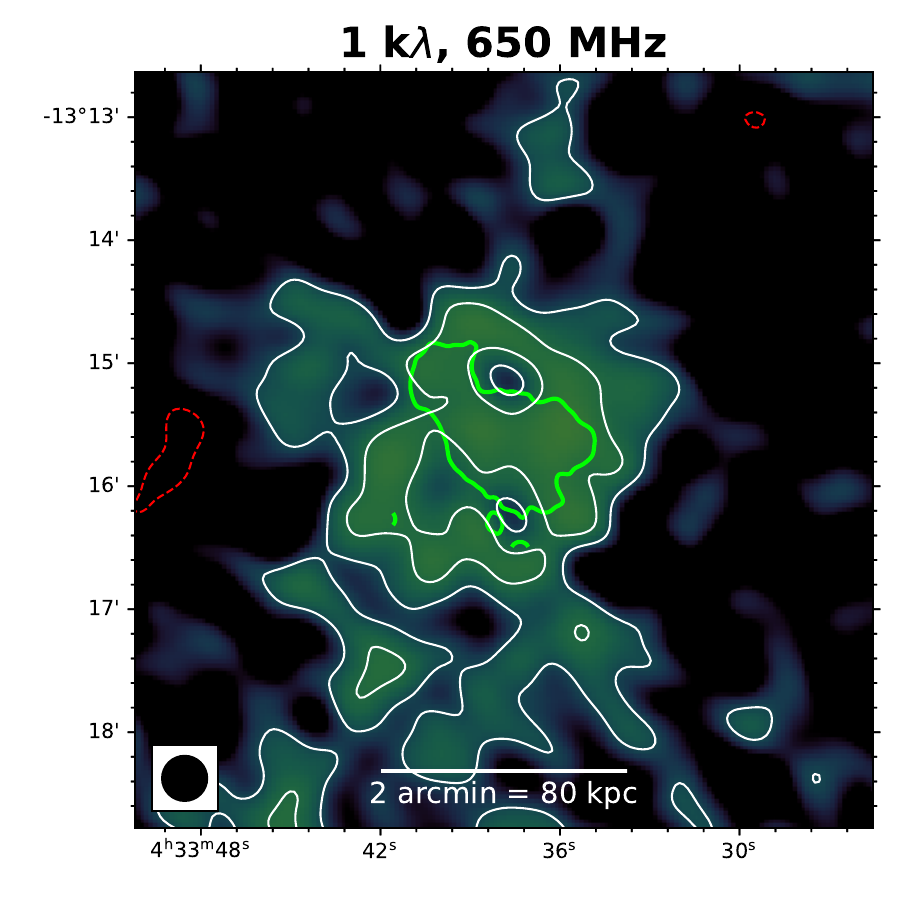}
    \includegraphics[width=0.24\linewidth]{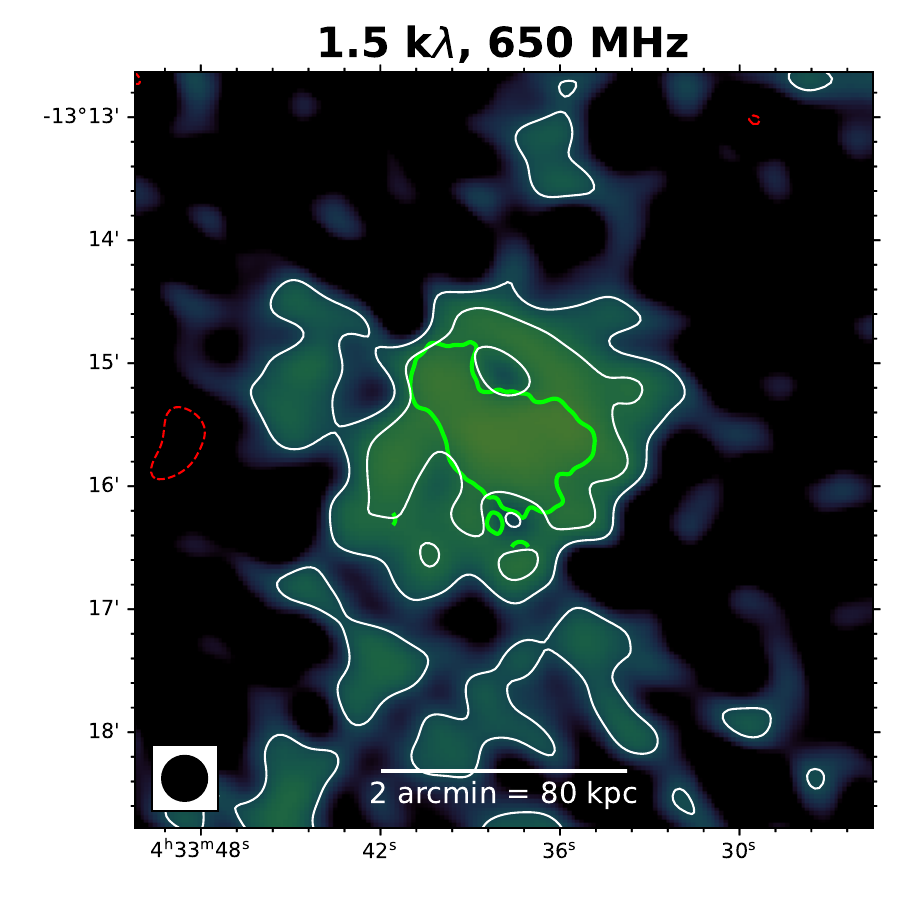}
    \includegraphics[width=0.24\linewidth]{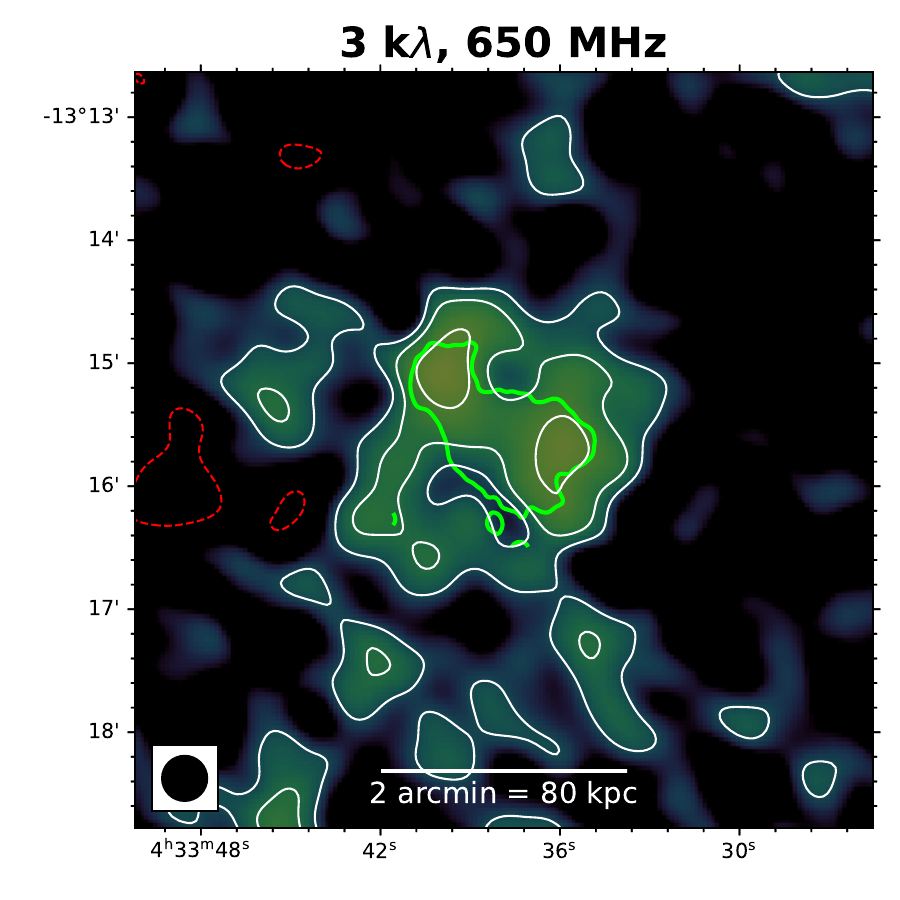}
    \includegraphics[width=0.24\linewidth]{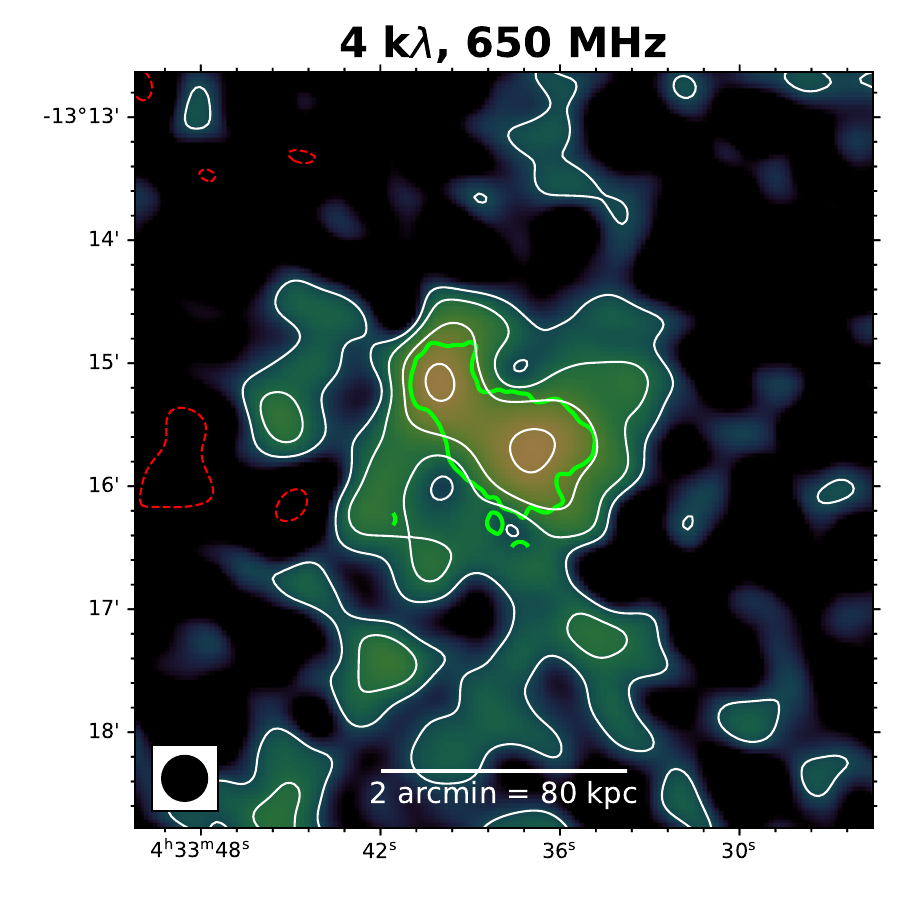}\\
    \includegraphics[width=0.24\linewidth]{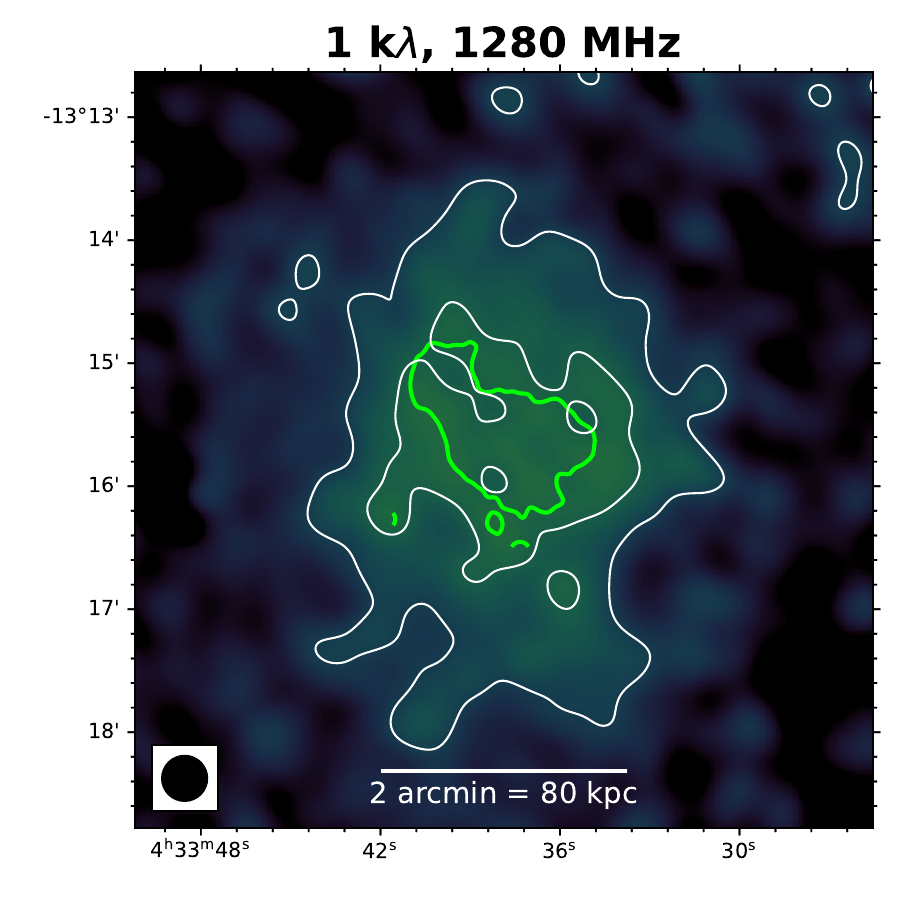} 
    \includegraphics[width=0.24\linewidth]{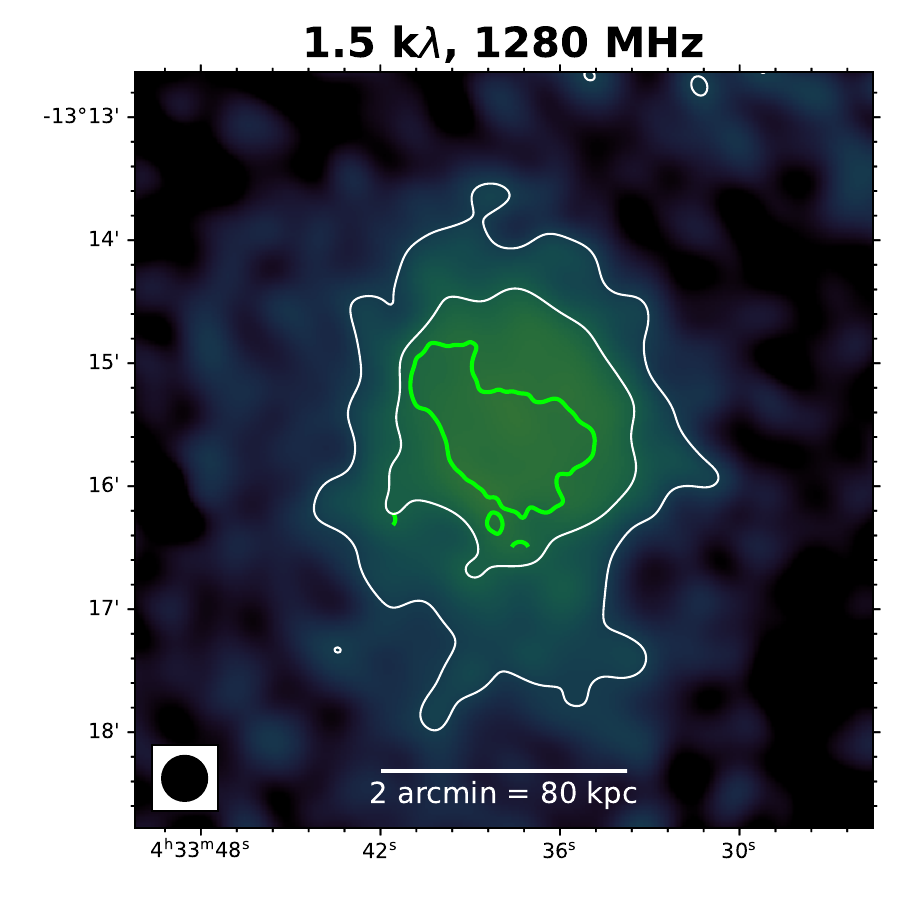} 
    \includegraphics[width=0.24\linewidth]{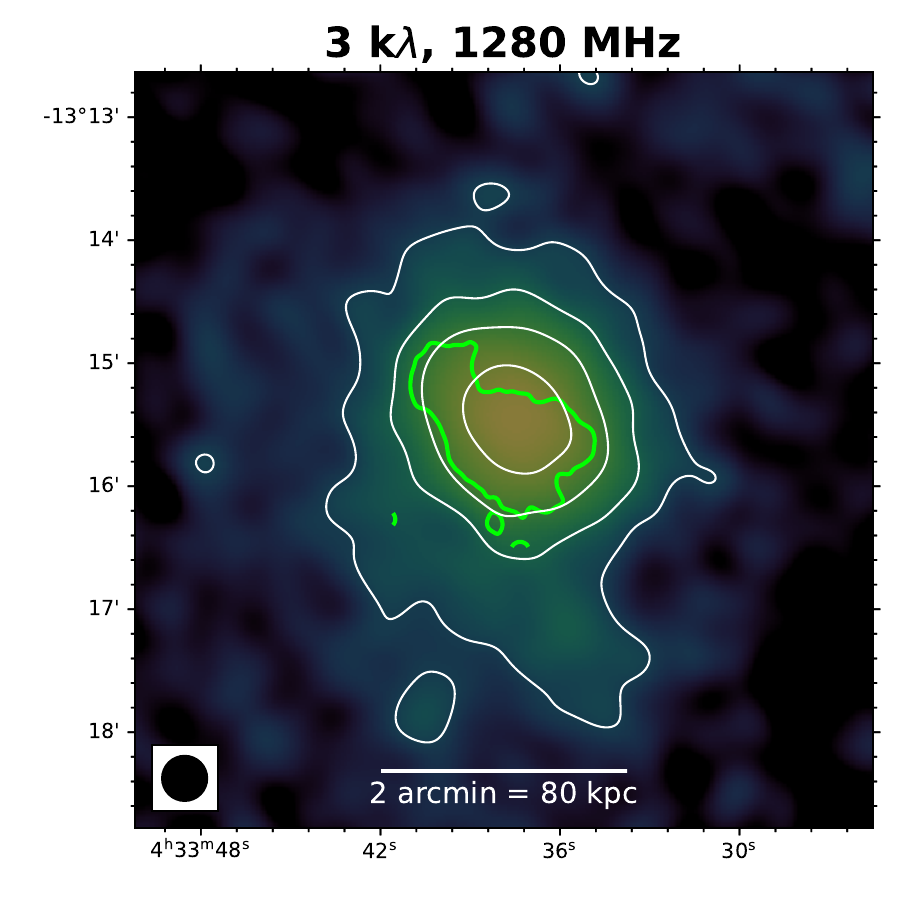} 
    \includegraphics[width=0.24\linewidth]{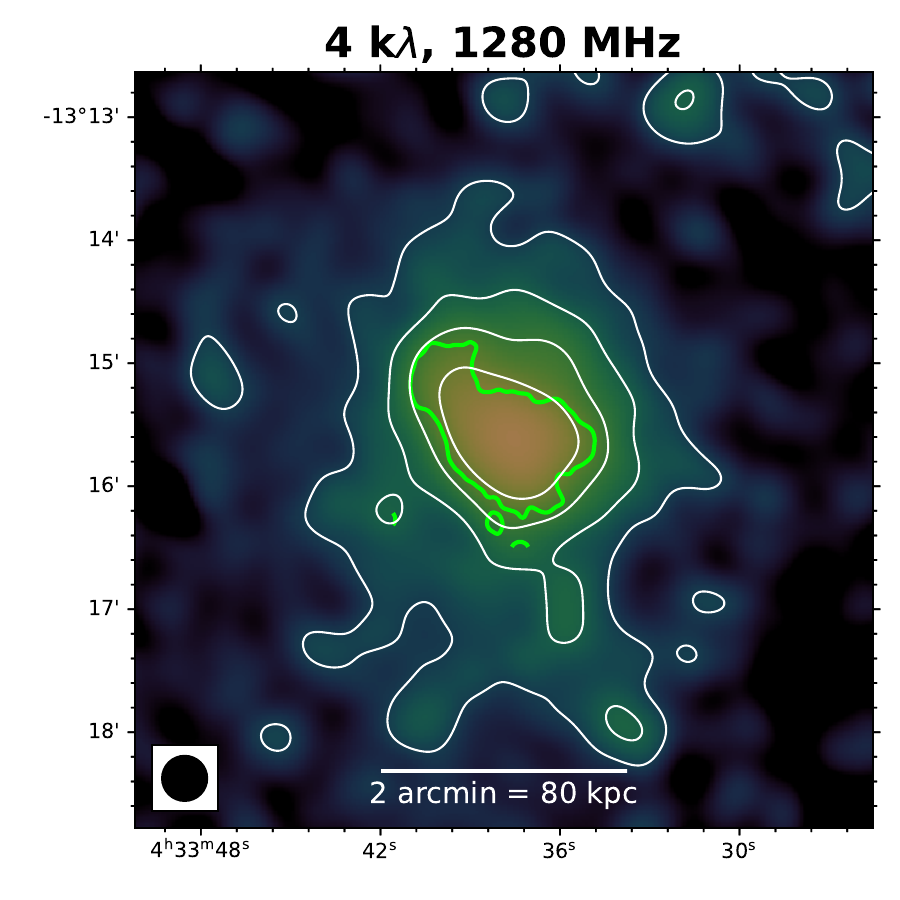}\\
    \includegraphics[width = 0.55\linewidth]{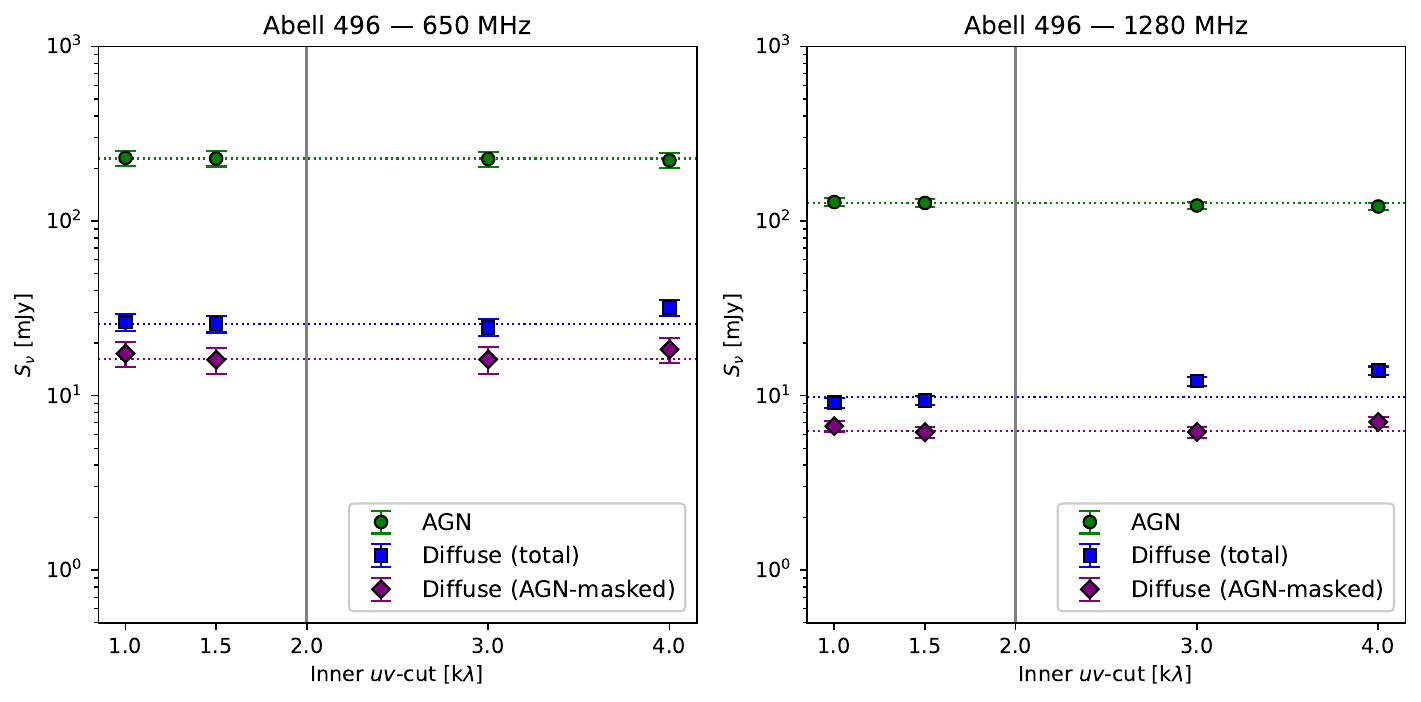}
    \\
    \includegraphics[width=0.55\linewidth]{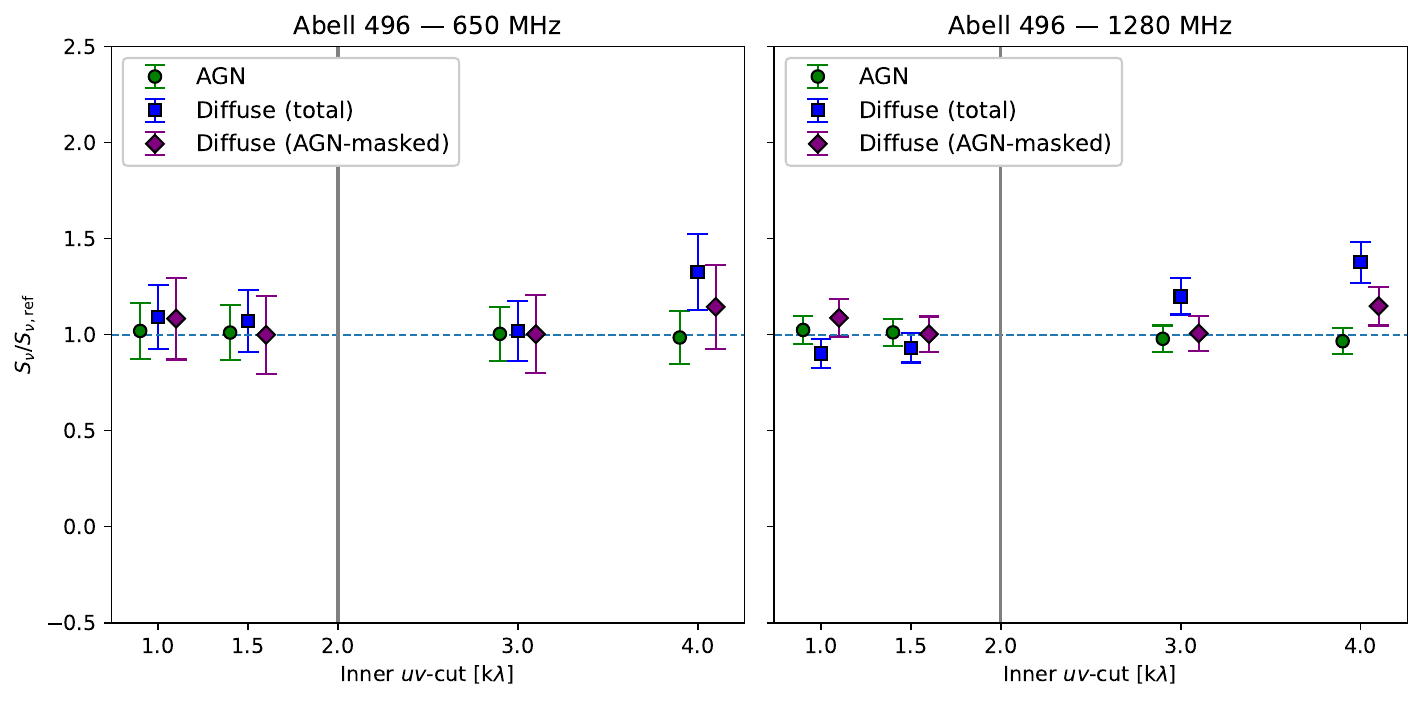}
    \includegraphics[width=0.275\linewidth]{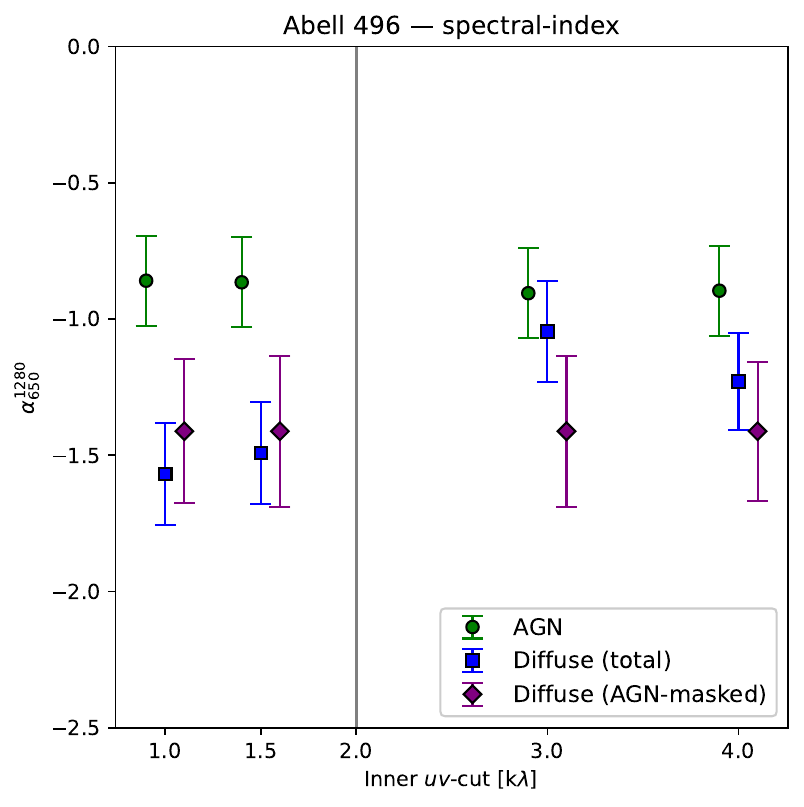}
     \caption{Results of testing different choices of minimum uv-cut for Abell~496. In the upper two rows we show images of the discrete-source subtracted diffuse emission in Abell 496 at 650 MHz (top) and 1280 MHz (middle), with panels from left to right corresponding to increasing minimum uv-cut (1~k$\lambda$, 1.5~k$\lambda$, 3~k$\lambda$, 4~k$\lambda$). White contours are drawn starting at $3\times\sigma_{\rm rms}$ (see Tab.~\ref{tab:radioimages}) and increase by a factor of two. In green we also show the outline of the central radio galaxy, from the discrete-source image at 1280 MHz shown in Fig.~\ref{fig:total_A496}. In the lower two rows we show the flux densities (top), the normalized flux densities (bottom left) and the spectral index (bottom right) of different components as a function of inner uv-cut. Flux densities are normalized by the values used in the main text of this work, corresponding to 2~k$\lambda$ (see Sect.~\ref{subsec:diffuseemission} and Tab.~\ref{tab:fluxtable}). In the $S_{\nu}$ versus uv-cut plots, the dotted horizontal lines represent the mean of the flux density measurements at 1.5, 2, 3 k$\lambda$. A small offset in the x-coordinate with respect to the reference value has been applied to the different components to help visualization.}
    \label{fig:uvcut-test-A496}
\end{figure*}

\begin{figure*}[htp!]
    \centering
    \includegraphics[width=0.24\linewidth]{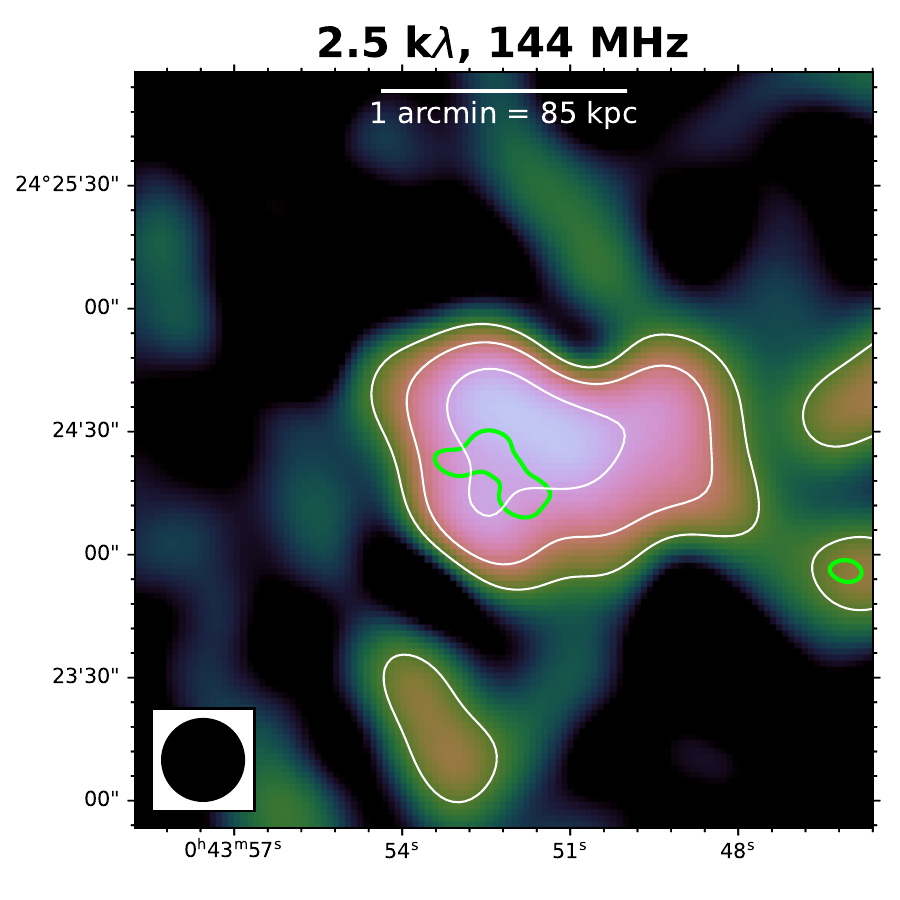}
    \includegraphics[width=0.24\linewidth]{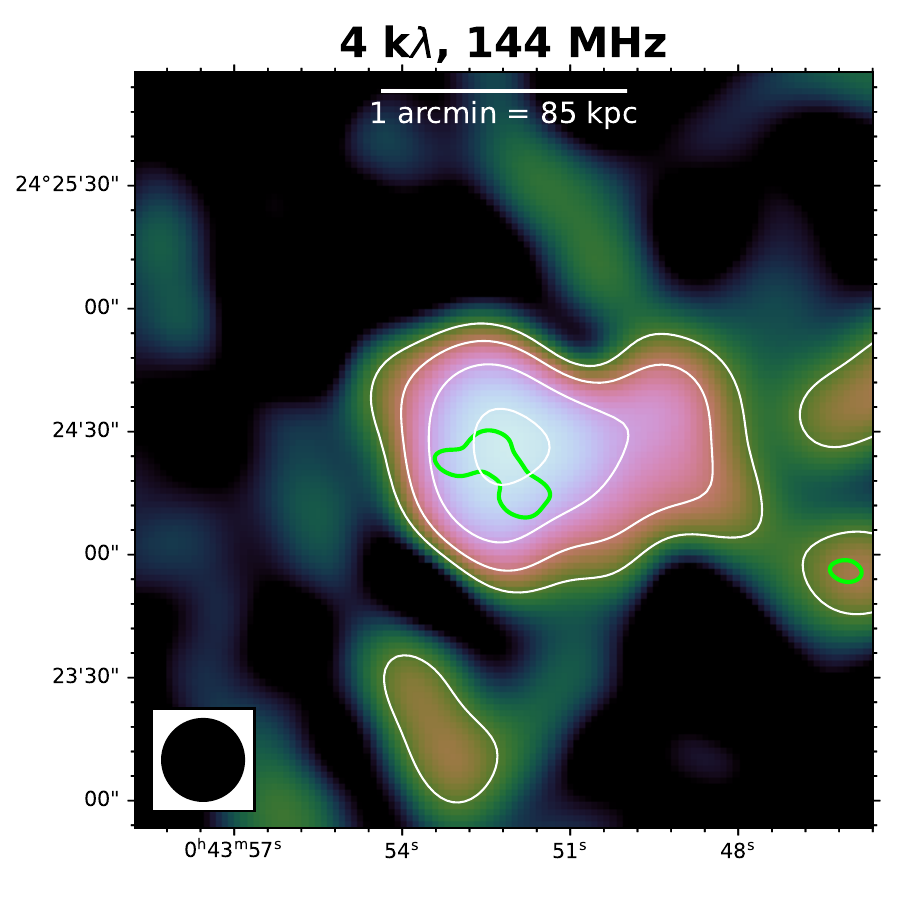}
    \includegraphics[width=0.24\linewidth]{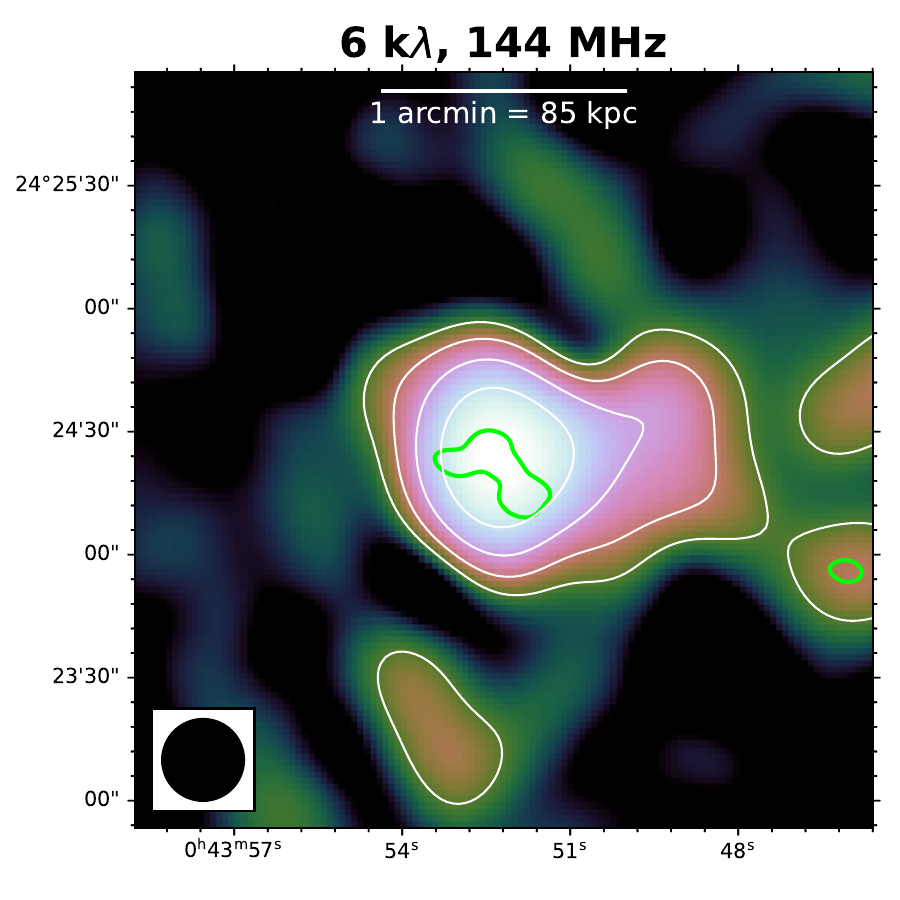}
    \includegraphics[width=0.24\linewidth]{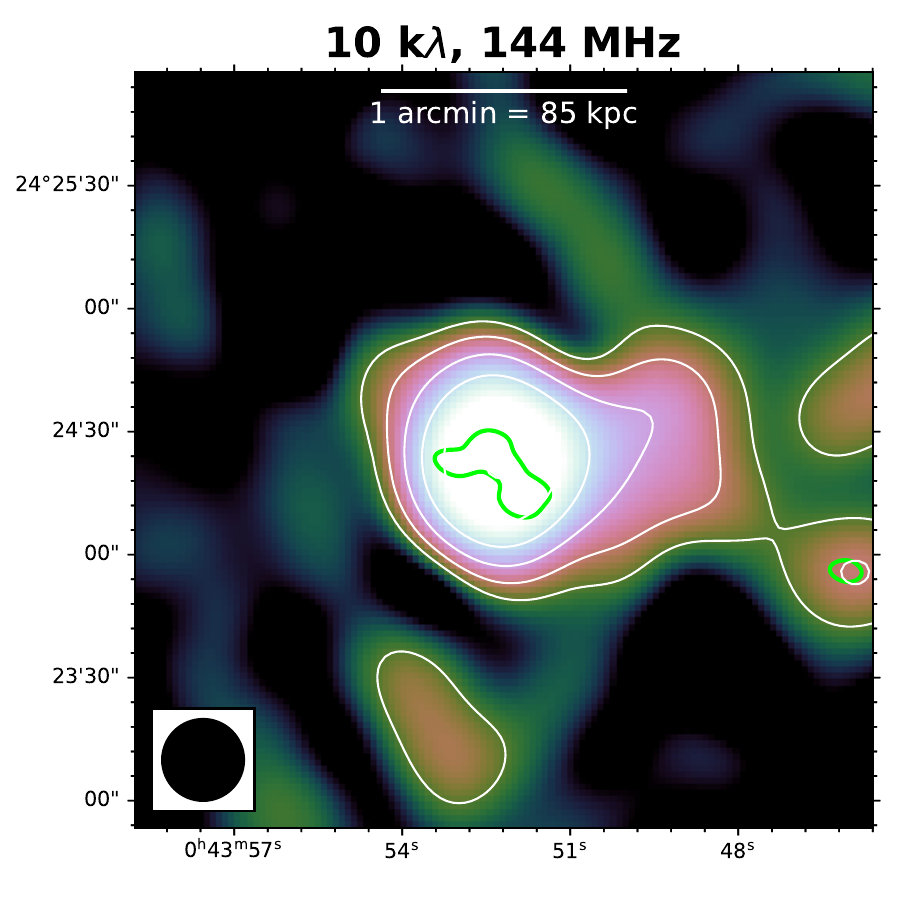}\\
    \includegraphics[width=0.24\linewidth]{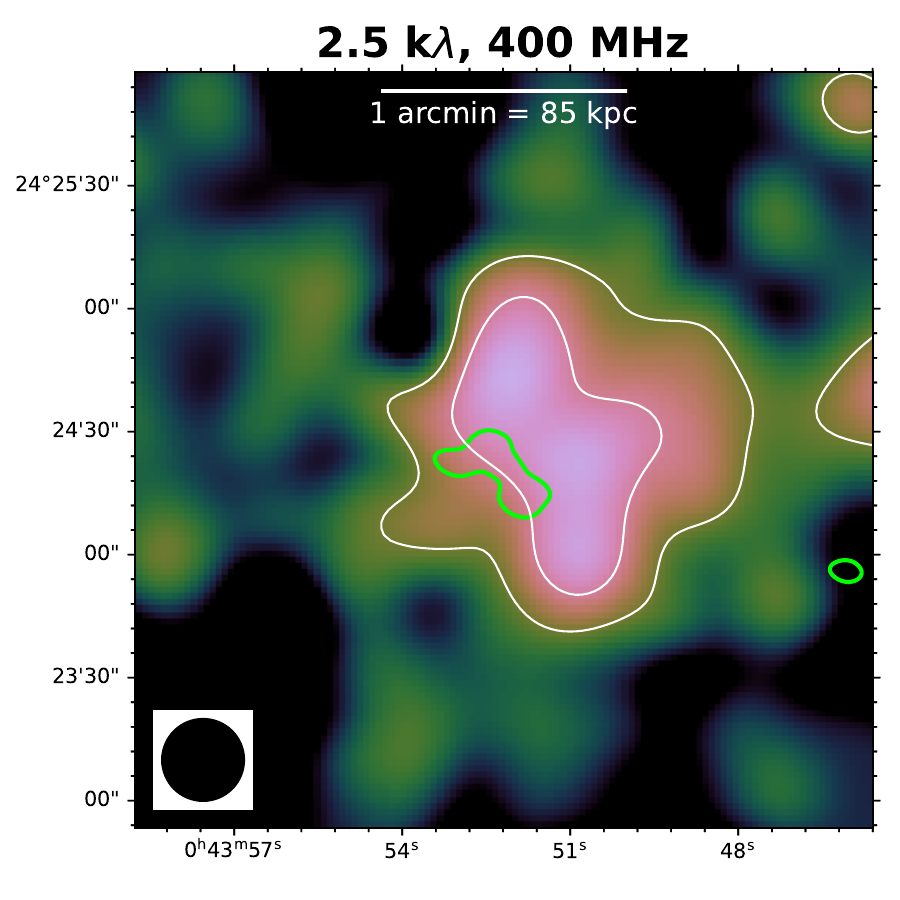} 
    \includegraphics[width=0.24\linewidth]{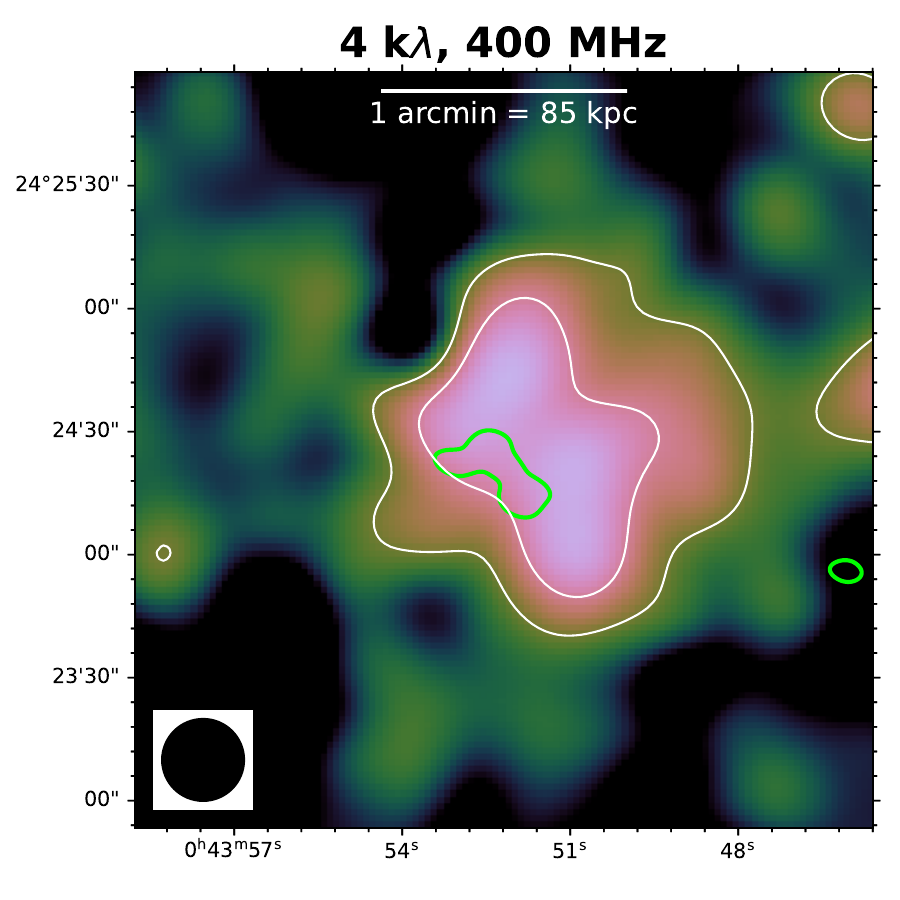} 
    \includegraphics[width=0.24\linewidth]{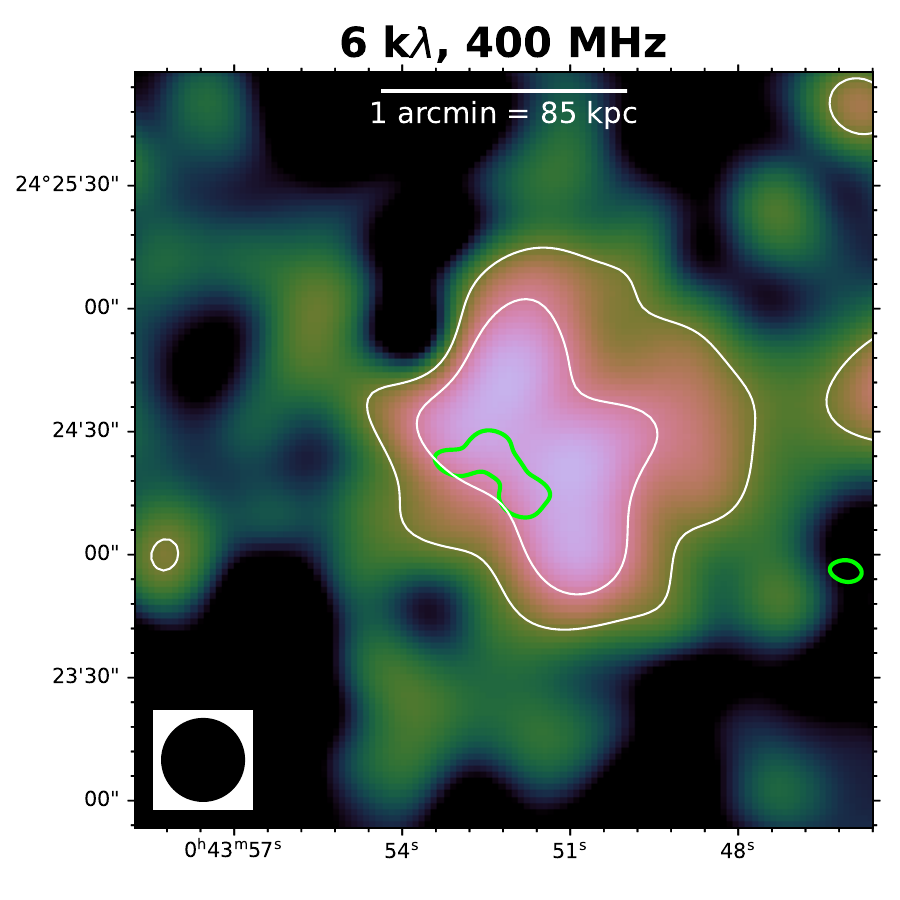} 
    \includegraphics[width=0.24\linewidth]{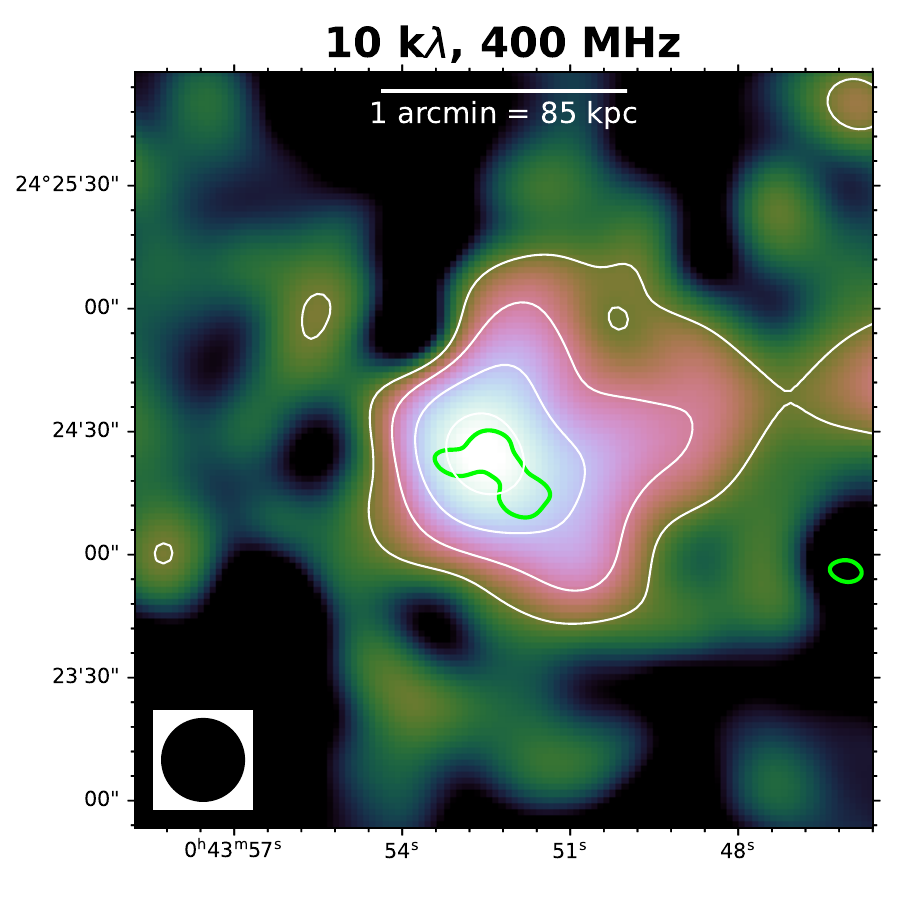}\\
    \includegraphics[width = 0.55\linewidth]{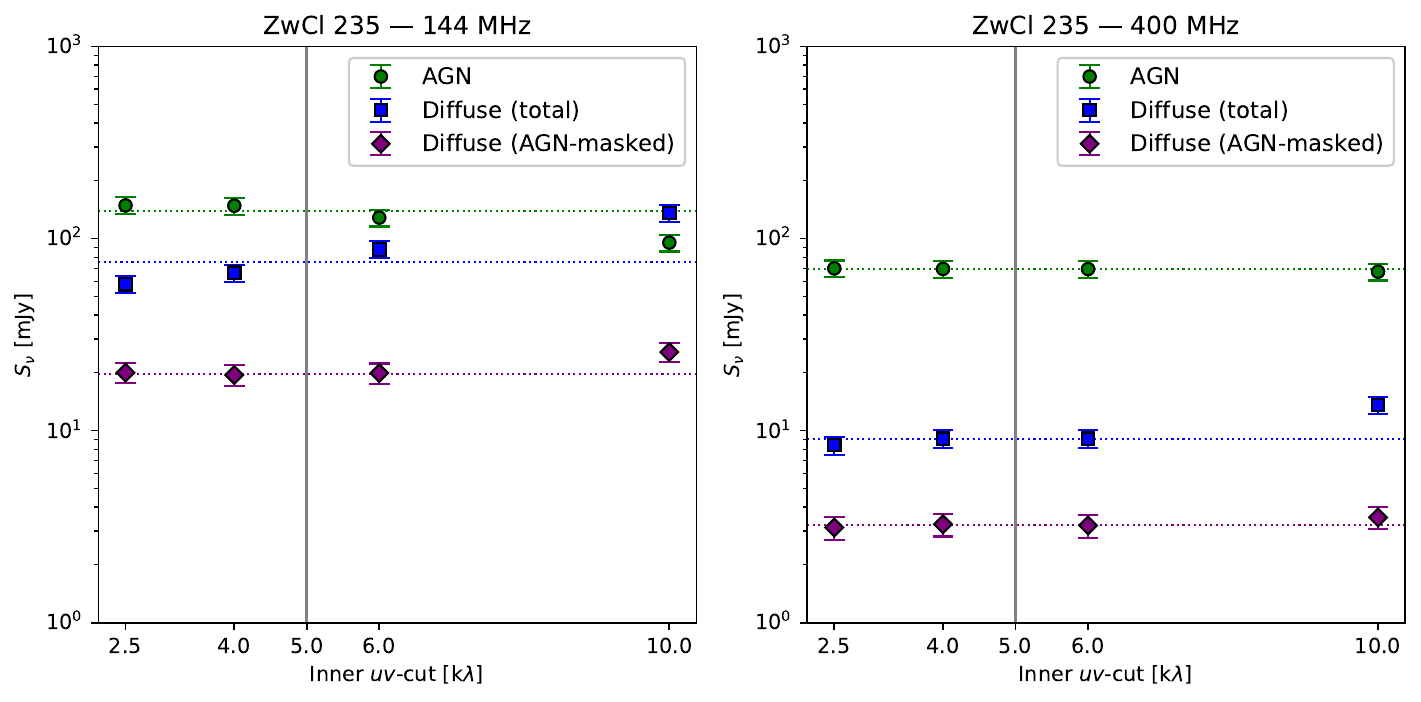}
    \\
    \includegraphics[width=0.55\linewidth]{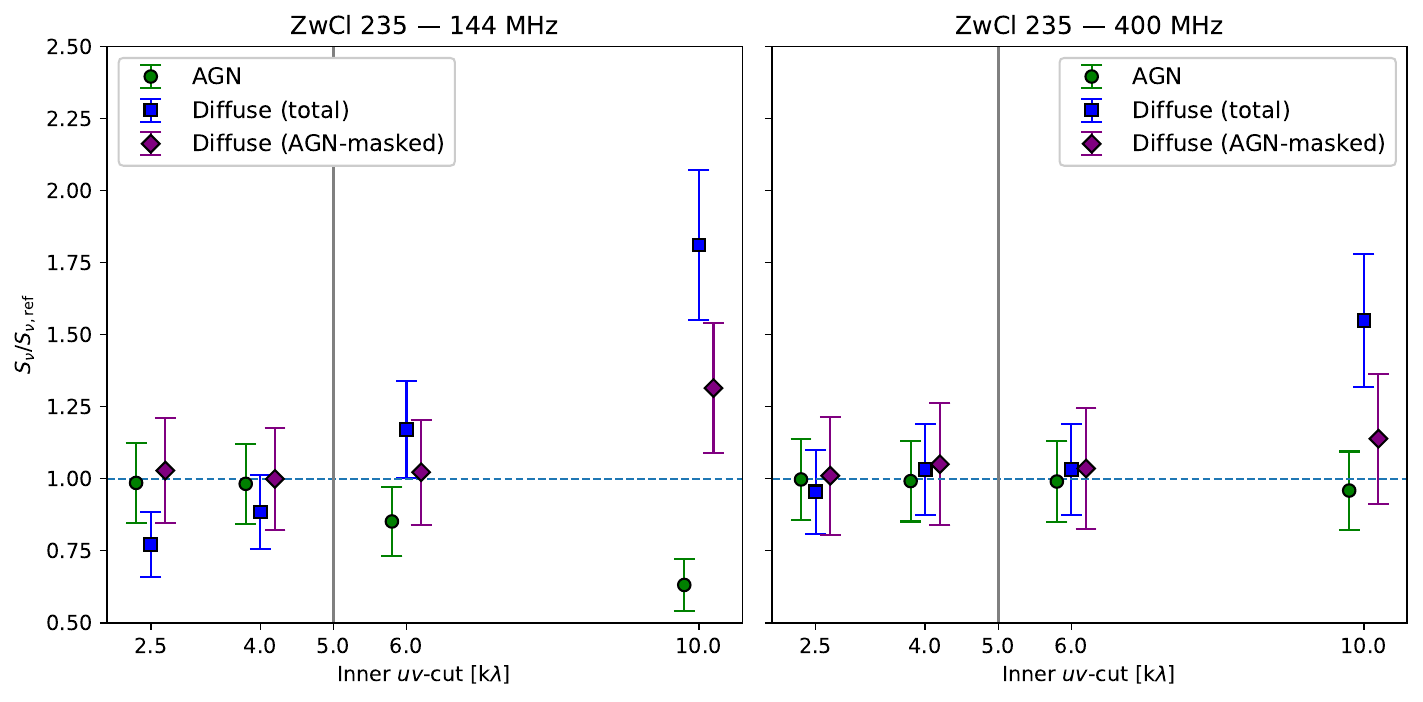}
    \includegraphics[width=0.275\linewidth]{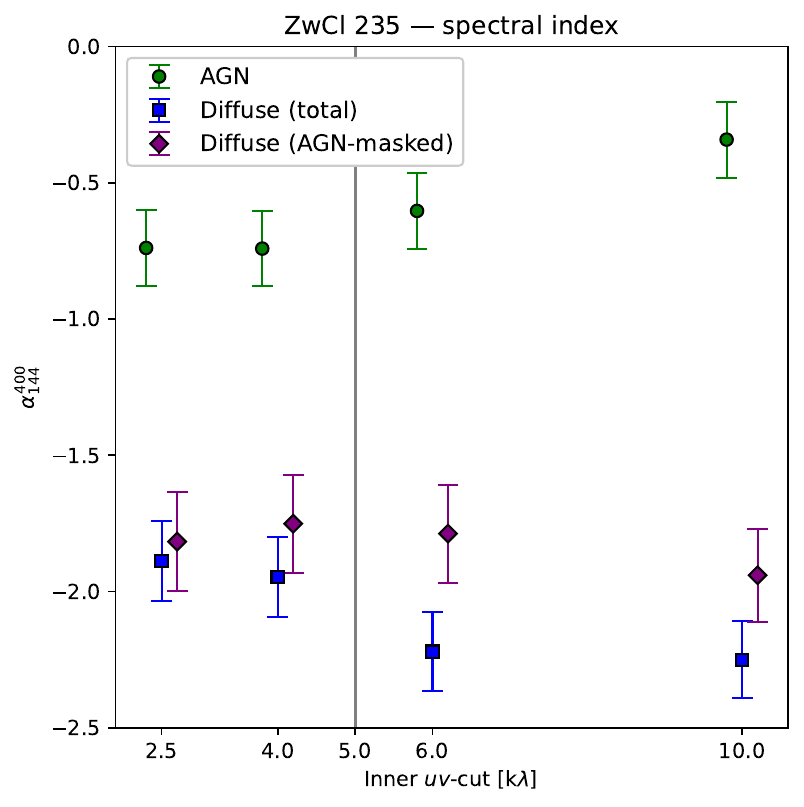}
    \caption{Results of testing different choices of minimum uv-cut for ZwCl~235. In the upper two rows we show images of the discrete-source subtracted diffuse emission in ZwCl~235 at 144 MHz (top) and 400 MHz (middle), with panels from left to right corresponding to increasing minimum uv-cut (2.5~k$\lambda$, 4~k$\lambda$, 6~k$\lambda$, 10~k$\lambda$). White contours are drawn starting at $3\times\sigma_{\rm rms}$ (see Tab.~\ref{tab:radioimages}) and increase by a factor of two. In green we also show the outline of the central radio galaxy, from the discrete-source image at 144 MHz shown in Fig.~\ref{fig:total_z235}. In the lower two rows we show the flux densities (top), the normalized flux densities (bottom left) and the spectral index (bottom right) of different components as a function of inner uv-cut. Flux densities are normalized by the values used in the main text of this work, corresponding to 5~k$\lambda$ (see Sect.~\ref{subsec:diffuseemission} and Tab.~\ref{tab:fluxtable}). In the $S_{\nu}$ versus uv-cut plots, the dotted horizontal lines represent the mean of the flux density measurements at 4, 5, 6 k$\lambda$. A small offset in the x-coordinate with respect to the reference value has been applied to the different components to help visualization.}
    \label{fig:uvcut-test-Z235}
\end{figure*}

The identification and characterization of the diffuse radio emission rely on the subtraction of compact and AGN-related emission before imaging the residual diffuse component (Sect.~\ref{subsubsec:subtraction}). Since the spatial scales of the central diffuse emission in both Abell~496 and ZwCl~235 are only a factor of $\sim3-5$ times larger than those of the AGN-connected structures (see Sect.~\ref{subsec:diffuseemission}), we performed a set of tests to assess how the flux densities and spectral indices of the recovered AGN emission and diffuse emission vary depending on the choice of the inner uv-cut (see \citealt{Botteon2018} for a previous similar test on the discrete source subtraction).

For Abell~496, we repeated the source-subtraction procedure using inner $uv$-cuts of 1, 1.5, 3, and 4~k$\lambda$, compared with the $>2$~k$\lambda$ cut adopted for the analysis in Sect.~\ref{subsec:diffuseemission}. We performed the same test for ZwCl~235 at 144 and 400~MHz, using inner $uv$-cuts of 2.5, 4, 6, and 10~k$\lambda$; the adopted value for the discrete-source model is $>5$~k$\lambda$. The resulting residual datasets were then imaged using the same tapering and weighting scheme adopted for the fiducial diffuse-emission images, convolving the final images to a common angular resolution of $22\arcsec$ (for Abell~496) and $20\arcsec$ (for ZwCl~235). The range of values adopted for this test allows us to assess the effect of both more extreme choices relative to the reference one (1 and 4 k$\lambda$ for Abell~496; 2.5 and 10 k$\lambda$ for ZwCl~235), and similar values to the one chosen for the paper (1.5 and 3 k$\lambda$ for Abell~496; 4 and 6 k$\lambda$ for ZwCl~235). We note that, for ZwCl~235, we performed the tests at 144 and 400~MHz, without including the LBA 54~MHz data. This is because the LBA data is more limited on the small angular scales relevant to the source-subtraction procedure, and we have verified that the relatively lower sensitivity at 54 MHz means that the changes produced by modest variations in the inner $uv$-cut are generally consistent within the corresponding flux density uncertainties. We therefore restrict testing to the LOFAR HBA  and uGMRT Band 3 observations.

Fig.~\ref{fig:uvcut-test-A496} and Fig.~\ref{fig:uvcut-test-Z235} show representative maps obtained from the different source-subtraction tests at 650 and 1280~MHz for Abell~496 and at 144 and 400~MHz for ZwCl~235, respectively. The corresponding integrated flux densities were measured over the same regions used for the fiducial analysis (Sect.~\ref{subsec:diffuseemission}, Fig.~\ref{fig:spix} and Tab.~\ref{tab:fluxtable}). The absolute flux densities, the flux densities normalized by the reference values at 2k$\lambda$ (Abell~496) and 5k$\lambda$ (ZwCl~235), and the spectral indices of different components are shown in the bottom panels of Fig.~\ref{fig:uvcut-test-A496} for Abell~496 and Fig.~\ref{fig:uvcut-test-Z235} for ZwCl~235. Such different components are, specifically, the central AGN, the total diffuse emission, and the diffuse emission after masking the central AGN region (as in Fig.~\ref{fig:spix} and Tab.~\ref{tab:fluxtable}). 

For Abell~496, the images shown in the upper two rows of Fig.~\ref{fig:uvcut-test-A496} show that residual AGN-related emission starts being visible at both 650~MHz and 1280~MHz for inner uv-min $\geq3{\rm k}\lambda$. The $S_{\nu}$ versus inner uv-cut, and the $S_{\nu}/S_{\nu,ref}$ versus inner uv-cut plots show that, at both frequencies, (a) the AGN flux densities remain relatively stable as a function of inner uv-cut, showing an overall mild decrease for increasing inner uv-cut; (b) the total flux density of the diffuse emission increases with increasing inner uv-cut, especially for inner uv-min $\geq3{\rm k}\lambda$; (c) the flux density of the diffuse emission after masking the region where the AGN is located is nearly constant as a function of inner uv-min, showing only a mild increase for the higher inner uv-min ($4{\rm k}\lambda$). Additionally, the spectral index versus inner uv-cut plots shows that (d) the AGN and \say{Diffuse (AGN-masked)} spectral indices remain constant as a function of inner uv-cut, while (e) the spectral index of the \say{Diffuse (total)} emission flattens for inner uv-cut $\geq3{\rm k}\lambda$, as expected if emission from the central AGN starts being excluded from the discrete sources model. These pieces of evidence are compatible with the images shown in the upper two rows, and suggest that, for inner uv-cut $<3{\rm k}\lambda$, contamination from any central AGN-related structure is strongly limited. 

For ZwCl~235, the images shown in the upper two rows of Fig.~\ref{fig:uvcut-test-Z235} provide a qualitatively similar picture. However, for ZwCl~235 the AGN emission (at 144 MHz) shows a more pronounced decrease in flux density with increasing inner uv-cut, which is reflected in its flatter spectral index for inner uv-cut $\geq6{\rm k}\lambda$ in the bottom right panel. This is probably due the 144~MHz image being dominated by the flux density of the lobes of the AGN, while the 400 MHz AGN flux density is composed of a lobes component and an unresolved \say{core} component, which is less sensitive to more drastic (larger) inner uv-cuts. We also see that the 2.5k$\lambda$ cut probably starts excising genuine diffuse emission at 144 MHz, as demonstrated by the central deficit in flux density in the top left panel of Fig.~\ref{fig:uvcut-test-Z235}. For the \say{Diffuse (total)} and the \say{Diffuse (AGN masked)} we observe the same trends highlighted for Abell~496 in terms of $S_{\nu}$, of $S_{\nu}/S_{\nu,ref}$, and of spectral index, with the difference being (a) the inner uv-cut at which differences start being visible (in this case, $\geq6{\rm k}\lambda$), and (b) the fact that the spectral index of the \say{Diffuse (total)} \textit{steepens} with increasing inner uv-cut - this can be explained as an effect of the relatively stronger decrease in AGN-related flux density at 144 MHz than at 400~MHz with inner uv-cut noted above, which leaves more contamination from AGN-related emission in the images at 144 MHz for an inner uv-cut of $10{\rm k}\lambda$ in the top and middle panels of Fig.~\ref{fig:uvcut-test-Z235}.

Overall, these tests demonstrate that the total flux densities, morphologies, and spectral properties reported in the main analysis are not driven by the precise choice of the adopted inner $uv$-cut, provided that we select a value below a specific scale. This scale seems to match the visually identified scale in Sect.~\ref{subsubsec:subtraction}, which was based on the properties and morphology of the AGN related emission in radio images from the literature and from this work. We therefore consider the adopted $uv$-cuts to provide a reasonable separation between the AGN-related and diffuse components for the purposes of our analysis.

Ultimately, we provide an estimate of the scatter (i.e., the systematic error) that arises from adopting slightly different choices of uv-cut. This was estimated by excluding the measurements obtained for the lowest and highest uv-cut (1 k$\lambda$, 4 k$\lambda$ for Abell~496 and 2.5 k$\lambda$, 10 k$\lambda$ for ZwCl~235), which were only included in the plots to illustrate the effect of extreme (and unsupported) choices of uv-cut. We find that the flux densities of the \say{Diffuse (total)} emission (which includes the region where the AGN is present) show a systematic scatter varying between 3\% and 10\% depending on frequency, while the flux densities of the \say{Diffuse (AGN-masked)} emission show a systematic scatter that is $<2\%$ at all frequencies. Since throughout the analysis of Sect.~\ref{sec:results} we conservatively consider the \say{Diffuse (AGN-masked)} measurements only, for which the statistical error ($\sim10\%$) is dominant over the systematic one ($<2\%$), the uncertainties on the flux densities reported in this work do not take into account this negligible systematic contribution. We note, however, that in the analysis of mini-halo scaling relations (Sect.~\ref{subsec:discscalingrelations}), we used the total mini-halo flux densities to calculate the diffuse radio power at 1.4~GHz of Abell~496 and ZwCl~235. This choice is motivated by two considerations. First, as discussed in Sect.~\ref{subsec:diffuseemission}, the \say{Diffuse (AGN-masked)} flux densities underestimate the total flux densities. Second, using the total flux densities facilitates comparison with the literature, where the region surrounding the central AGN is generally not masked, even in maps from which discrete sources have been subtracted. Furthermore, the additional systematic uncertainty associated with the choice of uv-cut used to subtract discrete sources is generally not accounted for in the literature. A systematic re-analysis of a uniformly studied sample is a future perspective of the MIMIC project.

\section{Tailed radio galaxies in Abell~496 and ZwCl 235}\label{app:tailed}

This Appendix presents radio images and main properties of peculiar tailed radio galaxies identified in Abell~496 and ZwCl~235 from the sensitive radio data presented in the main body of this paper. While a detailed discussion of these radio sources is out of the scope of this work, we summarize for completeness their main properties below.

\begin{figure*}[ht!]
    \centering
    \includegraphics[width=0.75\linewidth]{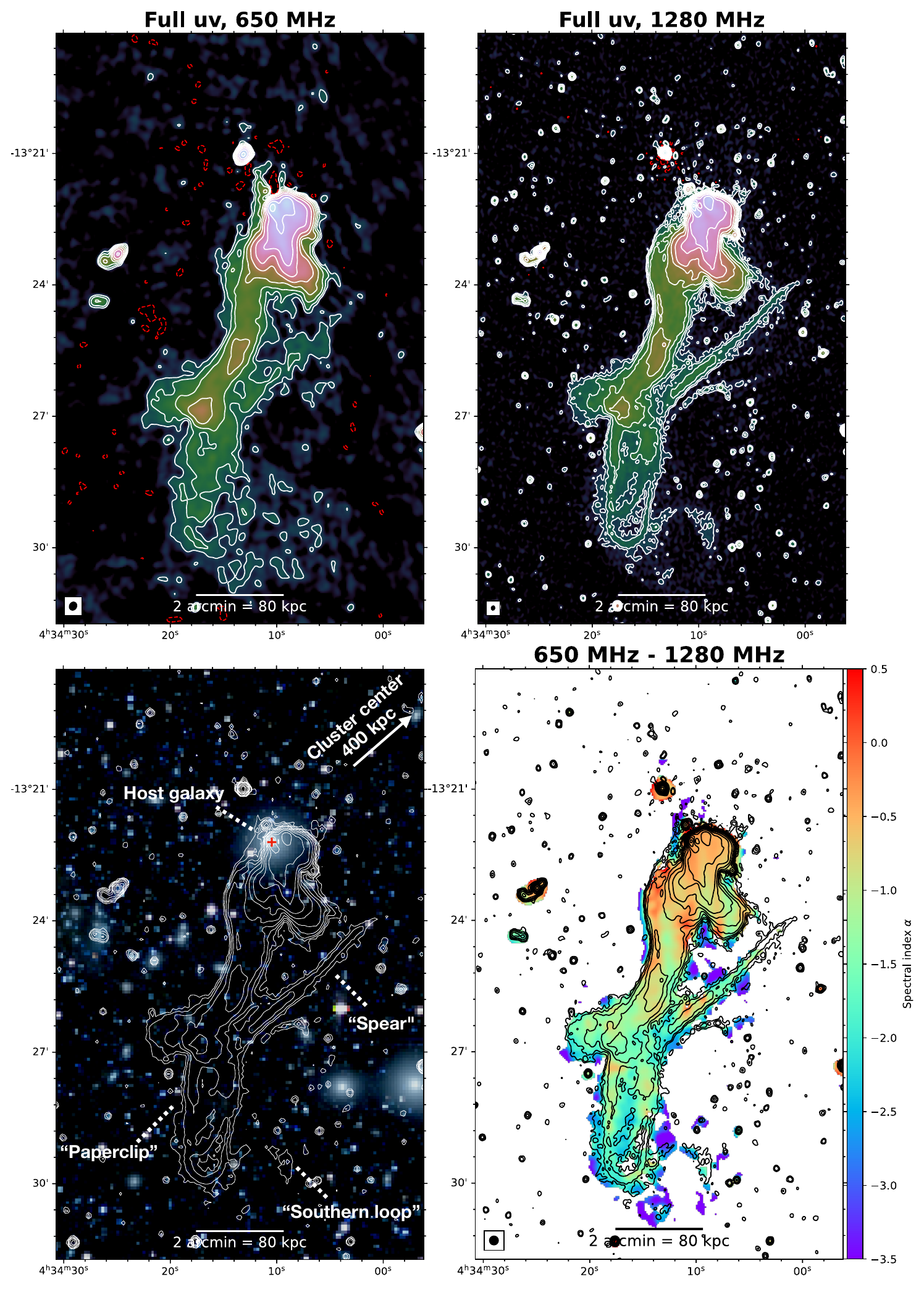}
    \caption{Multifrequency radio view of the 'Giant Paperclip' radio galaxy in Abell 496. {\it Top Left}: uGMRT Band 4 (550 MHz - 750 MHz) image of Abell 496, centered on PKS 0431-134. {\it Top Right}: MeerKAT L band (856 MHz - 1712 MHz) image of Abell 496, centered on PKS 0431-134. The angular resolution of the two images is reported in Tab. \ref{tab:radioimages}. In both panels, contours start at 3$\sigma_{rms}$ and increase by a factor of two (the $-3\sigma_{rms}$ contour is shown in dashed red). For the uGMRT Band 4 image, the rms noise is 0.08 mJy~beam$^{-1}$, while for the MeerKAT L band image, the rms noise is 0.005 mJy~beam$^{-1}$. {\it Bottom left:} Optical RGB image from the DESI survey, with radio contours overlaid from the MeerKAT L band image of the top left panel. Major features are labeled. The core of the host galaxy is marked with a red plus sign. {\it Bottom right:} Spectral index map between 650 MHz and 1280 MHz at an angular resolution of 12". Overlaid are the contours of the MeerKAT L band image of the top left panel.}
    \label{fig:tailedA496}
\end{figure*}
\begin{figure*}[ht!]
    \centering
    \includegraphics[width=\linewidth]{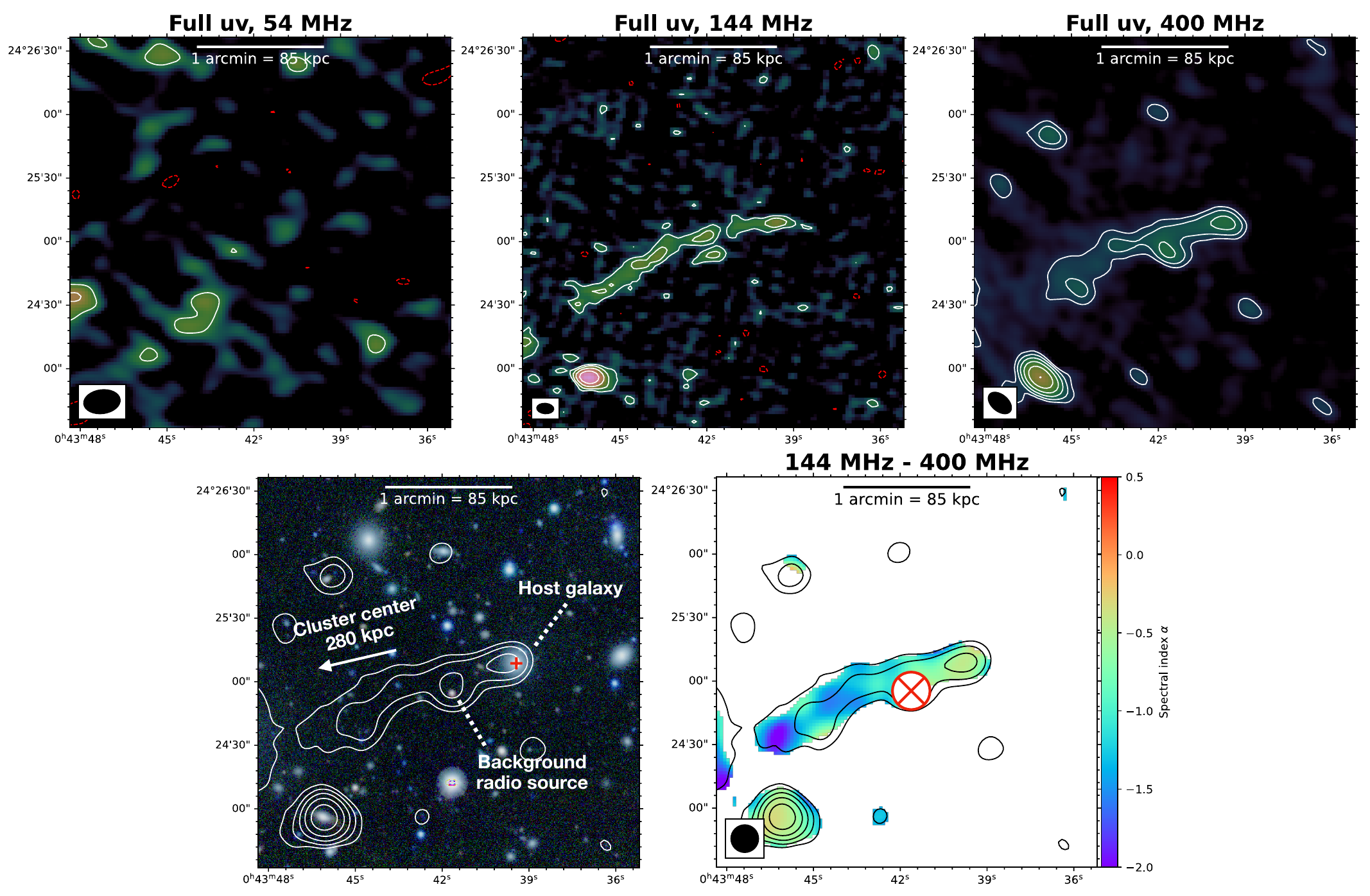}
    \caption{Multifrequency radio view of the tailed radio galaxy in ZwCl 235. {\it Left}: LOFAR LBA (42 MHz - 66 MHz) image of ZwCl 235, centered on LEDA 1706877. {\it Center}: LOFAR HBA (120 MHz - 166 MHz) image of ZwCl 235, centered on LEDA 1706877. {\it Right}: uGMRT Band 3 (300 MHz - 500 MHz) image of ZwCl 235, centered on LEDA 1706877. The angular resolution of the images is reported in Tab. \ref{tab:radioimages}. In the three panels, contours start at 3$\sigma_{rms}$ and increase by a factor of two (the $-3\sigma_{rms}$ contour is shown in dashed red). Note that the tailed radio galaxy is non-detected in the LOFAR LBA image. {\it Bottom left:} Optical RGB image from the DESI survey, with radio contours overlaid from the uGMRT 400 MHz image at 13" resolution used to compute the spectral index map (contours are drawn starting at 3$\sigma_{rms}$, with $\sigma_{rms} = 40\mu$Jy~beam$^{-1}$. Major features are labeled. The core of the host galaxy is marked with a red plus sign. {\it Bottom right:} Spectral index map between 144 MHz and 400 MHz at an angular resolution of 13". Overlaid are the contours from the uGMRT 400 MHz image shown in the bottom left panel. The region corresponding to the background radio source within the tail is masked.}
    \label{fig:tailedz235}
\end{figure*}
\subsection{The 'Giant Paperclip' in Abell~496}\label{subsubsec:paperclip}
From the wide-field MeerKAT image shown in red in Fig.~\ref{fig:largeA496Z235} (left panel), we identify a very large radio source on the south-east side of the cluster. This is a known tailed radio galaxy, PKS~0431$-$134 \citep{feretti1992}, associated with a cluster member of Abell~496. The optical counterpart of the tailed radio galaxy is LEDA~15544, at a redshift of $z = 0.03472$. Compared to the redshift of the brightest cluster galaxy (BCG) of Abell~496, $z = 0.03273$, this difference corresponds to a line-of-sight velocity offset of $V_{\rm los} = 580$~km~s$^{-1}$. Assuming an isotropic velocity distribution, such that $V_{\rm 3D} \approx \sqrt{3},V_{\rm los}$, we infer $V_{\rm 3D} \approx 980$~km~s$^{-1}$, corresponding to $\sim1.5$ times the velocity dispersion of Abell~496, $\sigma_v = 645 \pm 80$~km~s$^{-1}$ \citep{girardi1993}. This is likely an underestimate, as the large linear extent of the tail suggests that the majority of the motion is occurring in the plane of the sky; at the same time, the ICM itself in this region could be subject to bulk motions, which could influence the extent of the radio tail. The host elliptical galaxy is located at a projected distance of $10.3' \simeq 405$~kpc from the cluster center.
\par Although this tailed radio galaxy was already known, its full extent is revealed only by the new MeerKAT and uGMRT observations. As shown in Fig.~\ref{fig:tailedA496}, the radio tail displays a wealth of complex structures. Starting from the galaxy core, the emission extends southwards, forming multiple twists and protrusions on alternating sides. At a projected distance of $\sim200$~kpc (${\sim5\arcmin}$) from the optical host, the synchrotron emission develops into a distinctive 'paperclip' shape (especially in the higher resolution MeerKAT image), which motivates the name 'Giant Paperclip' for this radio source. Hints of an additional southern loop of emission are visible at $\sim300$~kpc (${\sim7.6\arcmin}$) from the galaxy centre. Another remarkable feature is a very thin ($\sim10~{\rm kpc}\sim15\arcsec$) and long ($\sim150~{\rm kpc}\sim230\arcsec$) filament, hereafter referred to as the ``spear'', which originates near the base of the paperclip and extends toward the north-west. As shown in the bottom-left panel of Fig.~\ref{fig:tailedA496}, none of these features is clearly associated with any other individual galaxy, indicating that they are most likely part of the tailed radio galaxy itself.
\par We present a spectral index map between 650~MHz and 1280~MHz, obtained by combining uGMRT Band~4 and MeerKAT L-band observations. The data were imaged at a common angular resolution of $12\arcsec$, using uniform weighting and an identical pixel scale. The resulting spectral index map, shown in the bottom-right panel of Fig.~\ref{fig:tailedA496}, was produced by measuring the spectral index only in regions detected above $3\sigma_{\rm rms}$ at both frequencies. The map reveals a clear spectral steepening with increasing distance from the optical host, from $\alpha \sim -0.5$ near the core to $\alpha \sim -2$ along the tail, as commonly observed in tailed radio galaxies (e.g. \citealt{bruno2024tail} and references therein). In addition, some regions at the southernmost end of the tail exhibit ultra-steep spectra. Interestingly, the thin filaments forming the paperclip and the spear display relatively flatter spectral indices than the surrounding diffuse emission. Similar structures have been reported in other systems (e.g. \citealt{derubeis2025}) and may indicate local compression of the magnetic field due to the shear stretching at the interface between the radio plasma and the ICM.

\subsection{The tailed radio galaxy in ZwCl~235}\label{subsubsec:tailedz235}
The 144~MHz and 400~MHz maps reveal a very long ($190~{\rm kpc}\sim122\arcsec$) and narrow ($9~{\rm kpc}\sim5.8\arcsec$) tail of synchrotron emission east of the main cluster. Its outermost tip sits on top of the elliptical galaxy LEDA 1706877 (see Fig. \ref{fig:tailedz235}), that \citet{rykoff2016} classified as a cluster member of ZwCl~235 (based on proximity and colors -- no spectroscopic redshift is available). Given its association with an elliptical member galaxy of ZwCl 235 and its elongated shape, we classify this radio source as a tailed radio galaxy. We note that there is an unrelated radio point-source at approximately the mid-point of the tail, coinciding with a background galaxy, that contaminates the radio emission (see Fig. \ref{fig:tailedz235}, bottom left panel). The host of this tailed radio galaxy is located at a projected distance of $3' \simeq 280$~kpc from the cluster centre.
\par Following the same approach as the one adopted for the Giant Paperclip in Abell~496 (see Sect.~\ref{subsubsec:paperclip}), we combine the 144 MHz LOFAR HBA and 400 MHz uGMRT Band 4 data to produce a spectral index map of the radio galaxy. The data were imaged at a common angular resolution of $13\arcsec$, using uniform weighting and an identical pixel scale. The resulting spectral index map, shown in the bottom-right panel of Fig.~\ref{fig:tailedz235}, was produced by measuring the spectral index only in regions detected above $3\sigma_{\rm rms}$ at both frequencies. The map reveals a clear spectral steepening with increasing distance from the optical host, from $\alpha \sim -0.5$ near the core to $\alpha \sim -1.5$ along the tail, again a typical feature of tailed radio galaxies (e.g. \citealt{bruno2024tail} and references therein).

\section{Complementary information}\label{app:samplefull}

In Tab.~\ref{tab:minihalo_scalings} we report the main properties of the mini-halos and their host clusters analyzed in Sect.~\ref{subsec:discscalingrelations} and used to analyze mini-halo scaling relations. Apart from Abell~496 and ZwCl~235, that were analyzed in this work, the other clusters form the sample analyzed in \citet{giacintucci2019}, that includes confirmed mini-halos only. From their work we adopt the values of cluster mass, $M_{500}$, mini halo radio power at 1.4 GHz, P$_{1.4}$, radio power of the BCG at 1.4 GHz, P$_{BCG,1.4}$, bolometric X-ray luminosity of the cool core within the central 70~kpc, $L_{X,70}$, and temperature of the cool core within the central 70~kpc, $T_{70}$. The integrated spectral index of the mini halo radio emission is reported for many systems in \citet{giacintucci2019}. Whenever an updated measurement is available from the literature, we adopt it (corresponding references are given in the table). 

\indent For Abell~496 and ZwCl~235, we have used the results from Sect.~\ref{subsec:diffuseemission} to derive the 1.4 GHz radio power of the mini-halos. Using the total flux density of the mini-halos at the highest frequency available -- $S_{1280\text{MHz}} = 10.1\pm0.5$~mJy for Abell~496 and  $S_{400\text{MHz}} = 8.8\pm0.9$~mJy for ZwCl~235 -- and extrapolating to 1.4 GHz using the spectral indices reported in Sect.~\ref{subsec:diffuseemission}, we derive radio powers of $P_{1.4} = (0.022\pm0.002)\times10^{24}$~W/Hz for Abell~496 and  $P_{1.4} = (0.013\pm0.003)\times10^{24}$~W/Hz for ZwCl~235. These are the least powerful confirmed mini-halos known so far. Using the {\it Chandra} data presented in \citet{Ubertosi2024c} and \citet{ubertosi23AA}, we estimated X-ray bolometric luminosities within 70~kpc of $L_{X,70} = (1.2\pm0.1)\times10^{44}$~erg/s for Abell~496 and $L_{X,70} = (1.0\pm0.1)\times10^{44}$~erg/s for ZwCl~235. We consider the mass M$_{500}$ from \citet{planck_2016_IIDatarelease} for Abell~496, while for ZwCl~235 we use the mass reported in \citet{piffaretti2011,mcxc_ii} (see Tab.~\ref{tab:mimics}). We also derive the radio power of the AGN in the BCG at 1.4 GHz. For Abell~496, we used the total flux density of the radio galaxy at 1.28 GHz as determined in Sect.~\ref{subsec:diffuseemission}, $S_{1280\text{MHz}} = 125.5\pm6.3$~mJy, and a global spectral index at GHz frequencies of $\alpha = -1.5$ \citep{Ubertosi2024c}, to estimate $P_{BCG,1.4} = (2.9\pm0.3)\times10^{23}$~W/Hz. For ZwCl~235, the integrated AGN spectral index of $-0.68\pm0.03$ (Sect.~\ref{subsec:diffuseemission}) estimated between 54~MHz and 400~MHz and the flux density at 400~MHz of $S_{400\text{MHz}} = 70.1\pm7.0$~mJy give a 1.4 GHz flux density of $S_{1.4\text{GHz}} = 30.3\pm4.1$~mJy. This is underestimating the total flux density of the BCG at 1.4~GHz (49.4~mJy, \citealt{Hogan15b}) due to the emergence of a flat-spectrum core-component that our low-frequency modeling does not account for. We thus use a 1.4 GHz flux density of $S_{1.4\text{GHz}} = 49.4\pm1.5$~mJy and a typical AGN spectral index of $\alpha = - 0.7$ to estimate a radio power of $P_{BCG,1.4} = (8.2\pm0.6)\times10^{23}$~W/Hz.

\begin{sidewaystable*}[ht!]
\centering
\renewcommand{\arraystretch}{1.3}
\caption{Properties of the mini-halos and host clusters used in the discussion of Sect. \ref{subsec:discscalingrelations}. }
\label{tab:minihalo_scalings}
\resizebox{\textwidth}{!}{%
\begin{tabular}{lccccccccccccccc}
\hline
Cluster &$M_{500}^{(\ast)}$ &$P_{1.4}$ &$\alpha$ &Ref. ($\alpha$) &$P_{BCG,1.4}$ &$L_{X,70}$ &$T_{70}$ & R$_{\rm MH}$ & $\log$J$_{R_{\rm MH}}^{R}$ & $\log$J$_{R_{\rm MH}}^{X}$\\
 &($10^{14}\,M_\odot$) &($10^{24}$ W Hz$^{-1}$) &&&($10^{24}$ W Hz$^{-1}$) &($10^{44}$ erg s$^{-1}$) &(keV) & (kpc) & (erg~s$^{-1}$~cm$^{-3}$~Hz$^{-1}$) & (erg~s$^{-1}$~cm$^{-3}$) \\
\hline
Phoenix & $12.6 \pm 2.0$ & $9.6 \pm 3.1$ & $-0.95 \pm 0.10$ & (1) & $73.9 \pm 7.4$ & $62.6 \pm 6.5$ & $8.8 \pm 0.4$ & 230 & $-40.19 \pm 0.14$ & $-26.14 \pm 0.05$\\
RX~J1347.5$-$1145 & $10.6 \pm 0.8$ & $26.7 \pm 2.1$ & $-0.98 \pm 0.05$ & (2) & $22.7 \pm 1.1$ & $40.9 \pm 4.3$ & $9.8 \pm 0.4$ & 320 & $-40.18 \pm 0.03$ & $-26.85 \pm 0.05$\\
RX~J1504.1$-$0248 & $7.0 \pm 0.6$ & $2.7 \pm 0.14$ & $-1.24 \pm 0.10$ & (3) & $5.7 \pm 0.3$ & $31.3 \pm 3.3$ & $4.9 \pm 0.3$ & 140& $-40.10 \pm 0.02$ & $-25.77 \pm 0.05$
\\
RBS~797 & $6.3 \pm 0.6$ & $2.2 \pm 0.24$ & $-1.02 \pm 0.05$ & (4) & $7.6 \pm 0.4$ & $18.9 \pm 1.9$ & $5.3 \pm 0.1$ & 120 & $-39.99 \pm 0.05$ & $-25.97 \pm 0.04$\\
Abell~1835 & $8.5 \pm 0.6$ & $1.19 \pm 0.25$ & -- & -- & $6.3 \pm 0.3$ & $16.8 \pm 1.7$ & $5.6 \pm 0.1$ & 240& $-41.16 \pm 0.09$ & $-26.82 \pm 0.04$\\
RXJ1532.9+3021 & $4.7 \pm 0.6$ & $3.35 \pm 0.17$ & $-1.20 \pm 0.03$ & (5) & $7.0 \pm 0.4$ & $15.9 \pm 1.6$ & $4.5 \pm 0.1$ & 100& $-39.57 \pm 0.02$ & $-25.78 \pm 0.04$\\
ZwCl~3146 & $6.7 \pm 0.8$ & $1.40 \pm 0.30$ & $-1.00 \pm 0.20$ & (5) & $0.8 \pm 0.2$ & $13.8 \pm 1.4$ & $4.7 \pm 0.1$ & 90& $-39.81 \pm 0.09$ & $-25.78 \pm 0.04$\\
Abell~2204 & $8.0 \pm 0.4$ & $0.54 \pm 0.05$ & -- & -- & $3.7 \pm 0.2$ & $13.1 \pm 1.3$ & $5.1 \pm 0.1$ & 50& $-39.45 \pm 0.04$ & $-25.14 \pm 0.04$\\
MACS~J0159.8$-$0849 & $6.9 \pm 1.0$ & $1.4 \pm 0.14$ & -- & -- & $20.3 \pm 0.1$ & $9.6 \pm 1.0$ & $5.9 \pm 0.3$ & 90& $-39.81 \pm 0.04$ & $-25.95 \pm 0.05$\\
MACS~J0329.6$-$0211 & $4.9 \pm 0.7$ & $2.84 \pm 0.30$ & -- & -- & $2.8 \pm 0.1$ & $9.5 \pm 1.0$ & $4.3 \pm 0.3$ & 70& $-39.17 \pm 0.05$ & $-25.65 \pm 0.05$\\
Abell~3444 & $7.6 \pm 0.5$ & $2.4 \pm 0.2$ & $-1.00 \pm 0.10$ & (6) & $0.28 \pm 0.01$ & $8.9 \pm 0.9$ & $5.0 \pm 0.2$ & 120& $-39.95 \pm 0.04$ & $-26.22 \pm 0.04$ \\
MS~1455.0+2232 & $3.5 \pm 0.4$ & $1.75 \pm 0.23$ & $-1.00 \pm 0.10$ & (7), (12) & $0.96 \pm 0.05$ & $8.6 \pm 0.9$ & $4.6 \pm 0.1$ & 120& $-40.08 \pm 0.06$ & $-26.13 \pm 0.05$\\
Abell~478 & $7.1 \pm 0.4$ & $0.32 \pm 0.06$ & -- & -- & $0.6 \pm 0.03$ & $7.8 \pm 0.9$ & $5.2 \pm 0.5$ & 160& $-41.20 \pm 0.08$ & $-26.60 \pm 0.05$\\
Abell~2029 & $6.8 \pm 0.2$ & $0.28 \pm 0.04$ & -- & -- & $7.4 \pm 0.4$ & $6.3 \pm 0.6$ & $6.6 \pm 0.1$ & 270& $-41.94 \pm 0.06$ & $-27.27 \pm 0.04$\\
RX~J1720.1+2638 & $6.3 \pm 0.4$ & $5.33 \pm 0.32$ & $-1.00 \pm 0.10$ & (5), (8) & $0.50 \pm 0.02$ & $6.2 \pm 0.6$ & $5.1 \pm 0.1$ & 140& $-39.80 \pm 0.03$ & $-26.51 \pm 0.04$\\
Abell~2667 & $6.8 \pm 0.5$ & $1.1 \pm 0.1$ & $-1.00 \pm 0.20$ & (9) & $2.1 \pm 0.1$ & $6.2 \pm 0.7$ & $4.6 \pm 0.2$ & 70& $-39.58 \pm 0.04$ & $-25.83 \pm 0.05$\\
RX~J2129.6+0005 & $4.2 \pm 0.6$ & $0.40 \pm 0.03$ & $-1.20 \pm 0.10$ & (9) & $3.1 \pm 0.2$ & $5.5 \pm 0.5$ & $5.1 \pm 0.2$ & 80& $-40.20 \pm 0.03$ & $-26.01 \pm 0.04$\\
Abell~S780 & $7.7 \pm 0.6$ & $1.7 \pm 0.2$ & $-1.20 \pm 0.20$ & (9) & $17.0 \pm 1.0$ & $5.9 \pm 0.6$ & $3.9 \pm 0.1$ & 50& $-38.96 \pm 0.05$ & $-25.54 \pm 0.04$\\
PSZ1~G139.61+24.20 & $7.1 \pm 0.6$ & $0.13 \pm 0.02$ & $-1.33 \pm 0.08$ & (9), (10) & $\leq0.01$ & $5.6 \pm 0.6$ & $6.2 \pm 0.5$ & 50& $-40.07 \pm 0.07$ & $-25.44 \pm 0.05$\\
Perseus & $6.1 \pm 0.6$ & $2.18 \pm 0.11$ & $-1.21 \pm 0.05$ & (5) & $13.4 \pm 0.1$ & $4.4 \pm 0.5$ & $4.1 \pm 0.1$ & 130& $-40.09 \pm 0.02$ & $-26.67 \pm 0.05$\\
Abell~907 & $5.2 \pm 0.5$ & $0.9 \pm 0.2$ & -- & -- & $4.1 \pm 0.4$ & $2.9 \pm 0.3$ & $4.8 \pm 0.1$ & 65 &$-39.57 \pm 0.10$ & $-26.12 \pm 0.04$\\
2A~0335+096 & $2.3 \pm 0.3$ & $0.06 \pm 0.01$ & $-1.20 \pm 0.10$ & (11) & $0.058 \pm 0.003$ & $2.3 \pm 0.2$ & $2.3 \pm 0.1$ & 70& $-40.85 \pm 0.07$ & $-26.26 \pm 0.04$\\
Ophiuchus & $12.4 \pm 1.2$ & $0.11 \pm 0.02$ & $-1.20 \pm 0.10$ & (12) & $0.064 \pm 0.003$ & $3.0 \pm 0.4$ & $8.8 \pm 1.1$ & 250& $-42.24 \pm 0.08$ & $-27.45 \pm 0.06$\\
\hline
Abell~496 & $2.7 \pm 0.2$ & $0.022 \pm 0.002$ & $-1.42 \pm 0.18$ & This work & $0.29 \pm 0.03$ & $1.2 \pm 0.1$ & $3.0 \pm 0.2$ & 70& $-41.29 \pm 0.04$ & $-26.55 \pm 0.04$\\
ZwCl~235 & $2.8 \pm 0.3$ & $0.013 \pm 0.003$ & $-1.85 \pm 0.03$ & This work & $0.82 \pm 0.06$ & $1.0 \pm 0.1$ & $2.7 \pm 0.1$ & 65& $-41.41 \pm 0.11$ & $-26.54 \pm 0.04$\\
\hline
\end{tabular}}
\tablefoot{The clusters are ordered by decreasing X-ray bolometric luminosity within 70~kpc, $L_{X,70}$. Apart from Abell~496 and ZwCl~235, the other clusters form the sample analyzed in \citet{giacintucci2019}. From their work, we adopt the values of cluster mass, $M_{500}$, mini halo radio power at 1.4 GHz, P$_{1.4}$, radio power of the BCG at 1.4 GHz, P$_{BCG,1.4}$, bolometric X-ray luminosity of the cool core within the central 70~kpc, $L_{X,70}$, temperature of the cool core within the central 70~kpc, $T_{70}$, and average radius of the mini-halo, R$_{\rm MH}$. The integrated spectral index of the mini halo radio emission is reported for many systems in \citet{giacintucci2014,giacintucci2019}, and whenever updated or reference measurements are available from the literature, we adopt those values (references are given in the fifth column). In the last two columns, we also report the radio and X-ray average emissivities within the volume encompassed by ${\rm R_{\rm MH}}$, derived as explained in Sect.~\ref{subsec:discscalingrelations} and Appendix \ref{app:sub:scalingxraylum}. 

$^{\ast}$: The masses M$_{500}$ (see \citealt{giacintucci2019}) are from \citealt{planck_2016_IIDatarelease} whenever available. Exceptions are: MACS J0329.6–0211, RX J1532.9+3021, Ophiuchus, Perseus, and MS 1455.0+2232, for which the mass is estimated based on scaling relations with the core-excised temperature in \citet{giacintucci2017};  Phoenix, for which the mass is from \citet{mcdonald2012nature}; and ZwCl~235, for which the mass is from MCXC \citep{piffaretti2011}. References: (1) \citet{timmerman2021}, (2) \citet{hoang2025}, (3) \citet{giacintucci2011}, (4) \citet{bonafede2023}, (5) \citet{giacintucci2014}, (6) \citet{giacintucci2024}, (7) \citet{riseley2022}, (8) \citet{biava2021_minihalo}, (9) \citet{giacintucci2019}, (10) \citet{savini2018}, (11) \citet{ignesti2022}, (12) \citet{giacintucci2025}.}
\end{sidewaystable*}

\subsection{Rescaling of the X-ray luminosity to the mini-halo radius and average emissivity calculation}\label{app:sub:scalingxraylum}

For a consistent comparison of the non-thermal radio power of mini-halos with the X-ray luminosity of the thermal gas (see Sect. \ref{subsec:discscalingrelations}) among systems with different mini-halo sizes, we rescaled the X-ray luminosity measured within a fixed aperture of 70~kpc, $L_{X,70}$, to the mini-halo radius, $R_{\mathrm{MH}}$. This was done considering that the X-ray surface brightness distribution of the cool-core clusters in our sample follows a double-$\beta$ model. In this case, the surface brightness is expressed as the sum of two components:

\begin{equation}
S(r) = S_1 \left[1 + \left(\frac{r}{r_{c_{1}}}\right)^{2} \right]^{-3\beta_1 + \frac{1}{2}}
+ S_2 \left[1 + \left(\frac{r}{r_{c_{2}}}\right)^{2} \right]^{-3\beta_2 + \frac{1}{2}} \, ,
\end{equation}

\noindent The projected luminosity within a radius $R$ is proportional to:

\begin{equation}
L(<R) \propto \int_0^R S(r)\, 2\pi r\, dr \, .
\end{equation}

\noindent Therefore, the X-ray luminosity within the mini-halo radius can simply  be obtained by scaling $L_{X,70}$ as:

\begin{equation}
L_{X,\mathrm{MH}} = L_{X,70} \times 
\frac{\int_0^{R_{\mathrm{MH}}} S(r)\, r\, dr}
{\int_0^{70\,\mathrm{kpc}} S(r)\, r\, dr} \, .
\end{equation}

\noindent This scaling was applied to the clusters considered in Sect.~\ref{subsec:discscalingrelations} using double-$\beta$-model parameters from the literature. Specifically, we considered the values reported in \citet{giacintucci2017} for RX~J1504.1-0248, Abell~3444, Abell~S780, PSZ1~G139.61+24.20, Phoenix; in \citet{1999ApJ...517..627M} for Perseus, Ophiuchus, Abell 2029, Abell~478, Abell~496; in \citet{shitanishi2018} for MACS~J0329.6-0211, RX~J1532.9+3021, Abell~2204, Abell~1835, RX~J1347.5-1145, RX~J2129.6+0005; in \citet{lancaster2011} for ZwCl~3146; in \citet{landry2013} for RX~J1720.1+2638, MS1455.0+2232; in \citet{Doria12} for RBS~797; in \citet{ubertosi23AA} for ZwCl~235; in \citet{mazzotta2003} for 2A~0335+096; in \citet{vikhlinin06} for Abell~907; in \citet{yu2011} for Abell~2667; in \citet{kotov2005} for MACS~J0159.8-0849. Subsequently, the X-ray average emissivities were computed by dividing the luminosities by the spherical volume encompassed by ${R_{\rm MH}}$:
\begin{equation}
    J_{\rm R_{\rm MH}}^{X} = \frac{L_{X,\mathrm{MH}}}{\frac{4}{3}\pi {\rm R_{\rm MH}^{3}}}\,\,
\end{equation}
\noindent Similarly, we computed the radio average emissivity by dividing the mini-halo radio powers at 1.4 GHz reported in Tab.~\ref{tab:minihalo_scalings} by the spherical volume encompassed by ${R_{\rm MH}}$:
\begin{equation}
    J_{\rm R_{\rm MH}}^{R} = \frac{P_{1.4}}{\frac{4}{3}\pi {\rm R_{\rm MH}^{3}}}\,\,
\end{equation}
The resulting radio and X-ray average emissivities are reported in the last two columns of Tab.~\ref{tab:minihalo_scalings}.

\end{appendix}

\end{document}